\documentclass[fleqn,usenatbib]{aa}  

\usepackage[switch]{lineno}  
\usepackage{graphicx}
\usepackage{txfonts}
\usepackage{orcidlink}
\usepackage{multicol}
\usepackage{float}

\title
{Dark from light (DfL): Inferring halo properties from luminous tracers with machine learning trained on cosmological simulations}

\subtitle
{I. Method, proof of concept \& preliminary testing}

\titlerunning
{Dark from Light (DfL): Method}

\author
{Asa F. L. Bluck\inst{1}\orcidlink{000-0001-6395-4504}, Joanna M. Piotrowska\inst{2}\orcidlink{0000-0003-1661-2338}, Paul Goubert\inst{1}, Roberto Maiolino\inst{3,4,5}\orcidlink{0000-0002-4985-3819}, Camilo Casimiro\inst{1}, \\Thomas Pinto Franco\inst{1} and Nicolas Cea\inst{1}}

\authorrunning
{Asa F. L. Bluck et al.}

\institute
{Stocker AstroScience Center, Department of Physics, Florida International University, 11200 SW 8th Street, Miami, FL, USA
\and
Cahill Center for Astronomy and Astrophysics, California Institute of Technology, Pasadena, CA, USA
\and
Kavli Institute for Cosmology, University of Cambridge, Madingley Road, Cambridge, CB3 0HA, UK
\and
Cavendish Laboratory Astrophysics Group, University of Cambridge, 19 JJ Thomson Avenue, Cambridge, CB3 0HE, UK
\and
Department of Physics and Astronomy, University College London, Gower Street, London WC1E 6BT, UK
\\ \email{abluck@fiu.edu}}

\begin{document}
%\linenumbers

% ABSTRACT:

\abstract
{We present Dark from Light (DfL) - a novel method to infer the dark sector in wide-field galaxy surveys, leveraging a machine learning approach trained on contemporary cosmological simulations. The aim of this algorithm is to provide a fast, straightforward, and accurate route to estimating dark matter halo masses and group membership in wide-field spectroscopic galaxy surveys. This approach requires a highly limited number of input parameters (R.A., Dec., $z$, and stellar mass plus uncertainties) and yields full probability distribution functions for the output halo masses of galaxies, groups, and clusters. To achieve this, we train a series of Random Forest (RF) regression models on the IllustrisTNG and EAGLE simulations at $z = 0 - 3$, which provide model-dependent mappings from luminous tracers to dark matter halo properties. We incorporate the individual regression models into a virial group-finding algorithm ({\small DfL}), which outputs halo properties for observational-like input data. We test the method at $z$ = 0 - 2 for both the EAGLE and IllustrisTNG models, as well as in a cross-validation mode (where one simulation is used to train the model and the other to test). We demonstrate that known halo masses can be recovered with a mean systematic bias of $\langle b \rangle = \pm 0.10\,$dex (resulting from simulation choice), a mean statistical uncertainty of $\langle \sigma \rangle = 0.12 \,$dex across epochs, and a central - (core) satellite classification accuracy of 96\%. We establish that this approach yields superior halo mass recovery to standard abundance matching applied to groups identified through a friends-of-friends algorithm. Additionally, we compare the outputs of {\small DfL} to observational constraints on the $M_* - M_{\rm Halo}$ relation from strong gravitational lensing at $z \sim 0$, demonstrating the promise of this novel approach. Finally, we systematically quantify how {\small DfL} performs on observational-like input data with varying stellar mass uncertainty and spectroscopic incompleteness, enabling robust error calibration in applications with observational galaxy surveys.}

\keywords{Galaxies: formation, evolution, environment; Dark Matter; Large Scale Structure}

\maketitle

`

\section{Introduction}
\label{sec1}

\noindent Dark matter remains a mystery in large part because, whatever its ultimate physical nature turns out to be, it does not interact with the electromagnetic field (e.g., \citealt{Frenk1988, Efstathiou1990, Navarro1996}). Even with the advent of gravitational wave and neutrino astronomy (e.g., \citealt{Mannheim2000, Abbott2016, Abbott2017, Ansoldi2018}), the electromagnetic spectrum provides the vast majority of information we have on the Universe. As such, looking for ways to infer the presence and properties of dark matter using luminous tracers has a long history (e.g., \citealt{Zwicky1933, Refsdal1964, Sunyaev1972, Rubin1978, Walsh1979, Hu1995, Warren2006, Xue2008, Cappellari2006, Cappellari2013}).

Today, the existence of dark matter is almost beyond serious scientific dispute, due primarily to precision constraints from the power spectrum of the cosmic microwave background (CMB; see \citealt{Spergel2003, Spergel2007, Komatsu2011, PLANCK2014, PLANCK2016, PLANCK2020}) and strong gravitational lensing measurements in galaxy clusters (e.g., \citealt{Turner1984, Schneider1992, Bolton2006}). Despite its elusive nature, dark matter provides the backdrop to modern cosmology and galaxy formation theory, driving the growth of cosmic structure and dictating the underlying environment in which baryons coalesce to become galaxies (see, e.g., \citealt{Cole2000, Springel2005, Bower2006, Bower2008, Henriques2015, Vogelsberger2014a, Vogelsberger2014b, Schaye2015, Nelson2018, Pillepich2018, Dave2019}).

The most accurate method to measure the total mass content in high mass groups and clusters is through strong gravitational lensing (e.g., \citealt{Turner1984, Schneider1992, Bolton2006, Hoekstra2013, Vegetti2014, Atek2018}). This approach has the least assumptions, primarily just general relativity (which is now tested to extreme precision, see \citealt{Shapiro1964, Bertotti2003, Everitt2011, Abbott2016}). Additionally, leveraging the assumption of virialization of galaxies and/or groups and clusters, in conjunction with Newtonian gravity (which is an excellent approximation of general relativity on these scales), a host of alternative dynamical methods have been developed. 

In disk galaxies, the rotation curve can be used to infer the mass within a given radius (e.g., \citealt{Rubin1978, Begeman1989, Persic1996, deBlok2008}). Similar techniques can also be used leveraging velocity dispersion in spheroids (e.g., \citealt{Erb2006, Auger2010, Wolf2010, Cappellari2013}). In groups and clusters, the satellite velocity dispersion can be used to trace the total mass of the system (e.g., \citealt{Hickson1992, Carlberg1996}). Additionally, measurements of the temperature of the circum-galactic medium (CGM) or the intra-cluster medium (ICM) from X-ray observations may be used to infer mass (e.g., \citealt{Evrard1996, Giodini2009, Pratt2010}). Furthermore, there are also methods to infer cluster masses from observations of the cosmic microwave background (CMB), leveraging the Sunyaev - Zeldovich effect (SZ effect: \citealt{Sunyaev1972, Carlstrom2002, Bocquet2019}).\footnote{Note that the above is by no means a complete list of methods to infer dynamical masses.}

Given the litany of dynamical techniques outlined above, one might wonder if there is any need for additional approaches. However, the need for an alternative approach is a simple consequence of all of the above techniques being limited in their application. For many science goals in extragalactic astrophysics one requires {\it complete} halo mass estimates for large galaxy surveys, which rules out the use of the above techniques (even in tandem). Ultimately, all of the above techniques are highly expensive - both in terms of observational time required on premier facilities and in terms of funding. These issues are further exacerbated by attempting to constrain dark matter masses at early cosmic times. At best, one can hope to obtain dynamical constraints on dark matter halo masses for a small subset of objects within a wide-field galaxy survey, and frequently none at all for high-$z$ surveys.

As a consequence of the above limitations, an entire industry of techniques has emerged over the past two decades to approximate the halo masses (and more generally the dark sector) in wide-field galaxy surveys. The two main variants are: (i) halo occupation distributions (HOD; see \citealt{Benson2000, Peacock2000, Berlind2002, Zheng2007, Croton2007}); and (ii) halo abundance matching (HAM; see \citealt{Vale2004, Kravtsov2004, Conroy2006, Moster2010, Guo2010, Behroozi2010, Trujillo2011, Klypin2011, Klypin2015}).

The first of these techniques, HOD, functions by constructing the two point correlation function around each galaxy in the survey as a function of physical (or comoving) distance, and matching this to the dark matter two point correlation function from a dark matter only simulation, analytical approximation, or gravitational lensing survey. This process is complicated by the need to assume (or infer) a bias - allowing for luminous and dark matter structures to have different clustering properties (see \citealt{Benson2000, Croton2007} for discussions). The bias must be learned either from simulations, which historically have been primarily semi-analytic models, or else from high quality gravitational lensing surveys matching the relevant parameter space (e.g., \citealt{Mandelbaum2006, Mandelbaum2016, Mandelbaum2018}). The difficulty with the latter approach is in obtaining strong constraints on the low-mass end of the relationship between luminous and dark matter, as well as challenges with extending to high-redshifts. Alternatively, the primary limitation with the former approach is the inevitability of incurring model bias in the halo estimation when using simulations.

The second of these techniques, HAM, essentially just matches the number density of galaxy stellar mass to the number density of halo mass, the latter historically taken from a dark matter only N-body simulation. This approach relies on the simple assumption that the most massive galaxies reside in the most massive haloes (see \citealt{Vale2004, Conroy2006, Moster2010}). The basic HAM technique has been improved with Sub-Halo Abundance Matching (SHAM; e.g., \citealt{Hearin2013, Chaves2016, Campbell2018}). This approach leverages information on the satellite population (in addition to centrals), using group stellar masses to constrain dark matter content through a similar number density matching assumption. In recent years, a further enhancement has been constructed, referred to as Conditional Abundance Matching (CAM; see \citealt{Hearin2014, Lehmann2017}). This approach utilizes parameters in addition to stellar mass to further constrain dark matter properties, usually in conjunction with Semi Analytic Models (SAMs). 

In addition to the rise of halo abundance matching and halo occupation distribution techniques, some authors have sought to directly obtain a simple mapping from stellar mass of galaxies (and/or groups and clusters) to dark matter halo masses leveraging the $M_{\rm Halo} - M_*$ relationship in SAMs (see, e.g., \citealt{Yang2007, Yang2009}). The approach of this paper is qualitatively most similar to this technique, but applied here to cosmological hydrodynamical simulations, incorporating machine learning to enable multi-parameter dark matter constraints within an iterative group finding algorithm.

Perhaps the most important theoretical contribution to galaxy formation over the past decade has been the emergence of cosmological hydrodynamical simulations (e.g., \citealt{Vogelsberger2014a, Vogelsberger2014b, Schaye2015, Nelson2018, Pillepich2018, Dave2019}). These simulations simultaneously model the gravitational and hydrodynamical physics of galaxy formation and evolution within a given background cosmology. It is natural, therefore, to improve both the HOD and (C/SH/H)AM techniques by leveraging these simulations, and several examples of this now exist in the literature (see, e.g., \citealt{Khandai2015, Artale2018, Cochrane2018}). 

Nonetheless, the basic assumptions underpinning both halo occupation distributions and abundance matching still exist in these approaches. In the case of abundance matching the essential assumption is that there is a one-to-one mapping between the number densities of either galaxies, groups, or a sub-set thereof and dark matter haloes. In the case of halo occupation distributions the essential assumption is that there is a simple bias which can be utilized to bring the halo and galaxy density distributions into accord. Yet, important open questions remain: Are these assumptions correct? And, to what extent are the halo properties derived from these approaches biased by their modeling choices? 

These questions have been brought into sharp focus by \cite{Hadzhiyska2020}, with numerous solutions proposed in \cite{Hadzhiyska2021, Hadzhiyska2022, Hadzhiyska2023a, Hadzhiyska2023b}. In summary, different galaxy clustering is expected for emission line and red galaxies in cosmological simulations, which brings into question the standard HOD approach. One solution is to use different biases for different classes of galaxies. In some ways this approach is similar to CAM in that the mapping between luminous tracers and the dark sector can be made more accurate by adding dependencies upon galaxy properties or classes, which can be carefully controlled for in the methodology. However, one potential issue with these approaches is that they impede the possibility of establishing independent connections between these properties and dark haloes in observations, since they are assumed in advance. As such, one of the first steps in the analysis of this work is to investigate precisely which parameters are effective at estimating halo masses in cosmological simulations and limit our mapping to only those which are essential. Ultimately, this is a trade-off between accuracy and independence in the output halo catalog.

In parallel with the theoretical advances leading to cosmological hydrodynamical simulations, a revolution in machine learning is now underway (e.g., \citealt{Pedregosa2011, Brink2013, Baron2019}), with many applications to extragalactic astrophysics (e.g., \citealt{Banerji2010,  Mucesh2021, Bluck2022, Piotrowska2022, Remy2023, Huertas-Company2024, Ho2024}). Consequently, it is now possible to learn arbitrarily complex mappings from luminous tracers to dark matter properties, utilizing multiple input parameters, even among inter-correlated variables. Hence, one need not restrict the mapping to individual number densities of a given parameter, most frequently stellar mass (as in AM), or to the identification of one or multiple biases (as in HOD). 

Moreover, the machine learning tools may be trained on sate-of-the-art cosmological hydrodynamical simulations, enabling one to fully account for the baryon - dark matter interaction in the relationship between dark matter and luminous tracers. In conjunction, these advances remove the need for simplifying assumptions in constructing a mapping from luminous tracers to dark matter properties. One interesting example of this approach is to use the morphologies of galaxies to constrain their dark matter properties in imaging surveys (see \citealt{Hahn2024}). Our approach should be seen as complementary to this work in that it is geared towards spectroscopic surveys without the need for high-resolution imaging. One advantage of our approach is that morphology remains independent of the estimated halo masses and hence one can look for connections between halo mass and galaxy morphology without prejudicing the result.

Ultimately, the overarching goal of this work is to leverage cosmological hydrodynamical simulations and modern machine learning techniques to construct a novel approach for solving the problem of which galaxies reside in which dark matter structures. One of the main advantages of this methodology is to rigorously determine the uncertainties of the estimated halo masses, and group properties. We also aim for scalability in the halo finder algorithm, enabling straightforward application to the vast galaxy surveys of the future. Our primary motivation is to provide a fast, efficient, and accurate halo finder and halo mass estimator which can be used at cosmic noon ($z \sim 1 - 2$) in the upcoming VLT-MOONRISE spectroscopic galaxy survey (see \citealt{Maiolino2020, Cirasuolo2020}).

To this end, we present {\it Dark from Light: Dfl}\footnote{\url{https://github.com/abluck/Dark-from-Light}}. DfL is a publicly available halo finder and halo mass estimator intended for application to wide-field spectroscopic galaxy surveys. DfL utilizes random forest regression trained on contemporary cosmological simulations (particularly, EAGLE: \citealt{Schaye2015, Crain2015, McAlpine2016}; and IllustrisTNG: \citealt{Marinacci2018, Naiman2018, Nelson2018, Pillepich2018, Springel2018}). The machine learning modules are incorporated into a physically motivated virial group finder, which simultaneously identifies clustering and creates a mapping from luminous tracers to the dark sector. Special care is made to ensure that this method can be applied to a wide range of redshifts. Moreover, DfL also characterizes both the random and systematic error associated with the method (including an estimate of uncertainty from model choice). This enables robust application to a host of contemporary cosmological and astrophysical problems.

The paper is structured as follows. In Section 2 we present the simulated data used in this work from EAGLE and IllustrisTNG. In Section 3 we present a thorough description of the halo finder pipeline. Initially, we explore which luminous tracers are most effective at constraining halo mass. We then detail the training of the machine learning regression models. Finally, we present a step-by-step explanation of the full pipeline. In Section 4 we test the performance of the halo finder pipeline at $z = 0 - 2$ in EAGLE and IllustrisTNG. We also test the performance in cross-validation mode (where one model is used in training and another for testing). Moreover, we compare the results from DfL to a more conventional method (abundance matching applied to total stellar masses output from a friends-of-friends, FOF, group finding algorithm), and compare the output $z=0$ halo mass - stellar mass relationship to observational constraints utilizing strong lensing. Finally, we systematically explore how stellar mass uncertainty and survey incompleteness impact halo recovery in DfL. In Section 5 we discuss the potential of DfL applied to VLT-MOONRISE data for a host of scientific applications. The main contributions of this paper are summarized in Section~6. 

We also provide three appendices to this work. Appendix A presents a detailed users guide for the public DfL code. Appendix B presents the full details of the FOF-AM approach used for comparison to DfL in Section 4. Appendix C lists the hyper-parameters used in the random forest modules incorporated within DfL, and in the ANN modules used to ensure the RF achieves an optimal regression solution.

Unless otherwise specified, we assume a spatially flat $\Lambda$CDM background cosmology with $\Omega_M = 0.3$, $\Omega_{\Lambda} = 0.7$, and $h \equiv H_0 / (100 \, {\rm km/s/Mpc}) = 0.7$.

\section{Simulations data}
\label{sec2}

\noindent In this work we employ two cosmological simulation suites to learn redshift and model dependent mappings from luminous tracers to the dark sector. Specifically, we utilize: (i) IllustrisTNG\footnote{IllustrisTNG data access: \url{www.tng-project.org/}} (hereafter TNG; \citealt{Marinacci2018, Naiman2018, Nelson2018, Pillepich2018, Springel2018}); and (ii) EAGLE\footnote{EAGLE data access: \url{http://icc.dur.ac.uk/Eagle/}} (\citealt{Crain2015, Schaye2015, McAlpine2016}). From each simulation we extract data from snapshots at $z = 0 - 3$ for all sub-haloes with stellar mass, $M_* > 10^{9.5} M_{\odot}$. Full information on the two simulations are provided in the above references. Additionally, we have prepared a docker\footnote{Simulations docker: \url{https://hub.docker.com/u/jpiotrowska}} to aid in data access and the application of these catalogs.

Both simulation suites model a $\Lambda$CDM background cosmology tracing both gravitational and fluid dynamical physics directly. Subgrid models are utilized to implement gas cooling, star formation, stellar evolution, supernova feedback, supermassive black hole seeding and growth, active galactic nucleus (AGN) feedback, and the formation of metals, among several other properties. The principal differences between the models are due to the choice of code used to solve the hydrodynamical fluid equations (explicitly, adaptive moving mesh in IllustrisTNG vs. smoothed particle hydrodynamics in EAGLE), and most importantly, the details of the subgrid recipes used to model feedback, especially from supermassive black holes and AGN.

\subsection{IllustrisTNG}

\noindent We utilize the IllustrisTNG-100-1 simulation (\citealt{Nelson2018, Pillepich2018}), which has a box size of $\sim$(100 cMpc)$^3$ and is run using the {\small AREPO} unstructured moving mesh code \citep{Springel2010}. This code has been updated to include magnetic fields in addition to gravity and hydrodynamics. The cosmological parameters are based on \cite{PLANCK2016}. Dark matter and stars are treated as particles, with particle masses of $M_{DM} = 7.5 \times 10^6 M_{\odot}$ and $\langle M_{\rm star} \rangle = 1.4 \times 10^6 M_{\odot}$, respectively. Baryons in gas are modeled using the adaptive moving mesh Voronoi binning approach (see \citealt{Springel2010}). 

Baryons transform into stars following a simple density threshold prescription (see \citealt{Pillepich2018}), based on the empirical \cite{Kennicutt1998} law, assuming a \cite{Chabrier2003} initial mass function (IMF). Dark matter haloes and sub-haloes (including galaxies) are identified through the {\small SUBFIND} algorithm \citep{Springel2001, Dolag2009}. This approach uses a linking-length friends-of-friends algorithm to associate dark matter, stars, and gas within gravitationally connected systems. 

In this work, the principal parameters of interest are dark matter halo mass (specifically, $M_{200}$) in addition to the stellar mass of galaxies, groups, and clusters. Additionally, we also consider the number of galaxies above a stellar mass threshold associated with each group. Hence, we are primarily interested in the relationship of stellar and halo properties. The main subgrid physics which can impact the stellar mass - halo mass relationship at the relatively high masses probed in this work is AGN feedback, which we discuss below.

IllustrisTNG-100-1 seeds supermassive black holes in low-mass dark matter halos and models their evolution through Eddington-limited Bondi-Hoyle accretion \citep{Hoyle1939, Bondi1944}. This simulation incorporates three distinct modes of AGN feedback (see \citealt{Weinberger2017, Weinberger2018}): (i) quasar mode (which operates at high Eddington ratios); (ii) kinetic mode (which operates at low Eddington ratios); and (iii) radiative ionization (operational at all Eddington ratios). Among these, only the kinetic mode has a significant impact on quenching within the simulation (see \citealt{Weinberger2017, Zinger2020, Piotrowska2022}), and hence leaves a major imprint on the stellar - halo mass relationship, which is critical for this work.

In the kinetic mode, which is only operational at $M_{BH}~\gtrsim~10^8 M_{\odot}$, a fraction of the accreted rest energy is converted into the kinetic energy of gas cells surrounding the black hole in a stochastic fashion (see \citealt{Weinberger2017}). Although the direction of the momentum kick is random, this is constrained to be isotropic over long time scales, preserving conservation of momentum in the simulation. This kinetic feedback induces turbulence and outflows in the interstellar medium, and additionally provides long-term heating of the circum-galactic medium through gas percolation and shocks. In conjunction, kinetic mode AGN feedback shuts down star formation in the highest mass galaxies, leading to curvature in the fundamental $M_* - M_{\rm Halo}$ relationship.

\subsection{EAGLE}

\noindent We utilize the EAGLE-RefL0100N1504 simulation \citep{Schaye2015, McAlpine2016}, which has a box size of (100 cMpc)$^3$ and is run using an updated version of the GADGET-3 smoothed particle hydrodynamics (SPH) code (\citealt{Springel2005});  Dalla Vecchia et al. in prep.). This simulation adopts a cosmological model based on \cite{PLANCK2014}, assuming a spatially flat $\Lambda$CDM background universe, dominated by mass at early cosmic times and dark energy at late cosmic times. EAGLE models dark matter and baryons (in all phases) as particles with $M_{\rm DM} = 9.7 \times 10^6 M_{\odot}$ and an initial baryonic mass of $M_{\rm b} = 1.81 \times 10^6 M_{\odot}$, respectively. 

Baryons transform into stars following a metallicity based criterion with a pressure dependent star formation rate (see \citealt{Schaye2015}), which is ultimately based on the empirical \cite{Kennicutt1998} law. Dark matter haloes and sub-haloes (including galaxies) are identified through the {\small SUBFIND} algorithm \citep{Springel2001, Dolag2009}, as in TNG. As with TNG, the main subgrid physics which can impact the stellar mass - halo mass relationship at high masses is AGN feedback. Stellar and supernova feedback have a big impact on star formation for low mass galaxies, but by $M_* \sim 10^{9.5} M_{\odot}$ these effects become marginal, due to the gravitational potential of galaxies far exceeding the binding energy required to hold on to supernova ejecta. As such, we focus on the AGN feedback prescription here.

In EAGLE, AGN feedback serves as the primary mechanism for quenching massive galaxies, and hence the major driver of the $M_* - M_{\rm Halo}$ relation at high masses. Supermassive black holes are seeded in low-mass dark matter halos and grow through Eddington-limited Bondi-Hoyle accretion \citep{Hoyle1939, Bondi1944}, as in TNG. However, AGN feedback is modeled in EAGLE using a single mode, where a fraction of the accreted rest mass energy is converted into thermal energy of neighboring gas particles. Energy exchange between the black hole accretion disk and the galaxy is implemented via thermal injections in a stochastic, burst-like manner (see \citealt{Crain2015}). 

AGN feedback in EAGLE initiates quenching by driving ISM outflows, and helps maintain quiescence in massive galaxies by heating the circum-galactic medium (CGM) from shocks induced by the outflows. However, we have previously noted that the long-term quiescence of massive galaxies in EAGLE is less effective than in TNG (see \citealt{Piotrowska2022}). Nonetheless, EAGLE does provide a reasonable approximation of the multi-epoch stellar mass functions (see \citealt{Schaye2015}).

\subsection{Why two simulations?}

\noindent In principle one can train a machine learning algorithm to learn how to estimate dark matter properties from luminous tracers in a single simulation (as in, e.g., \citealt{Yang2009, Hahn2024}). We opt not to do this due to remaining uncertainty in the correct form of the baryonic physics of galaxy formation and evolution.

Gravitational physics is extremely well understood on the scales of galaxies, groups, and clusters. Additionally, the cosmological parameters governing the background expansion of the Universe are also very well constrained (e.g., \citealt{PLANCK2014, PLANCK2016, PLANCK2020}). Hence, a single model would likely be sufficient for investigating the formation of dark matter haloes in isolation. However, in this work we seek to estimate dark matter properties from luminous tracers, which are all baryonic in nature. More precisely, all of the luminous tracers in this work are of stellar origin. Hence, the baryonic physics - including gas cooling, star formation, and, crucially, feedback - is critical to the required mapping of luminous tracers to dark matter properties.

Unlike the background cosmology and gravitational physics, the physics of baryons in galaxy evolution is much less well constrained (see, e.g., \citealt{Somerville2015} for a review). To combat this issue, we elect to incorporate two successful\footnote{This is meant in the sense that both simulations accurately reproduce the observed stellar mass functions of galaxies across a wide range in cosmic time, making them highly suitable to learn a light-to-dark mapping from (see, e.g., \citealt{Schaye2015, Nelson2018, Pillepich2018}).} simulations of galaxy evolution with very different subgrid recipes and hydrodynamical solvers. This enables us to begin to quantify the impact of model choice on the inferred dark matter halo properties. Ultimately, this will yield an initial estimate of the bias induced on halo mass recovery from the choice of the (relatively) poorly constrained baryonic physics. In future work we plan to extend this framework to incorporate essentially all public cosmological simulations, enabling users to have complete flexibility with the application of DfL. In this first work in the series, we aim more modestly for a first proof of concept of this approach with two popular simulations.

\section{Methods}
\label{sec3}

\subsection{Defining the problem}

\noindent In essence we seek a mapping from observable luminous tracers, which are relatively straightforward to obtain in extant and upcoming wide-field galaxy surveys, to dark matter halo mass. To be useful, this mapping must yield accurate halo masses within some well defined random uncertainties, and have constrainable biases. Hence, schematically, we look for a mathematical relation of the form:

\begin{equation}
\{ \mathrm{Luminous \,\, Tracers} \} \rightarrow \{ \mathrm{Dark \,\, Matter \,\, Properties} \}.
\end{equation}

As a starting point, one must employ known dark matter halo masses for a representative sample of haloes, and hence galaxy, group, and cluster masses. At cosmic noon this is only possible in simulations at present. Direct measurements of halo masses (e.g., from strong gravitational lensing or density - temperature profiles) are limited to very high mass haloes at these epochs, and hence cannot be used to learn a general mapping across the full diversity of halo types. This poses an immediate problem in that any mapping will be model dependent. This furthermore implies that the method will most probably be biased relative to the correct galaxy formation model of the Universe. To combat these issues we employ two very different cosmological simulations and rigorously compare the predicted dark matter halo properties from each to ascertain how impactful the model choice is in the mapping from luminous tracers to the dark sector. 

The next issue is how to optimally learn the mapping from light to dark in a given simulation. In the simplest case of one luminous tracer, a straightforward non-linear fit could be made for each appropriate epoch and galaxy type. However, in general this is not necessarily optimal. To form a more general mapping one needs to consider multiple input parameters, which are in general inter-correlated. This leads to a highly complex mathematical problem, which is not easily tractable with conventional fitting techniques. Conversely, this problem is ideally suited to machine learning.

\subsection{A machine learning solution}

\noindent Machine learning describes the capacity of a certain class of algorithms to adapt to inputs, `learning' to solve complex problems without explicit instructions. In supervised learning, known solutions are made available to the machine learning system to aid the algorithm in assigning specific weights (or hyper-parameters) which minimize the inaccuracy of the task at hand. There are a number of machine learning algorithms which are potentially suitable to solving the mapping from luminous tracers to dark matter properties, described above. These range from deep learning with artificial neural networks to more straightforward approaches like decision trees and random forests. 

Additionally, one must specify whether classes or numbers are desired as the output. Since dark matter halo masses form a continuum, it is clear in this case that we require a numerical output and hence we must utilize some form of supervised machine learning with regression (see, e.g.,  \citealt{Pedregosa2011}).

One possibility is `deep learning' with multi-layered, fully connected artificial neural networks (ANN, e.g., \citealt{Ilbert2006, Dieleman2015, Teimoorinia2016}). This type of machine learning is highly effective at learning non-linear mappings between input data and a desired output. However, the key weakness of these techniques is interpretability. Due to multiple inputs with non-linear activation functions between layers in an ANN, it is extremely hard to reverse engineer the trained network to understand why it works. For many applications this is not especially problematic. However, if one wishes to gain insight from the machine learning technique, artificial neural networks are frequently highly limited.

In the present application, our first goal is to learn precisely which parameters are most effective at making the mapping from light-to-dark, and, crucially, limit their number as much as possible. The latter is a result of our ultimate goal to have the final dark matter halo catalogs be as independent as possible from other measurable parameters. This goal is not well suited to deep learning. Fortunately, in preliminary testing we have found that ANN is no more effective than the more transparent Random Forest technique for this particular problem. As such, we continue with this more simplistic approach. 

Moreover, the RF technique has numerous advantages over the ANN approach. First, the RF enables extraction of feature importance, which we utilize in Section 3.3 to investigate the best parameters to incorporate into DfL. Furthermore, the RF offers a natural way to extract uncertainties (via the distribution of individual tree predictions) and propagate these through the DfL pipeline efficiently (which is discussed in Section 3.5.7). Whilst there are methods which enable full PDF error propagation through ANN with Bayesian neural networks (see, e.g., \citealt{Chua2020, Hortua2023, Jones2024}), this is much more complex, slower to train and apply, and in this case not necessary to yield optimal performance.

\subsubsection{The random forest method}

\noindent Random forests are formed from a collection of decision trees with differences between them enforced through bootstrapped random sampling of the input data, and (optionally) through partial sampling of the input features. A single decision tree is formed in a deterministic manner from a set of input data, a desired output parameter, and a performance indicator to minimize. 

In our application, the input data will be luminous tracers that are known, or expected, to reveal information on the dark sector, and the output parameter will be halo mass. The performance indicator, or loss-function, is chosen to be the mean square error, explicitly:

\begin{equation}
\mathrm{MSE} \equiv \frac{1}{N} \sum_{i=1}^{N}{(Y_i - \hat{Y}_i)^2}
\end{equation}

\noindent where, $N$ is the total number of predictions made, $Y_i$ is the true value of each prediction, and $\hat{Y}_i$ is the predicted value from the random forest, or partial step in each decision tree in training. The MSE quantifies how accurate the predictions of a given quantity are, treating over- and under-estimates as equally erroneous.

Each tree in the random forest has branches structured as follows. The single variable (referred to as a feature), and boolean operation using that variable, which minimizes the MSE in the combined daughter nodes is chosen at each parent node. For instance, this may be stellar mass with a cut at $M_* > 10^{10} M_\odot$. After splitting the data with this criterion and this feature, the mean halo mass for each subset is computed, using the known halo masses in the training data, and the MSE performance is computed. For each branch the process continues until either every final (leaf) node contains just one training datum, or else a predetermined limit is reached, employed to prevent over-fitting the data. This is continued for each tree in the forest and the final predictions are taken as the average over all of the individual tree predictions.

Due to the relative transparency of the random forest architecture, it is possible to extract precisely how much value each input parameter has to solving the regression problem. This may be quantified as the relative importance of each variable as follows:

\begin{equation}
I_R(k) = \frac{1}{N_{\mathrm{trees}}} \sum_{\mathrm{trees}} \bigg( \frac{\sum_{nk} \, N(nk) \, \Delta \mathrm{MSE}\,(nk)}{\sum_{n} N(n) \, \Delta \mathrm{MSE}\,(n)} \bigg)
\end{equation}

\begin{figure*}
\begin{centering}
\includegraphics[width=0.45\textwidth]{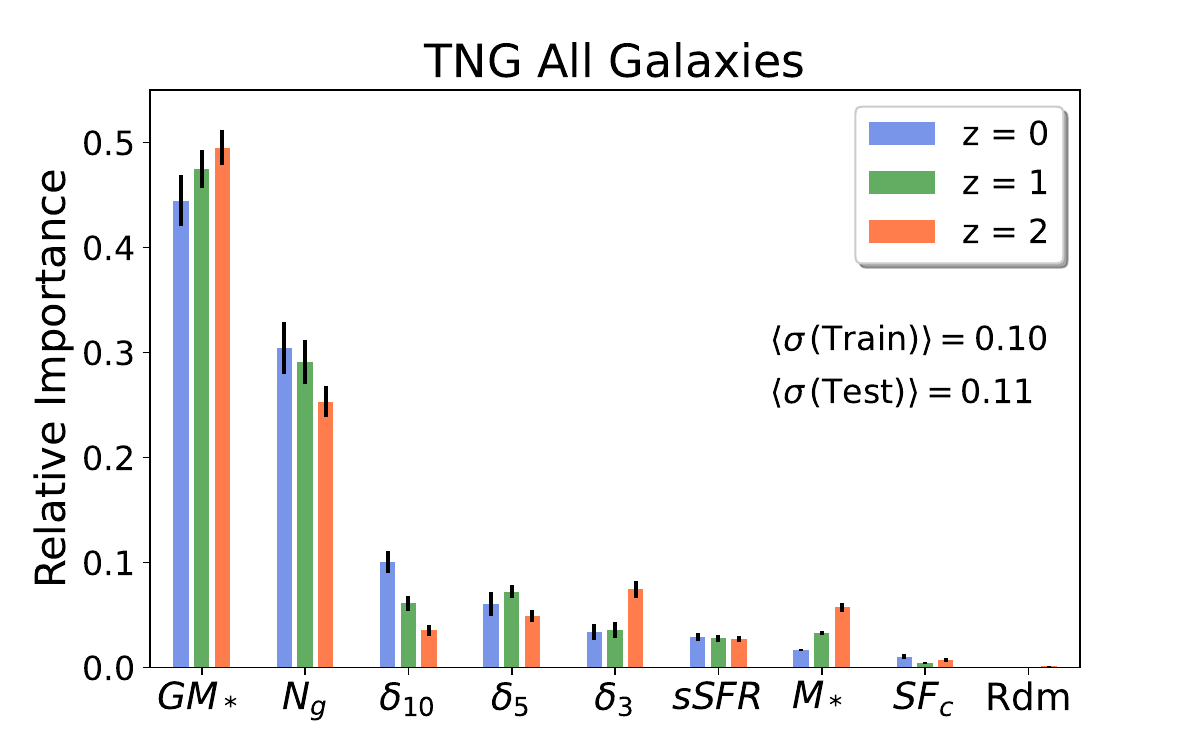}
\includegraphics[width=0.45\textwidth]{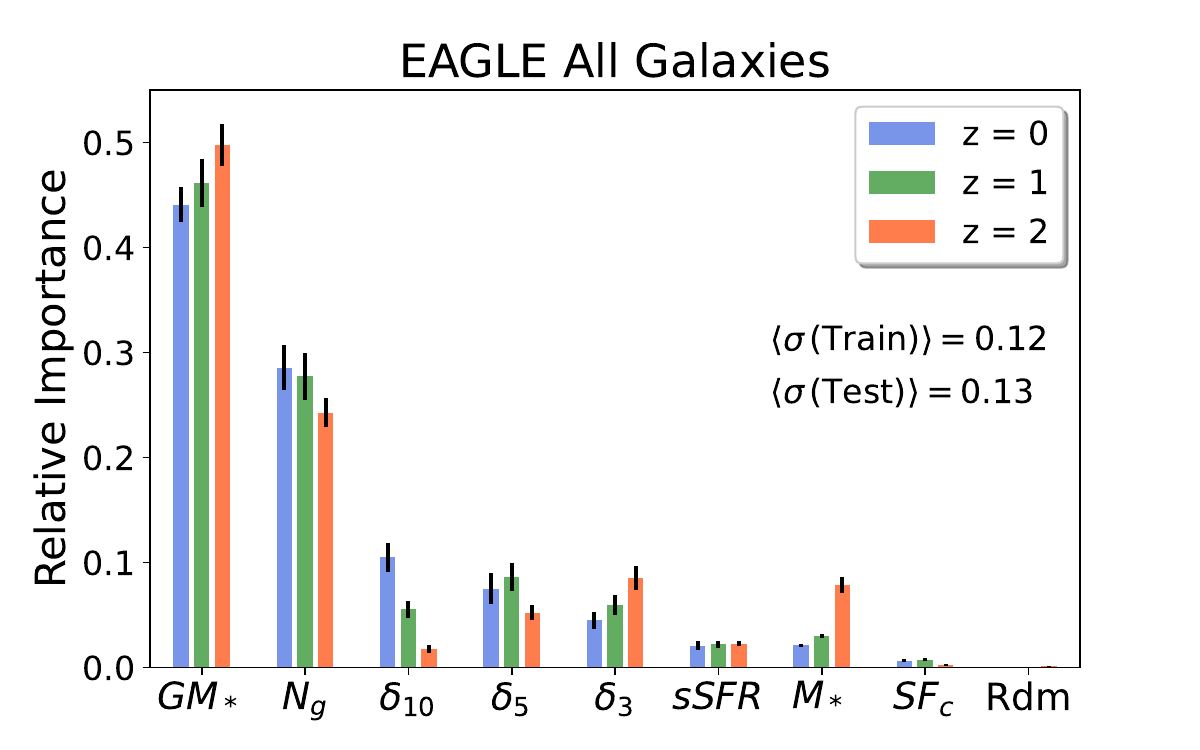}
\includegraphics[width=0.45\textwidth]{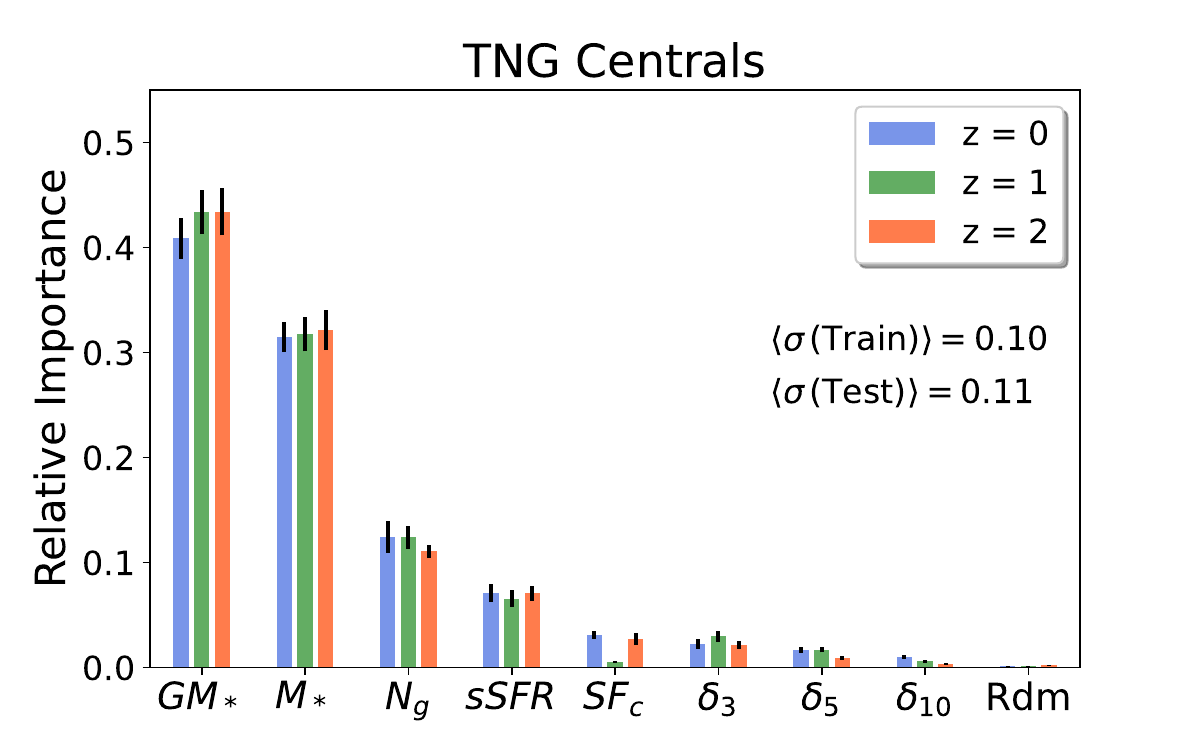}
\includegraphics[width=0.45\textwidth]{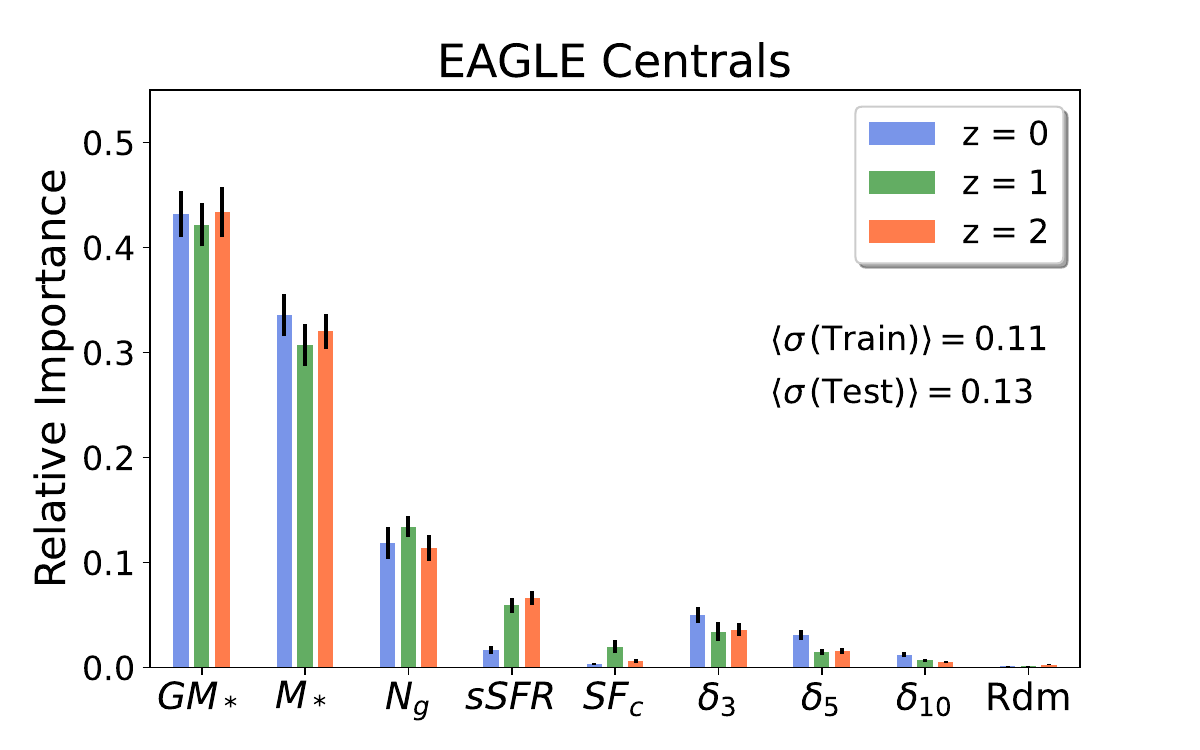}
\includegraphics[width=0.45\textwidth]{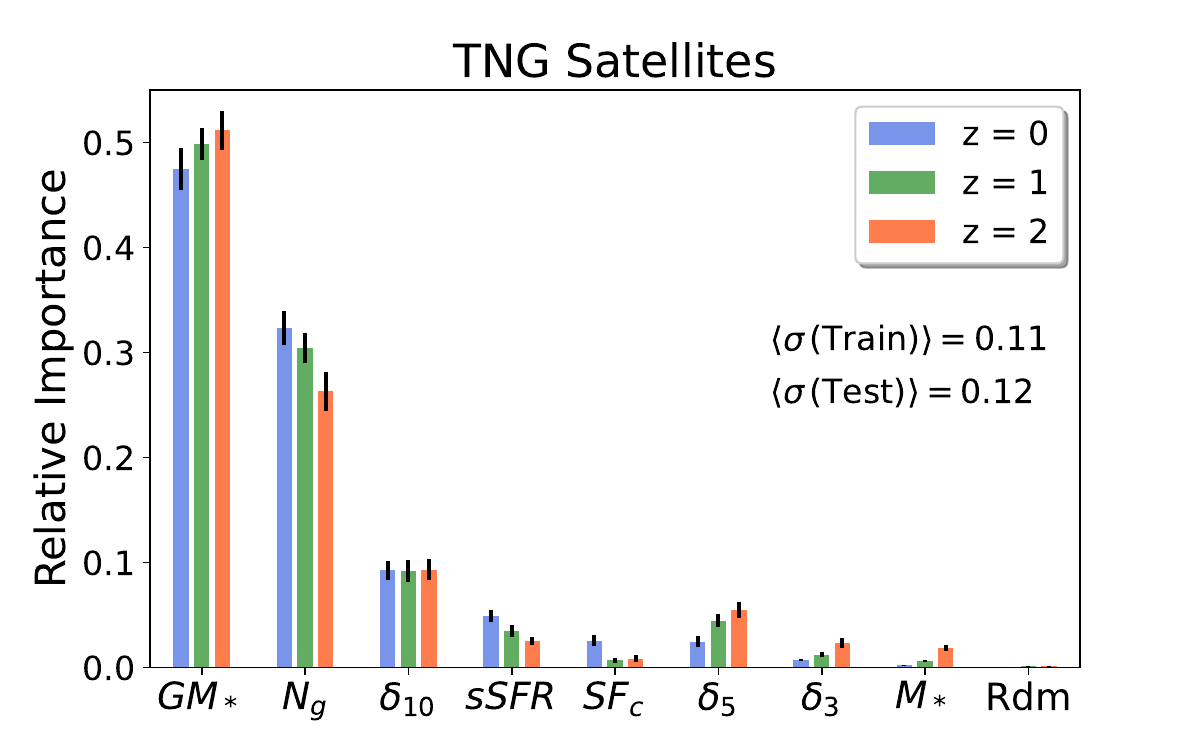}
\includegraphics[width=0.45\textwidth]{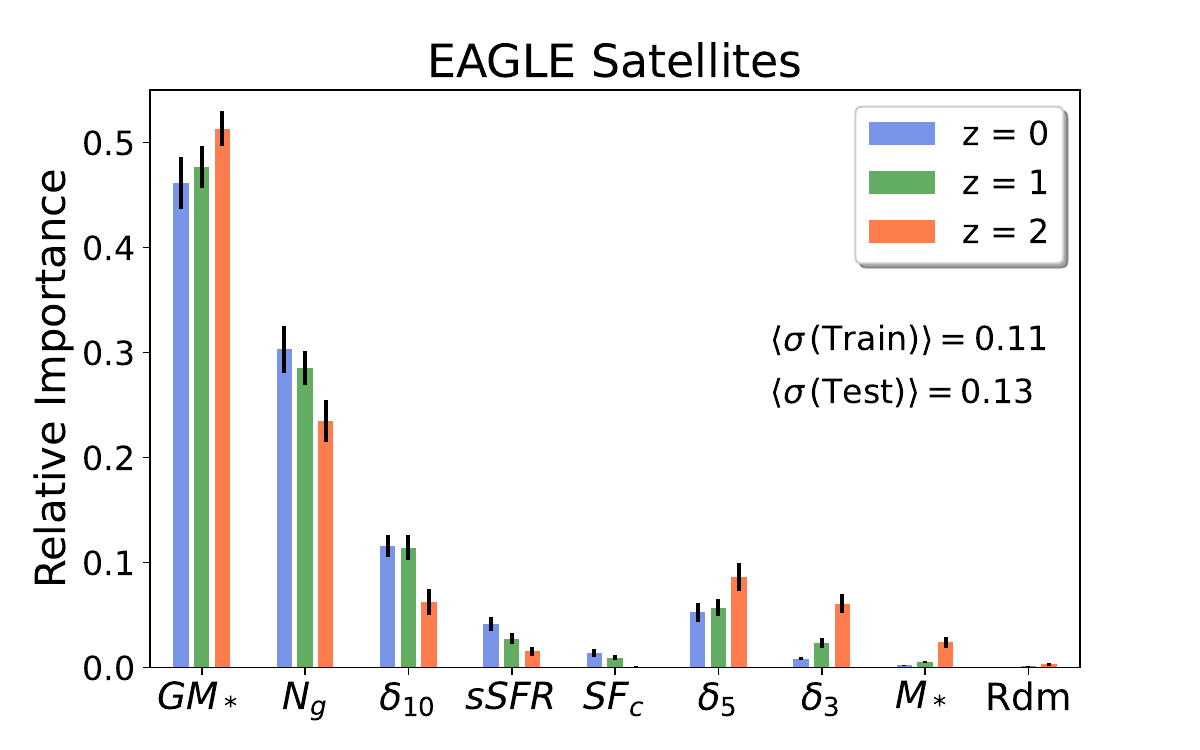}
\caption{Random forest regression analyses showing the predictive importance of various observable galaxy and environmental parameters for inferring dark matter halo mass, explicitly $M_{200}$ (as listed along the x-axis; see Section 3.3 for definitions). The left column shows results from the TNG simulation, with the right column showing results for the EAGLE simulation. Results for all galaxies are shown in the top row, results for central galaxies are shown in the middle row, and results for satellite galaxies are shown in the bottom row. On each panel, results are shown separately for three redshift snapshots ($ z = 0, 1, 2$). The uncertainties on the relative importance are computed as the standard deviation across 100 independent runs. The overall average performance of the random forest regressor ($\sigma \equiv \sqrt{\rm MSE}$) is shown on each panel for both the training and independent testing samples (each containing 50\% of the full dataset). For all populations of galaxies, and at all epochs studied, the total group stellar mass is seen to be the most predictive parameter for inferring halo masses in both EAGLE \& TNG. The second most important parameter depends on galaxy class, with the number of galaxies in the group being the second most important parameter for all galaxies and satellites, and stellar mass being the second most important parameter for centrals. All other investigated parameters (local densities, star formation tracers, etc.) are substantially less predictive than the former three parameters. Note that there is a high degree of similarity between the simulations and redshift snapshots, which implies that these results are not strongly model dependent and do not vary significantly with cosmic time.}
\label{fig1}
\end{centering}
\end{figure*}

\noindent where in the numerator the sum is made over all nodes in a given tree which utilize variable, $k$, and the denominator sums over all nodes in the tree, regardless of which features are used to make the cuts. The change in MSE ($\Delta$MSE) is weighted by the number of data which reach that individual node ($N$). This gives the importance of any given variable, $k$, to that tree. Finally, the overall importance is taken as the average importance across all of the trees within the forest.

In this work we generate our Random Forest models using {\small RandomForestRegressor} from the {\small SciKit-Learn} {\small PYTHON} package (see \citealt{Pedregosa2011}\footnote{SciKit-Learn: https://scikit-learn.org/stable/}. All hyper-parameters are listed in Appendix C to enable reproducibility of our results.

To prevent over-fitting, we systematically vary the {\it `min-samples-leaf'} (MSL) hyper-parameter to ensure an optimal regression performance in unseen validation data, but, crucially, not the training data. Following \cite{Bluck2022}, we also impose a limit on the similarity of performance between training and testing data sets. Here we require that $\Delta \sigma < 0.03$ dex (where, $\sigma \equiv \sqrt{\rm MSE}$, and MSE is defined in eq. 2). 

We also allow only partial access to features in any given tree, using the {\it `sqrt'} mode for the {\it `max-features'} hyper-parameter. Whilst this is not the optimal route to establish causality (see \citealt{Bluck2022}), it is the most accurate way to train the random forest architecture (see, e.g., \citealt{Breiman2001}), limiting the impact of erroneous values in any one feature in the overall estimation of halo masses.

\subsection{What are the best luminous tracers of halo mass?}

\noindent We start by investigating which parameters are most useful for predicting dark matter halo masses in the two simulations utilized in this work, EAGLE \& TNG. This will inform us as to the optimal parameters to incorporate into our halo finder pipeline.

In Fig. 1, we show the results from numerous random forest regression runs in the TNG (left column) and EAGLE (right column) simulations. Each random forest run predicts halo masses (explicitly, $M_{200}$) from a wide set of luminous tracers. The decision trees are trained with: stellar mass ($M_*$); group total stellar mass ($GM_*$, evaluated within $R_{200}$); the total number of galaxies in the group above $M_* > 10^{9.5}$ ($N_g$, evaluated within $R_{200}$); the galaxy over-density at the 3rd, 5th and 10th nearest neighbor (evaluated in 2D for an observational-like parameter); and two star formation tracers: sSFR (= ${\rm SFR} / M_*$), and  the star forming class (discussed below). 

The local galaxy over-densities are computed as:

\begin{equation}
\delta_N \equiv \log{(\Sigma_N)} - \langle \log{(\Sigma_N)} \rangle_{z}
\end{equation}

\noindent where,

\begin{equation}
\Sigma_N = \frac{N}{\pi D_N^2}
\end{equation}

\noindent and where $D_N$ is the physical distance to the $N$th nearest galaxy neighbor, evaluated in 2D observer-like space. We also require a velocity difference along the line of sight (i.e., perpendicular to the observer-like plane) of $| \Delta V | < 1000$ km/s. The over-density (eq. 4 above) is then taken as the difference between each galaxy's individual local surface density and the geometric mean of all surface densities in the simulation at the same redshift snapshot as the galaxy in question (as in \citealt{Bluck2020b, Goubert2024}).

Additionally, we add information on the star forming state of galaxies. We include the specific star formation rate (defined above) and the star forming class (${\rm SF}_c$), defined by identifying galaxies which are forming stars with sSFR lower than one order of magnitude of the peak of the star forming distribution at that epoch (as in \citealt{Bluck2023}). Specifically, star forming galaxies are assigned a number of zero, and quenched galaxies a number of one, with the criterion for being quenched being:

\begin{equation}
{\rm sSFR}(z) < {\rm sSFR}_{\rm peak}(z) - 1 {\rm dex}.
\end{equation}

\noindent Finally we also include a random number (Rdm) to aid in interpreting whether any importance may be spurious.

In the top row of Fig. 1 we present results for all galaxies treated together, in the middle row we show results for central galaxies, and in the bottom row we show results for satellite galaxies. Within each panel, we present the results separately for three redshift snapshots in the simulations, $z = 0, 1, 2$. The parameters are listed along the $x$-axis of each plot, ordered by how important they are for predicting halo mass in the regression analysis for that class of galaxy in TNG at $z = 0$. The $y$-axes in Fig. 1 display the relative importance of each feature, as defined in eq. 3 above. This quantifies which parameters are most effective for predicting dark matter halo masses in simulations.

For every galaxy class, redshift snapshot, and in both simulations, the group total stellar mass is consistently revealed to be the most important (and hence predictive) luminous tracer of dark matter halo mass. Therefore, this is clearly a parameter we want to include in our halo finder pipeline. However, it is not immediately obvious how to compute this directly from survey data. Obviously, one must first have a group defined. Of course, one could use a group finding algorithm such as a friends-of-friends (FOF, see, e.g., \citealt{Press1982, Davis1985, Springel2001a, Knabe2011}) in order to identify the groups first. But our present goal is to simultaneously find halo masses and group information, enabling the interrelated information on clustering and mass to interact. To help with this problem, we next look at the second most predictive parameter in various classes.

For galaxies as a whole, and for satellites, the number of galaxies in the group is the second most predictive parameter of halo mass. Whilst potentially useful once groups are defined, this still does not help us to identify groups, or motivate their expected size. For central galaxies, however, the stellar mass of the galaxy is the second best parameter, with a very respectable importance score. This is interesting because if we can identify central galaxies we can immediately get a reasonable estimate of halo mass and, moreover, use this to motivate the expected group size. Via iteration one can then ascertain the group parameters ($GM_*$ and $N_g$) and improve the halo mass estimate further. This is, of course, expected from the abundance matching approach (see \citealt{Moster2010, Behroozi2010}). However, as we will see, the stellar mass of central galaxies alone is not an optimal predictor of group halo mass.

All other parameters considered in Fig. 1 are of much less predictive power than the three discussed above. Furthermore, we have determined that the improvement in overall performance (total MSE) is only negligibly affected by removing these parameters (see Table 1). Thus, even though they do technically add something the regression analysis, their value is extremely limited. Including parameters with very little predictive power only makes the pipeline outputs harder to use in practice. Moreover, using the minimum number of parameters to achieve the desired mapping results in having the final halo mass estimates independent of more galaxy properties, which is especially valuable for many applications with star formation tracers. 

Hence, as a result of the above discussion on Fig. 1, we opt to utilize the following luminous tracers to infer dark matter halo masses in our pipeline: (i) total group stellar mass ($GM_*$), (ii)~total number of group members ($N_g$), (iii) central stellar mass ($CM_*$), and (iv) redshift ($z$). The final parameter allows for the possibility of significant evolution in the relationships between luminous and dark matter throughout an observational survey.

\begin{figure*}
\begin{centering}
\includegraphics[width=0.95\textwidth]{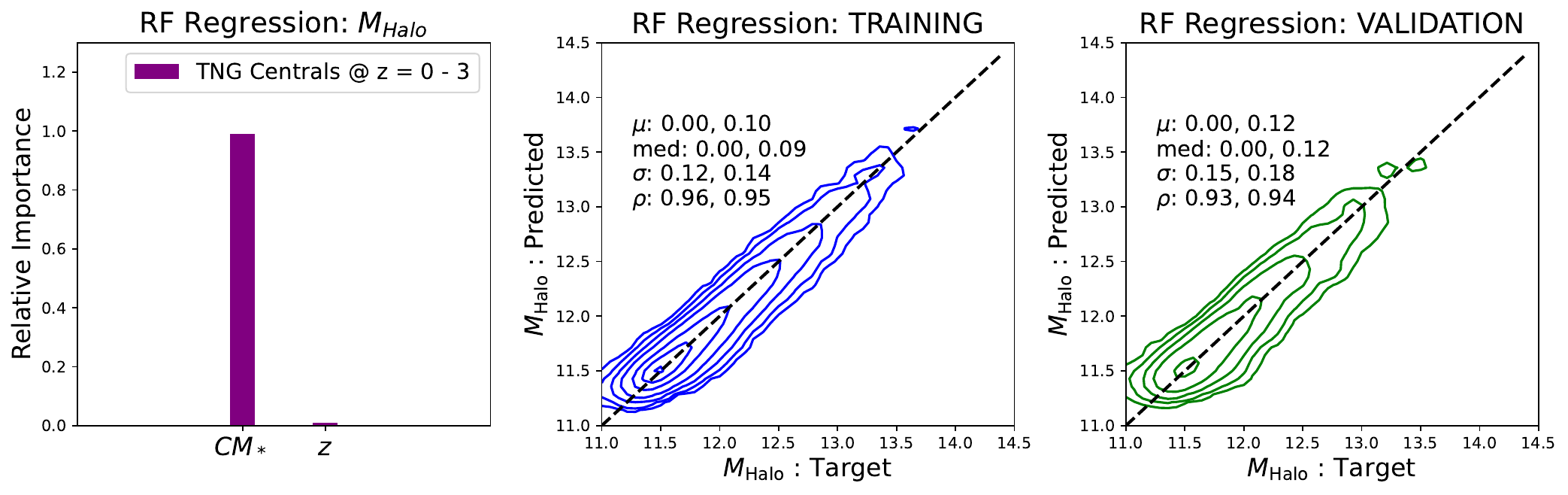}
\includegraphics[width=0.95\textwidth]{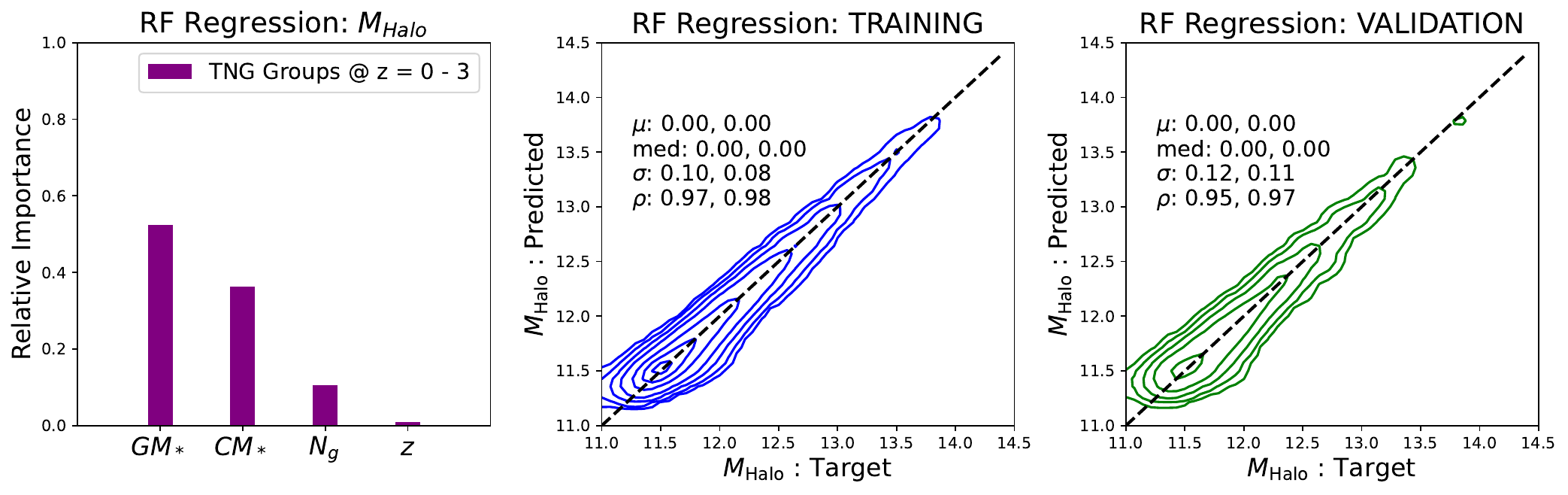}
\caption{Training \& validation of the RF regression models in the TNG simulation at $z = 0 - 3$. The top panels show the results from the central-only RF model and the bottom panels show the results from the full group model. The left-hand panels show the distribution in relative importance for the parameters used in training. This includes central galaxy stellar mass ($CM_*$) and redshift ($z$) for the central-only run, and adds group stellar mass ($GM_*$) and the number of galaxies in the group ($N_g$) in the case of the full group run. See Section 3 for full definitions of these parameters. The center panels show the recovery of halo mass in the training stage and the right-hand panels show the recovery of halo mass in the unseen validation sample (which is used to optimize the performance). The mean ($\mu$) and median (med) offset, standard deviation ($\sigma$), and Spearman rank correlation strength ($\rho$) are displayed for all groups (left) and multi-galaxy groups (right, after comma) for both training and testing. There is a modest improvement in performance in the full group RF model compared to the central-only model for the full data set. Crucially, however, the improvement is much more significant for the multi-galaxy groups. Hence, there is a clear need to incorporate the group information in order to have unbiased and more accurate halo mass estimation for groups and clusters. }
\label{fig2}
\end{centering}
\end{figure*}

\begin{table*}
\begin{center}
\caption{Performance comparison between Random Forest (RF) and Artificial Neural Network (ANN) methods.}
\label{tab1}
\begin{tabular}{ l c c c c c c } 
 \hline \\
Simulation  & ANN All & ANN Group & ANN Central & RF All  &  RF Group & RF Central   \\\\
  \hline\hline \\
TNG (All)    & (0.00, 0.12)  & (0.00, 0.12)  &  (0.00, 0.14)    & (0.00, 0.11)    & (0.00, 0.12)  & (0.00, 0.15)      \\ 
EAGLE (All)  & (0.00, 0.12)  & (0.00, 0.13)  &  (0.00, 0.14)    & (0.00, 0.12)    & (0.00, 0.13)  & (0.00, 0.15)     \\\\

TNG ($N_g > 1$)     & (0.00, 0.12)   & (0.00, 0.11)  & (0.13, 0.16)    & (0.00, 0.11)    & (0.00, 0.11)    & (0.12, 0.18)     \\ 
EAGLE ($N_g > 1$)   & (0.01, 0.12)   & (0.00, 0.12)  &  (0.13, 0.16)   & (0.00, 0.12)    & (0.00, 0.12)  & (0.15, 0.17)    \\\\
 \hline
\end{tabular}
\end{center}
Notes: The performance is given in each cell position as the bias followed by the dispersion (defined as in Fig. 2): ($b \, , \, \sigma$) [dex]. Note the similar performance between ANN and RF methods. Moreover, note the significant improvement when moving from central to group parameters (especially for multi-parameter groups, as seen in Fig. 2), but the lack of additional improvement moving to the `all' parameter set (as seen in Fig. 1). This serves to justify our choice in parameters and machine learning method.
\end{table*}

\subsection{Training the RF regression models}

\noindent From each simulation we construct three sub-samples. First, we make a simple cut to the simulation boxes at $X = 50$\,cMpc. This approximately splits each simulation in half. The first half ($X~<~50$\,cMpc) is used to train and validate the RF models, as outlined in this section. Within this half, training operates on a randomly selected 70\% of the available simulated groups. The remaining 30\% of groups are used to validate the RF models to ensure that we do not over-fit the data (as discussed below). The second half of each simulation ($X > 50$\,cMpc) is reserved for testing the DfL algorithm in Section 4. We emphasize that no individual groups or galaxies used in either training or validation of the RF models are used to test the performance later on in the paper. Additionally, for the testing data we restrict to an observational-like 2D+$z$ space from the full 6D phase space of each simulation (see Section 3.6).

We adopt a two stage approach for estimating halo masses, and more generally halo properties, from luminous tracers. First, we identify systems which are most likely centrals (discussed below in Section 3.5) and leverage the strong relationship between central stellar mass and halo mass in simulations to gain a first estimate of the halo mass, and hence the virial radius and velocity of the group. This part is very similar to the standard abundance matching procedure, with a redshift dependence added. We then use this information to identify likely group members (discussed at length in Section 3.5). 

The preliminary group identification enables us to estimate the total group stellar mass and number of galaxies within the group, which are then used in addition to the central's stellar mass and redshift to refine the prediction for the halo mass in the second stage.  In DfL, we iterate this procedure to find the optimal size of the group and halo mass estimate. The full details of this pipeline are discussed in Section 3.5, but first we show examples of the two versions of the random forest regression models which are incorporated into this strategy.

\subsubsection{For centrals:}

\noindent In Fig. 2 (top panels) we show results from an example trained random forest regression model to predict halo masses from central stellar mass and redshift alone in the TNG simulation. In this case, almost all feature importance is given to stellar mass, with a small importance found in redshift (as indicated in the left-hand panel). The weak importance of redshift is interesting, especially given that both the halo and stellar mass functions evolve significantly as a function of cosmic time (see, e.g., Fig. B2). However, the impact of varying redshift on the $M_* - M_{\rm Halo}$ relationship is actually quite weak in both TNG and EAGLE. At a fixed stellar mass, the expected halo mass does not change dramatically with redshift. Conversely, the probability of reaching a given stellar (and halo) mass does change significantly with redshift, but in a manner which keeps the relationship close to constant.

In training, the regressor is capable of learning a bias-free mapping from stellar mass to halo mass, with a statistical uncertainty of $\sigma = 0.12$ dex (see center panel). We next test the performance of this model on unseen data from the same simulation. Explicitly, we split the input simulation data into 70\% training and 30\% testing, to check for over-fitting. For the testing data the mapping is still essentially bias free, but the random uncertainty on the halo mass increases to $\sigma = 0.15$ dex (see left-hand panel). This is still a reasonably accurate recovery of halo mass, given a perfect knowledge of the central's stellar mass. Indeed, this is a comparable (or slightly improved) performance compared to that achievable in this simulation through abundance matching (see Appendix B, especially Fig. B2).

However, when one considers the performance of the central-only RF model on multi-galaxy groups (statistics presented to the right of commas on the center and right panels of Fig. 2), it is clear that there is a problem. Multi-galaxy groups are biased in their halo masses to $b = +0.12$ dex, and have a significantly worse statistical uncertainty of $\sigma = 0.18$ dex. This strongly motivates us to incorporate group information into DfL, in addition to that of the central galaxy. 

We perform an analogous training of central RF models for the EAGLE simulation. The plots from this simulation are not shown here for the sake of brevity, but the performance statistics achieved are presented alongside TNG in Table 1. Additionally, in Table 1 we compare the performance from RF to an ANN approach, which demonstrates that the former is as successful as the latter for solving this problem (with significant additional benefits, as outlined in Section 3.2).

\begin{figure*}
\begin{centering}
\includegraphics[width=0.95\textwidth]{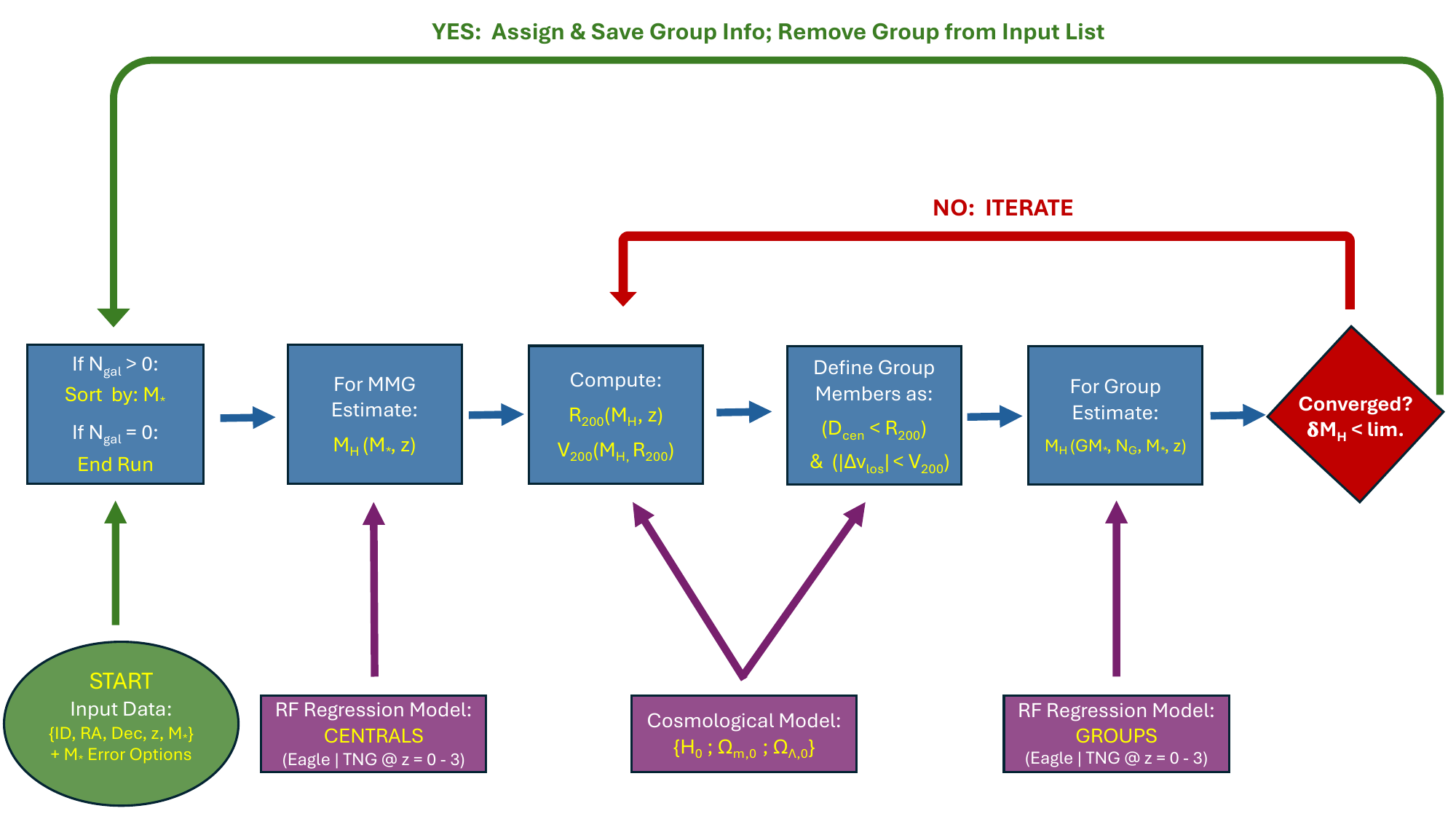}
\caption{Flow chart depicting the DfL pipeline, used to assign galaxies to haloes. The pipeline requires as input data only a unique galaxy identifier (ID), coordinates (R.A., Dec.), spectroscopic redshift ($z$), and a stellar mass estimate ($M_*$), plus optional uncertainties, for each galaxy in the survey. Additionally, one may specify the preferred cosmological parameters and the preferred simulation to use (here either TNG or EAGLE). First, the number of galaxies in the survey remaining to be assigned to a group is checked ($N_{\rm gal}$), provided this is greater than zero the remaining galaxies are sorted by stellar mass and the most massive galaxy in the survey is taken to be a central. Second, the halo mass of the central is estimated by applying the central galaxy regression model (see Section 3.4.1). Third, the virial radius and virial velocity of the halo are computed using the initial halo mass estimate, redshift, and the specified cosmology. Fourth, all galaxies within one virial radius and one virial velocity of the central galaxy are added to the proto-group. Fifth, the halo mass is re-estimated using the full regression model (see Section 3.4.2), here incorporating the total stellar mass of the group and the number of galaxies in the group with $M_* > 10^{9.5} M_\odot$, in addition to the central's stellar mass and redshift. Finally, the two halo mass estimates are compared. If they are within a variable tolerance limit of each other (default value of lim. = 0.05\,dex), the group is fixed and removed from the input data set. The pipeline then continues from the start for the most massive galaxy remaining in the survey. If the estimates are not converged, the virial radius and velocity are updated with the new halo mass estimate and the group is redefined accordingly. Through iteration the halo may grow or shrink in phase space until convergence is reached (or else an optional limit is exceeded, default value: $N_{\rm lim.} = 10$). Once converged, the pipeline begins again from the start on the reduced data set until every single galaxy within the survey is classified (as either a central or a satellite) and the halo properties are provided for each group.}
\label{fig3}
\end{centering}
\end{figure*}

\subsubsection{For groups:}

\noindent In Fig. 2 (bottom panels) we show the results from an example random forest regression model trained on groups from the TNG simulation at $z = 0 - 3$. Here we utilize: $GM_*$, $N_g$, $CM_*$ and $z$ to train the model. The relative importance of each parameter is shown in the left-hand panel. Here the group total stellar mass clearly outperforms central stellar mass, further justifying this second stage in the halo mass estimation. 

In the center panel we show the performance of halo mass recovery in the training data from TNG. This yields an essentially bias free result with a random uncertainty of $\sigma = 0.1$ dex. On the right-hand panel, we show the performance in unseen validation data from TNG. Compared to the results from centrals alone, we note an improved accuracy in halo mass recovery in the unseen data to $\sigma =$ 0.12 dex (TNG). More importantly, we find that the group RF model can reproduce the halo masses of both single and multi-galaxy groups with no bias, and achieves a much improved statistical uncertainty in the multi-galaxy case compared to the central RF model. Taken together, this strongly motivates us to incorporate group finding into the assignment of halo masses, which is the essential idea behind the DfL pipeline.

We perform an analogous training of group RF models for the EAGLE simulation. The plots from this simulation are not shown here for the sake of brevity, but the performance statistics achieved are presented alongside TNG in Table 1. Additionally, in Table 1 we compare the group performance from RF to an ANN approach, which demonstrates that the former is as successful as the latter for solving this problem. See Appendix C for full details on the ANN structure used here as a test.

\subsection{The halo finder algorithm: DfL}

\noindent The DfL halo finder pipeline is the main contribution of this work. It provides a fast, accurate, and straightforward method to assign galaxies from a wide-field spectroscopic survey into groups and clusters, whilst also estimating the dark matter halo masses and virial radii, velocities, and temperatures of each group. The pipeline provides a classification of central and satellite galaxies, and hence also enables one to calculate the 2D physical distances of each group member to its central.

The pipeline takes as input only the following five parameters for each galaxy in the survey: (i)~unique galaxy ID; (ii)~R.A.; (iii)~Dec.; (iv)~spectroscopic redshift ($z$); and (v)~an estimate of stellar mass ($M_*$). Additionally, one may specify the uncertainty on the stellar mass as either a single number (which is interpreted as the 1$\sigma$ width of a Gaussian probability distribution) or else a full probability distribution for each stellar mass value. Furthermore, the desired cosmological parameters (explicitly, $H_0$ and $\Omega_{M,0}$) may be set by the user. The pipeline assumes a spatially flat $\Lambda$CDM cosmology, and hence $\Omega_{k,0} = 0$ and $\Omega_{\Lambda,0} = 1 - \Omega_{M,0}$. The pipeline also makes use of the trained multi-snapshot RF regression models for centrals-only and for groups (outlined in Section 3.4).

We illustrate the logical flow of the DfL pipeline by a flow-chart in Fig. 3. Hereafter, in this section, we present each step in some detail. We recommend readers to follow along using the flow-chart to aid in understanding.

\subsubsection{Step 1:}

\noindent First, the number of galaxies remaining to be sorted into groups is checked. Provided this is greater than zero the pipeline begins (or continues, since it is a loop). Thereafter, the galaxies in the input catalog are sorted by stellar mass. The most massive galaxy in the survey is taken to be a central galaxy. Given that central galaxies are defined in this work to be the most massive galaxy in a given group, the most massive galaxy in the survey is highly likely to be a central. 

There are only two ways this could be untrue. First, the most massive survey galaxy (the MMG) may be part of a group which extends beyond the survey limit and hosts an even more massive galaxy than the MMG. Second, the uncertainty on the stellar mass estimate could result in the actual stellar mass being lower than some other group member. We quantify both of these possible issues in what follows. 

For the use on observational data, we recommend selecting a contiguous area of a wide-field survey with a buffer zone of at least 1\,Mpc (physical) around the target region. The halo finder will target only galaxies within the inner region for halo assignment, but have access to the wider survey to search for potential neighbors. Through this method we have found that less than 1\% of MMGs are erroneously identified as central galaxies due to edge effects. For the impact of stellar mass uncertainty on the overall results we defer to Section 4.5, where this is thoroughly analyzed and discussed.

\subsubsection{Step 2:}

\noindent For the most massive galaxy, we apply the central galaxy random forest regression model (see Section 3.4.1). This yields an estimate of halo mass, explicitly $M_{200}$, i.e. the total mass contained within $R_{200}$ (see next step for further details). This part of the pipeline would function just as well with a simple non-linear fit, or indeed with abundance matching. However, for consistency with what follows (and to enable the benefit of utilizing the individual tree predictions for error analysis) we also use random forest regression here.

\subsubsection{Step 3:}

\noindent Using the initial estimate of the dark matter halo mass, we utilize the spectroscopic redshift of the MMG and a chosen cosmology to infer the virial radius and velocity of the dark matter halo (explicitly, $R_{200}$ and $V_{200}$). This is an important step in the pipeline so we describe it in some detail here.

The dark matter halo mass is taken to be $M_{200}$ in the simulations used for training the regression models. This is defined as the mass contained within the radius, $R_{200}$, at which the average density within this limit is 200 times the critical density of the Universe at this epoch (see, e.g., \citealt{Mo2010}). Explicitly,

\begin{equation}
M_{200} \equiv 200\, \rho_{\mathrm{crit.}}\,(z) \times \frac{4}{3} \pi R_{200}^3
\end{equation}

\noindent where, from a simple rearrangement of the second Friedmann equation for cosmology, the critical density is given as:

\begin{equation}
\rho_{\mathrm{crit.}} = \frac{3H(z)^2}{8 \pi G} = \frac{3 H_0^2 \, (\Omega_{M,0}\, (1+z)^3 + \Omega_{\Lambda,0})}{8 \pi G},
\end{equation}

\noindent where $H(z)$ is the redshift dependent Hubble parameter. In the second step above we expand $H(z)$ in terms of the current Hubble parameter value ($H_0$), using the assumption of a spatially flat two component Universe, containing only matter (quantified by the $z=0$ contribution to the critical density of matter, $\Omega_{M,0}$) and dark energy (quantified by the $z=0$ contribution to the critical density of dark energy, $\Omega_{\Lambda, 0}$). This enables the virial radius to be computed as:

\begin{equation}
R_{200} = \bigg( \frac{G M_{200}}{100 H(z)^2}  \bigg)^{1/3}  = \bigg( \frac{G M_{200}}{100 H_0^2 (\Omega_{M,0} (1+z)^3 + \Omega_{\Lambda,0})}  \bigg)^{1/3}.
\end{equation}

\noindent We then compute the virial velocity as the circular velocity at $R_{200}$ around the halo. Explicitly,

\begin{equation}
V_{200} = \sqrt{\frac{G M_{200}}{R_{200}}}.
\end{equation}

\noindent In Fig. 4 we illustrate the relationships between halo mass and  virial radius, velocity, and temperature for a variety of redshifts. Note that whilst the virial temperature ($T_{200}$) is not used by our pipeline, it is an output parameter. This is computed from setting the thermal energy equal to the mean kinetic energy per particle in the plasma contained within the virial radius. Explicitly, the virial temperature is given as:

\begin{equation}
T_{200} = \frac{\mu m_p V_{200}^2}{2 K_{B} } \,\,,
\end{equation}

\noindent where $\mu$ is the mean atomic mass per particle. This is a free parameter in our pipeline, but we take as a default value $\mu = 0.6$, which is accurate for a primordial fully-ionized plasma.

\begin{figure*}
\begin{centering}
\includegraphics[width=0.95\textwidth]{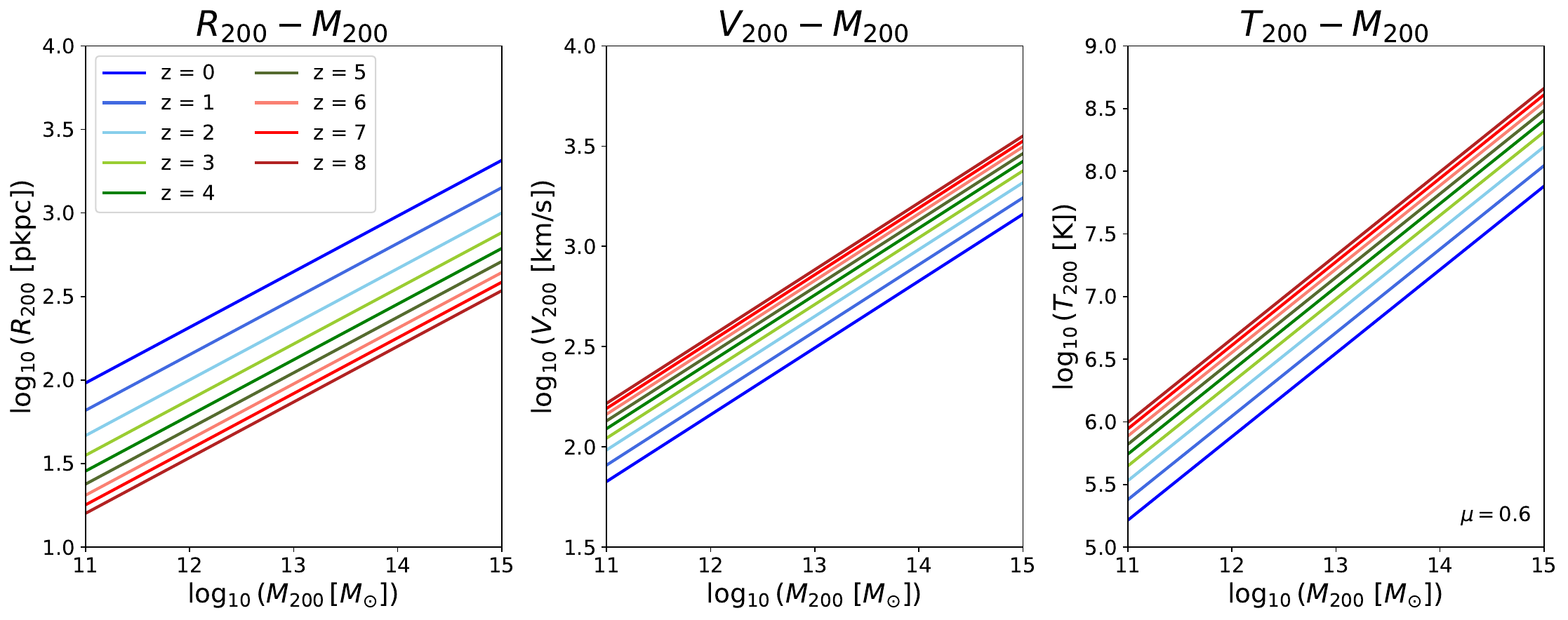}
\caption{The dependence of virial radius ($R_{200}$, left panel), virial velocity ($V_{200}$, center panel), and virial temperature ($T_{200}$, right panel) on halo mass ($M_{200}$) and cosmological redshift ($z$, see legends). Each panel shows results for four orders of magnitude in halo mass and across 13 Gyrs of cosmic time (see legends). These panels illustrate how an estimated value of halo mass can be used to infer the physical and phase-space extent of groups and clusters. Additionally, these plots illustrate how various important halo properties may be inferred directly from halo mass estimates and a redshift measurement. The cosmological parameters used in this figure are as follows: $\Omega_M = 0.3, \Omega_{\Lambda} =  0.7, H_0 = 70\, {\rm [km/s \, Mpc^{-1}]}$. For the virial temperature, we additionally assume a mean particle mass of $0.6\,m_p$, which is correct for a primordial plasma. However, in reality this must be updated by constraints on the individual group/ cluster metallically, especially at lower redshifts. }
\label{fig4}
\end{centering}
\end{figure*}

\subsubsection{Step 4:}

\noindent This step uses the derived halo properties of the previous step to identify group members. In the training of the regression models from simulations, we define the core group as:

\begin{equation}
D_{c, \, 3D} < R_{200}
\end{equation}

\noindent where, explicitly:

\begin{equation}
D_{c, \, 3D} \equiv \sqrt{(X_i - X_c)^2 + (Y_i - Y_c)^2 + (Z_i - Z_c)^2}
\end{equation}

\noindent and $(X,Y,Z)$ indicate the coordinates of arbitrary galaxies in the survey (with subscripts, $i$) and the central in question (with subscript, $c$). Both are given in physical units, i.e. in the rest-frame of the simulation snapshot. 

In observational, or observational-like data (which are used the test the pipeline), we measure the angular separation between any two galaxies using the haversine formula. Explicitly, this is defined as:

\begin{center}
\begin{equation}
\Delta \chi =
\end{equation}
\end{center}

$2 \sin^{-1}{\Bigg\{ \sqrt{\sin^2{\bigg(\frac{\theta_i - \theta_c}{2}\bigg) + \cos{(\theta_i)} \cos{(\theta_c)} \sin^2{\bigg(\frac{\phi_i - \phi_c}{2}\bigg)}}} \,\, \Bigg\}} $ \\

\noindent where, $\theta$ indicates declination (Dec.) and $\phi$ indicates right ascension (R.A.), with subscript `$i$' indicating any arbitrary galaxy in the survey and subscript `$c$' indicating the central galaxy in question. We next convert this to a physical distance at the epoch of the MMG. Explicitly we compute:

\begin{equation}
D_{c,\, 2D} = \Delta \chi \, D_A(z_c)    
\end{equation}

\noindent where, $D_A(z_c)$ is the angular diameter distance evaluated at the redshift of the central galaxy, which is defined as: $D_A(z) = D_C(z)/(1+z)$, where $D_C(z)$ is the comoving radial distance. This is defined explicitly as:

\begin{equation}
D_C\,(z) \equiv \frac{c}{H_0} \int_{0}^{z} \frac{dz'}{E(z')}  
\end{equation}

\noindent where,

\begin{equation}
E(z) \equiv \sqrt{\Omega_{M,0} \, (1+z)^3 + \Omega_{\Lambda,0}}  
\end{equation}

\noindent for a spatially flat $\Lambda$CDM universe.

We next compute the line-of-sight velocity difference between each galaxy and the central as a function of the difference in their observed redshifts. Explicitly, we compute: 

\begin{equation}
\Delta V_{\rm los} = \frac{c \, |\Delta z|}{(1+\bar{z})} \,\, \mathrm{for } \,\, \Delta V_{\rm los} \ll c
\end{equation}

\noindent where $\Delta z = (z_i - z_c)$, and $\bar{z}$ is taken as the mean redshift of the two galaxies, rather than the (unknown) true cosmological redshift. Note that for physically associated galaxies, the velocity difference is always much lower than the speed of light and hence the above expression is sufficient for our purposes.

Using the above distances and velocity offsets, we define the core group in the pipeline as:

\begin{equation}
(D_{c, \, 2D} < R_{200}) \,\, \& \,\, (\Delta V_{\rm los} < V_{200}).
\end{equation}

\noindent The galaxies which meet these criteria are used in the next step to infer $M_{200}$ on the basis of the total group stellar mass, the total number of galaxies within the group, and the central galaxy stellar mass and redshift (as before).

We also construct an extended group region within both the simulations and observational-like data, which will in principle contain both satellites and non-satellites. This is motivated in part by the literature on the splash-back radius, which demonstrates that satellite galaxies may extend beyond the virial radius out to $\sim 2\,R_{200}$ (e.g., \citealt{Adhikari2014, More2015, More2016}). Additionally, the impact velocity from infinity at the virial radius is $V_{\rm imp} = \sqrt{2}\,V_{200}$ (a classic result in Newtonian mechanics), and hence some true satellites will be moving at velocities exceeding virial. Moreover, three body interactions and non-virial group mergers can both lead to enhanced super-virial velocities. Unfortunately, this region of phase space will also contain chance alignments of true centrals as well. We test the accuracy of central - satellite classification via this method in Section 4.

Explicitly, in simulations, we define the extended satellite region as:

\begin{equation}
D_{c, \, 3D} < 2R_{200}.
\end{equation}

\noindent In the rare case of satellites in the simulations residing at $D_{c, \, 3D} > 2R_{200}$, we treat these objects as independent centrals and utilize their sub-halo mass as their group halo mass. These cases ($< 10\%$ of satellites) are likely not truly connected to the group at the snapshot in question, despite being identified as satellites via the {\small SUBFIND} algorithm. For example, in \cite{Goubert2024} we find that `satellites' beyond $\sim 1.5 \, R_{200}$ behave identically to isolated centrals in both EAGLE and TNG across a wide variety of parameters, despite being identified as part of a group or cluster.

For observations (and observational-like simulated data), we define the extended satellite region as:

\begin{equation}
(D_{c, \, 2D} < 2R_{200}) \,\, \& \,\, (\Delta V_{\rm los} < 2V_{200}).
\end{equation}

\noindent This essentially provides a buffer region around each group. The extended satellite region is not used to infer halo masses directly, but it is used to further isolate each group from truly independent structures. Hence, the extended satellite population is removed with the core group after each completed iteration.

In summary of Step 4, based on the initial halo estimate we now have a proto-group defined, with both core and extended satellite populations, in addition to the central galaxy.

\subsubsection{Step 5:}

\noindent Utilizing the proto-group, i.e. the IDs of galaxies within the core group identified in Step 4, we evaluate the total group stellar mass and the total number of galaxies within the group (with $M_* > 10^{9.5} M_\odot$). These parameters are then included with the central galaxy's stellar mass and redshift as input to to the group RF regression model (see Section 3.4.2). This yields a new estimate of $M_{200}$, now based on the group properties (not the properties of the central alone).

\subsubsection{Step 6:}

\noindent The two halo mass estimates (from central-only and the full group) are then compared. If they are sufficiently similar in value ($| \delta M_{\rm Halo} | < \zeta $) the group is fixed and the pipeline returns to Step~1, with this group removed from the input data. The tolerance limit ($\zeta$) is a user variable parameter in DfL. Through direct experimentation we have found that an optimal value is reached between $\zeta = 0.01 - 0.05$\,dex (depending on simulation and model pairing). However, no results are strongly dependent on this parameter within this range.   

Note that after removing the first group (that of the most massive member of the survey), the pipeline is free to begin again with the next most massive galaxy. Provided this galaxy should not have been included with the first group, the next most massive galaxy is also highly likely to be a central by the same logic as outlined in Step 1. As such, the pipeline will continue iterating until every single galaxy in the survey is assigned to a group, and a halo mass (plus virial parameters) are estimated. Note that groups may contain only one galaxy, i.e. for isolated centrals. Indeed, this is the most common group type.

It is important to appreciate that it is highly unlikely that the next most massive galaxy ought to have been included in the previous group because it is (by definition) at a distance greater than two virial radii projected, and/ or at a velocity difference of greater than two virial velocities line-of-sight from the first group's central galaxy.

On the other hand, if the two halo mass estimates disagree by more than the tolerance, the pipeline returns to Step 3. That is, a new virial radius and virial velocity is inferred from the updated halo mass estimate. This enables a new definition of the group to be made and hence a further halo mass estimation, based upon the updated group membership information.

Once again, the two most recent halo mass estimates are compared, and if convergence is still not reached the pipeline iterates again back to Step 3. This is allowed to go on for up to $N_{\rm lim}$ iterations. This parameter is specified by the user. The default value is a maximum of ten iterations, which we find to be sufficient to reach convergence in $\sim$ 99\% of cases. As such, the groups may grow or shrink in phase space based on the distribution of galaxies around the MMG. Once convergence is reached, the pipeline removes this group form from the survey and returns to Step 1 to continue for the next group in the survey (as outlined above). Once the number of galaxies remaining in the sample reaches zero, the DfL pipeline ceases running and outputs the group catalogs with halo mass estimates.

\subsubsection{Uncertainty on halo mass predictions}

\noindent The RF technique is inherently probabilistic and hence offers a natural way to model and propagate uncertainties throughout the DfL pipeline. DfL enables three levels of error propagation from the input stellar masses, which we outline here:

Case 1: No stellar mass uncertainties. DfL does not require uncertainties on the input stellar masses in order to run. However, we strongly recommend to provide uncertainties in order to get the best results out of the pipeline. In the case of no input uncertainties, the uncertainties on halo mass are derived directly and solely from the individual decision trees within the random forest. A total of 250 decision trees are utilized in the determination of each halo mass estimate. Each tree is different due to bootstrapped random sampling of the training dataset, and through random partial access to input features (in the group model). The normalized distribution of halo mass predictions is taken to be a probability distribution function (PDF) of the output halo mass. This is saved to a FITS file by DfL along with the point halo mass prediction (the geometric mean of the PDF) and the 1 and 2 $\sigma$ uncertainties, calculated in a non-parametric manner as the 16 - 84 and 2.5 - 97.5 percentiles of the data, respectively. This procedure quantifies the uncertainty inherent from the mapping from luminous tracers to dark matter halo mass in the simulation alone.

Case 2: Point stellar mass uncertainties. In the case of a single numerical value for the uncertainty on stellar masses, DfL treats this as the standard deviation of a Gaussian centered on the stellar mass input value. A variable number ($N_{\rm MC}$) of random draws from this Gaussian is taken during the processing of each galaxy. The default number of random draws is 100. The propagation of uncertainties is then implemented via a Monte Carlo approach. For each of the random draws in stellar mass a halo mass distribution function is created from the individual decision tree outputs (as in Case 1). All of the tree predictions from all of the random draws are concatenated, yielding an output distribution in halo mass predictions with $N_{\rm trees} \times N_{\rm MC}$ (i.e., a default of 25k) individual values. This is then normalized and treated as the PDF output for each halo mass estimate.

Case 3: Full stellar mass PDFs. DfL enables (and we strongly encourage) users to input full PDFs of stellar masses when available (see Appendix A for practical details). In the presence of a full PDF, the pipeline randomly samples the PDF exactly as in Case 2 for the simple Gaussian case, but now with the benefit of a more accurate characterization of the uncertainties on stellar mass. As in Case 2, an estimate of halo mass from each of the 250 trees in each RF model run is created for each of the $N_{\rm MC}$ random draws of the stellar mass PDF. These are then concatenated together and normalized to yield the final output PDF for each halo mass estimate, in addition to its default point value and uncertainties (exactly as in in Case 1).

In Fig. 5 we show example output PDFs for the halo mass predictions, operating with a bespoke (randomly generated) input PDF for the stellar mass inputs. The dashed vertical lines indicate the DfL halo mass prediction and the solid vertical lines indicate the simulation truth. In this figure three random PDFs are drawn at low mass (blue), intermediate mass (green), and high mass (red) from TNG at $z = 1$.

\begin{figure}
\begin{centering}
\includegraphics[width=0.5\textwidth]{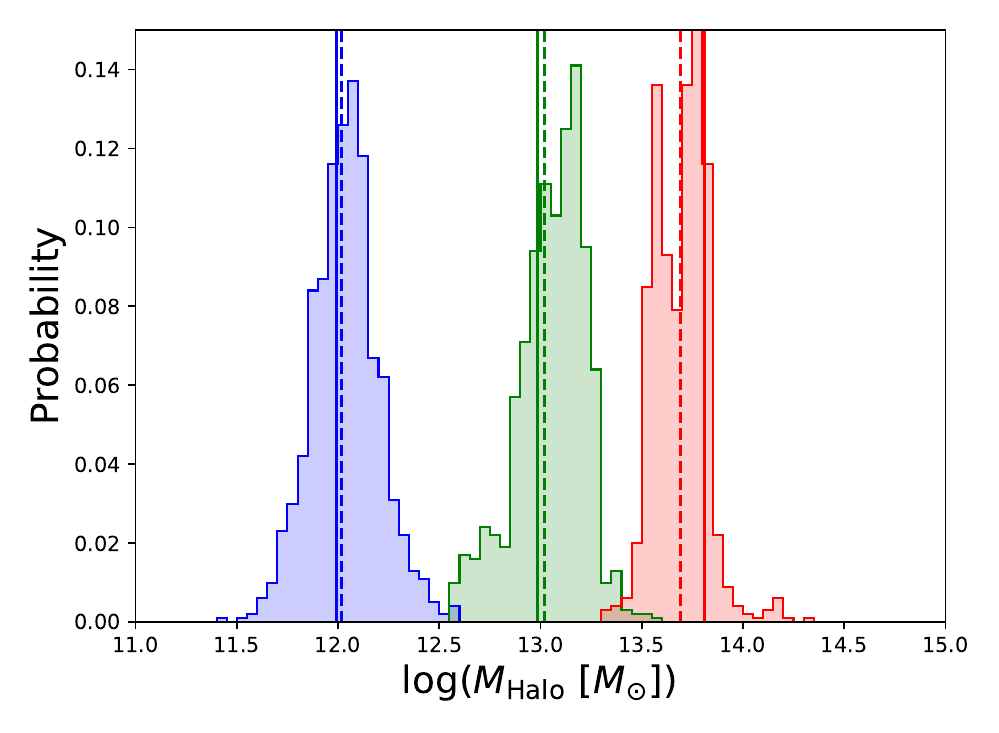}
\caption{Example of output halo mass PDFs from {\small DfL} from the TNG model applied to the TNG testing data at $z = 1$. The shaded histograms show the probability distributions of the halo mass for a randomly selected low, intermediate and high mass group (colored blue, green and red, respectively). The true halo mass from the simulation is indicated by a vertical solid line, with the {\small DfL} geometric mean output from the pipeline indicated by a vertical dashed line. Obviously, the PDF can be used to extract the mode, linear mean, or any other indicator for the best mass prediction for $M_{200}$, as desired.}
\label{fig5}
\end{centering}
\end{figure}

\subsubsection{Note on the structure of the pipeline}

\noindent The pipeline is essentially powered by two nested while loops programmed in {\small PYTHON-3}. The first loop operates while there are unclassified galaxies into groups within the survey (illustrated by the green arrow in Fig. 3). Within this, the second loop operates while there is a lack of convergence in the halo mass estimators for a single proto-group (illustrated by the red arrow in Fig. 3). In conjunction, the pipeline takes a relatively simple set of observational input data and assigns a physically motivated group structure to the entire galaxy survey. In the remainder of this paper we carefully test the performance of this method for estimating the dark sector in observational-like simulated data. In Appendix A we provide a user guide to DfL, enabling straightforward application to wide-field galaxy surveys.

\subsection{Observational-like simulated data: The testing sample}

\noindent In order to test the DfL pipeline, we construct an observational-like dataset from each simulation utilizing an unseen testing sample. The testing sample is defined as all data for which: $X > 50\,$cMpc. This represents approximately half of the TNG and EAGLE simulations. Crucially, this sub-sample is not used in either training or validating the RF models for DfL. As such, this data is completely unseen by the machine learning prescriptions prior to testing. Moreover, we also transform the testing sample into an observational-like sample by restricting the dimensionality of the data from 6D phase space to 2D+$z$ observer space (discussed in detail below). Hence, the purpose of the testing sample is: (i) to have an independent dataset unseen by the machine learning models in training and validation, with which to test the performance of DfL; and (ii) to have an observational-like data set with which to test the impact of observational limitations on the recovery of halo masses with DfL.

For the testing sample, we randomly project each simulation snapshot into a 2D plane (hereafter labeled as $X \& Y$). The $X$ and $Y$ coordinates are then converted to sky coordinates by applying the angular diameter distance and placing onto the sky at an appropriate location. Explicitly, we position the edge of each simulation snapshot 2D plane ($X = Y = 0$) at R.A. = 0 and Dec.~=~0. 

We then make the simplifying assumption that the sky is flat on the small angular scales subtended by the relatively small simulation boxes (when placed near the celestial equator). We have tested that at the scale of the most massive clusters this simplification incurs $<1\%$ distance errors at the lowest redshift positioning (taken as $z = 0.1$)\footnote{Obviously, one cannot literally place a galaxy at $z = 0$, since this would mean it resides on the Milky Way. To combat this technicality we position the $z = 0$ snapshot data at $z = 0.1$. This is only used to enable the conversion from simulation to observer-like distances and velocities, and is not used to modify the cosmological parameters. Hence, more concretely, we test the $z = 0$ epoch at a $z = 0.1$ distance. No changes are needed for the other redshift snapshots used in this work.}. At higher redshifts the difference becomes completely negligible.

Explicitly, we compute the following mappings:

\begin{equation}
\theta_i = \theta_0 + \frac{Y_i}{D_A(z_i)}    
\end{equation}

\begin{equation}
\phi_i = \phi_0 + \frac{X_i}{D_A(z_i) \cos{(\theta_i)}}    
\end{equation}

\noindent where $\phi$ represents R.A. and $\theta$ represents Dec.. In practice we set the zero points to zero, and the cosine function evaluates to very close to unity. As mentioned above, this assumes orthogonality between sky coordinates on the scales of relevance in the simulations. This is a very reasonable assumption in practice here due to the relatively small comoving box size of the simulations, though (of course) this is not true in general\footnote{More quantitatively, note that the harversine formula (eq. 14) reduces to: $\Delta \chi \approx \sqrt{ (\theta_2 - \theta_1)^2 + (\phi_2 - \phi_1)^2 \cos(\langle \theta_{12} \rangle)} $, using the small angle formula. At the celestial equator the cosine term equals unity and orthogonality is achieved. Using eqs. 22 \& 23, this yields: $D_{c,2D} = \Delta \chi D_A(z) = \sqrt{X^2 + Y^2}$, as desired.}.

One advantage of the simplifying approach outlined above is that the $Z$ coordinate may now be taken as perpendicular to the `sky plane'. We use this to construct a pseudo cosmological recession velocity (and hence redshift), which is then added to the real peculiar line-of-sight velocity in the simulation to construct an observational-like redshift. Explicitly, we construct the observer-like cosmological redshift as:

\begin{equation}
z_{\rm cos} = z \, (D_c\,(z_{\rm snapshot}) + Z_i\,(1+z_{\rm snapshot}))    
\end{equation}

\noindent where we evaluate redshift as a function of comoving radial distance by numerically solving the defining integral equation (see back to eq. 16). The true comoving distance is taken as the actual comoving distance to the snapshot redshift plus the comoving radial coordinate of the galaxy. Note that $Z_i$ is given in physical coordinates, which is why it is multiplied by the $(1+z)$ factor. We take this redshift to be equal to the snapshot redshift because: (i) this is approximately true; and (ii) the cosmological redshift is not defined prior to this computation. The above expression thus accounts for the positioning of galaxies within the simulation box in the assignment of a cosmological redshift to each galaxy. 

Next, we add in the contribution from the line-of-sight peculiar velocity of the galaxy, to better approximate a true observer-like scenario. Explicitly, we compute the observation-like redshift as:

\begin{equation}
z_{\rm obs} = z_{\rm cos} + z_{\rm pec} 
\end{equation}

\noindent where,

\begin{equation}
z_{\rm pec} = \frac{V_{Z} \, (1+z_{\rm snapshot})}{c}
\end{equation}

\noindent and $V_Z$ is the $Z$-component physical velocity of each galaxy in the rest-frame of the simulation snapshot's redshift. Note that when placed into eq. 18 (used in the pipeline, see above), for any two galaxies with the same cosmological redshift, eq. 26 yields the rest-frame velocity difference, as desired. This is accurate provided $z_{\rm obs} \approx z_{\rm cos} \approx z_{\rm snapshot}$, which is invariably the case in these data.

Finally, the actual stellar mass of each galaxy is given to the pipeline as an input. Initially, this is used to test the recovery of halo masses under a perfect knowledge of the stellar component. However, later in the next section we test the impact of varying uncertainty in stellar mass estimates, in order to assess halo recovery under realistic galaxy survey conditions. Additionally, we also test the impact of incompleteness in galaxy detection on the recovery of halo masses (see Section 4.5).

Crucially, no information on the halo masses of galaxies, or membership of galaxies to groups, is provided to the pipeline. Nonetheless, this information is used post analysis to check the performance of the pipeline in estimating these properties. This is why we test the pipeline on observational-like simulated data (where the true halo properties are known) instead of real observational data (where the true halo properties are unknown). That said, we do additionally test the recovery of the observed $M_* - M_{\rm Halo}$ relationship in DfL from gravitational lensing data later (see Section 4.4).

Finally, for testing, we remove edge-effects from the observation-like simulated data by selecting galaxies which reside at least 1\,cMpc from the nearest edge of the survey (which represents a distance approximately equal to the largest cluster virial radius at all epochs). We recommend that observational applications mimic this approach. This essentially prevents edge effects from impacting the results of the halo finder pipeline. Note that galaxies beyond this limit may be used to infer group properties, but these are only trusted if the central lies within the inner contiguous region (and hence all group members may in principle be identified).

\section{DfL performance results}
\label{sec4}

\noindent In this section we test the performance of the DfL halo finder pipeline on observational-like simulated data from TNG and EAGLE at example redshifts of $z = 0, 1, 2$. First, we consider the performance of the pipeline when using the appropriate model and data pairing (Section 4.1). In addition to quantifying the precision of halo mass estimates under different circumstances, we also quantify the accuracy of central - satellite classification and determine the fidelity of the overall halo mass function as recovered through the pipeline. Second, we consider the impact of deliberately applying a different model to the data to test the impact of model choice on the recovery of halo masses (Section 4.2). Third, we compare the performance results from DfL to abundance matching applied to the groups found from a friends-of-friends algorithm (Section 4.3). Fourth, we compare the output $M_* - M_{200}$ relation from DfL at $z = 0$ to observational constraints from strong gravitational lensing (Section 4.4). Finally, we test the impact on the recovery of halo masses with DfL of systematically varying stellar mass uncertainty and survey incompleteness (Section~4.5).

\subsection{Single model performance}

\begin{figure*}
\begin{centering}
\includegraphics[width=0.95\textwidth]{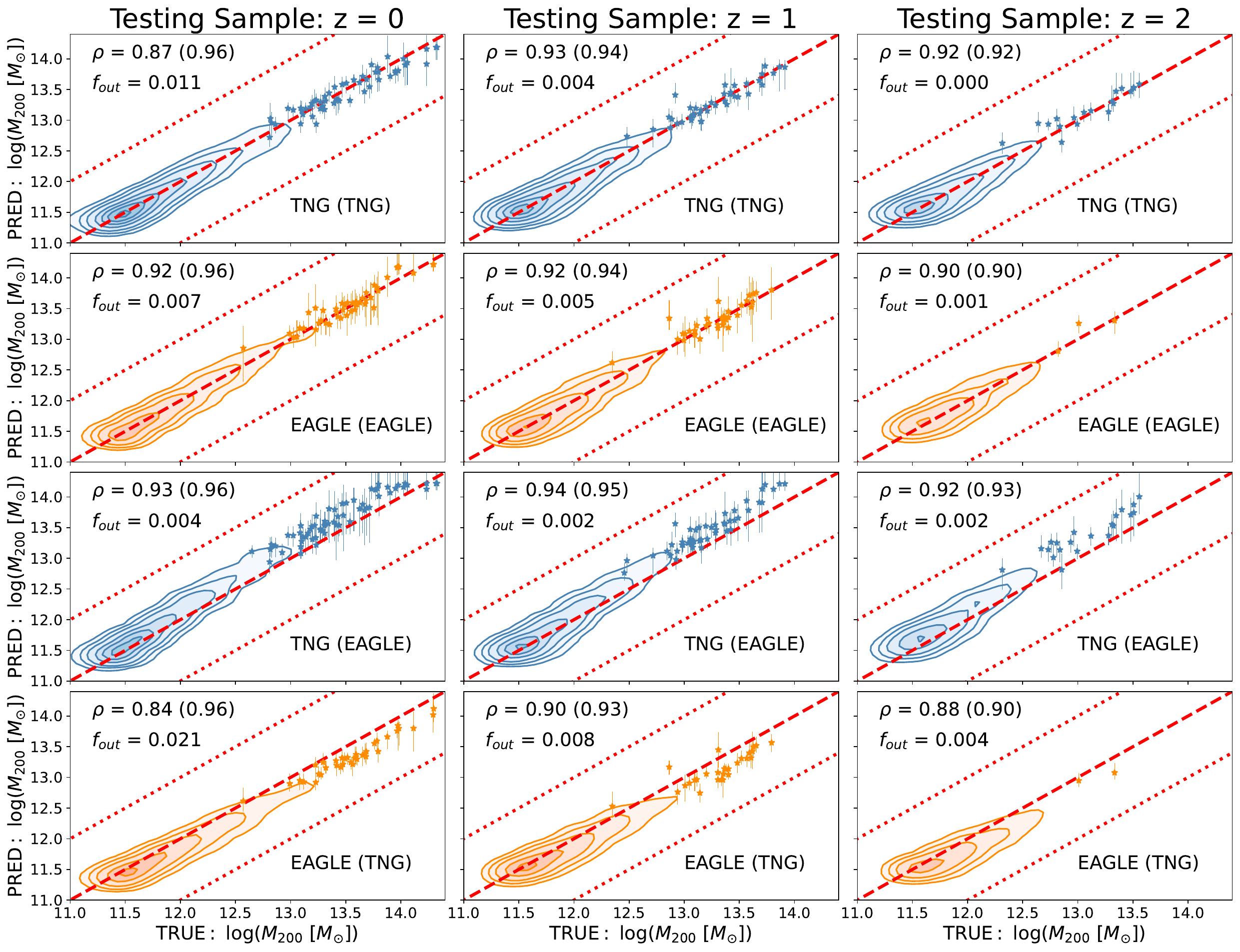}
\caption{Halo mass recovery performance with DfL I: {\small DfL} Predicted vs. simulation truth $M_{\rm 200}$ density contours for all output galaxy groups (including isolated galaxies). In addition to contours, we also show as star-shaped points (with error bars) the location of the largest groups/ clusters at each epoch (those with $N_g > 5$). This enables one to assess the recovery of this small but important population, which are not well represented by the density contours due to these being dominated by the most numerous lower-mass haloes. The figure is structured as follows: Columns indicate redshift snapshots (as labeled by the titles), and rows indicate data - model pairings (as indicated on the panels). The top row shows results for TNG data using the TNG regression model (trained and validated on different subsets of the data). The second row shows the equivalent results for EAGLE data using an EAGLE trained model. The third and final rows show the results for TNG data using an EAGLE model and vice-versa, respectively. On each panel the one-to-one relation is indicated by a dashed red line, and $\pm$1 dex is indicated by dotted red lines to help indicate scale. Also on each panel the Spearman correlation strength ($\rho$) and the fraction of catastrophic outliers ($f_{\rm out}$, i.e., those with $|\delta M_{\rm Halo}| > 1$ dex) are shown. In parenthesis, we additionally show the correlation strength excluding the very rare case of catastrophic outliers. Note that visually the recovery of halo masses from luminous tracers is excellent when the appropriate model and data pairing is used, but that a modest bias is induced when the model and the data are mismatched. }
\label{fig6}
\end{centering}
\end{figure*}

\begin{figure*}
\begin{centering}
\includegraphics[width=0.95\textwidth]{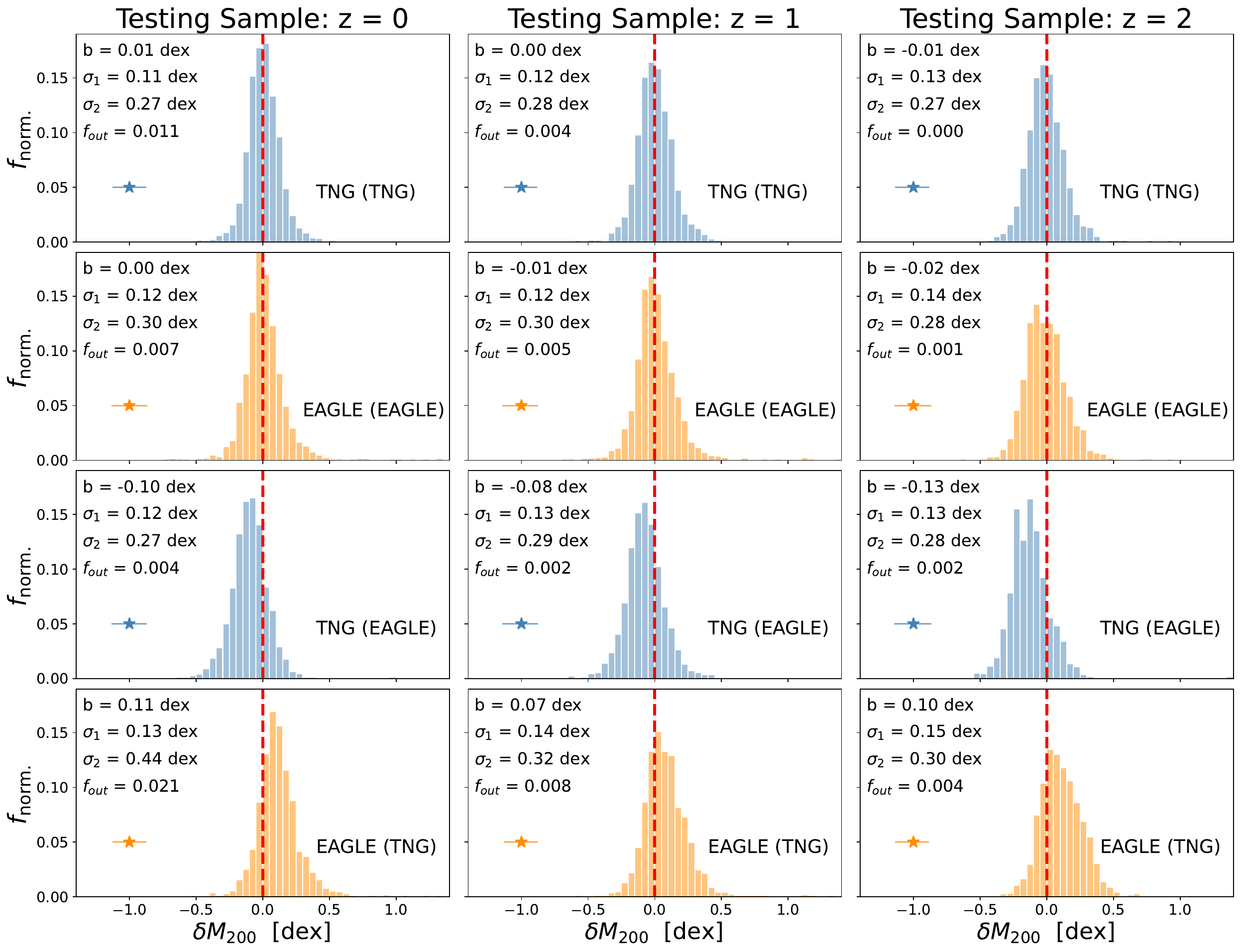}
\caption{Halo mass recovery performance with DfL IIa: Distributions in the offset from true-to-predicted $M_{200}$ values ($\delta M_{200} \equiv \, M_{200} \, ({\rm true}) -  \, M_{200} \, ({\rm predicted})$). The structure of this figure is identical to Fig. 6. On each panel a dashed red line shows the perfect regression value of $\delta M_{200} = 0$, and the mean uncertainty on halo masses output from {\small DfL} is indicated by a horizontal error bar. Additionally, the bias (median offset), full 1$\sigma$ dispersion ($\sigma_1$, containing 68\% of the data set), 2$\sigma$ dispersion ($\sigma_2$, containing 95\% of the data), and the fraction of catastrophic outliers (those with $|\delta M_{200}| > 1$\, dex) are indicated on each panel. For the cases of data and models being matched (top two rows), the pipeline recovers halo masses with essentially no bias ($|b| \leq 0.02$ dex) and an average uncertainty of $\langle \sigma_1 \rangle = 0.12$ dex. To test the impact of model uncertainty, in the lower two rows we show the $\delta M_{200}$ distributions for TNG data with an EAGLE model, and vice-versa. Here an average bias of $\langle b \rangle$ = -0.10 dex is found by applying the EAGLE model to TNG data, and an average bias of $\langle b \rangle$ = +0.10 dex is found by applying the TNG model to EAGLE data. The overall random uncertainty remains essentially unchanged.}
\label{fig7}
\end{centering}
\end{figure*}

\begin{figure*}
\begin{centering}
\includegraphics[width=0.95\textwidth]{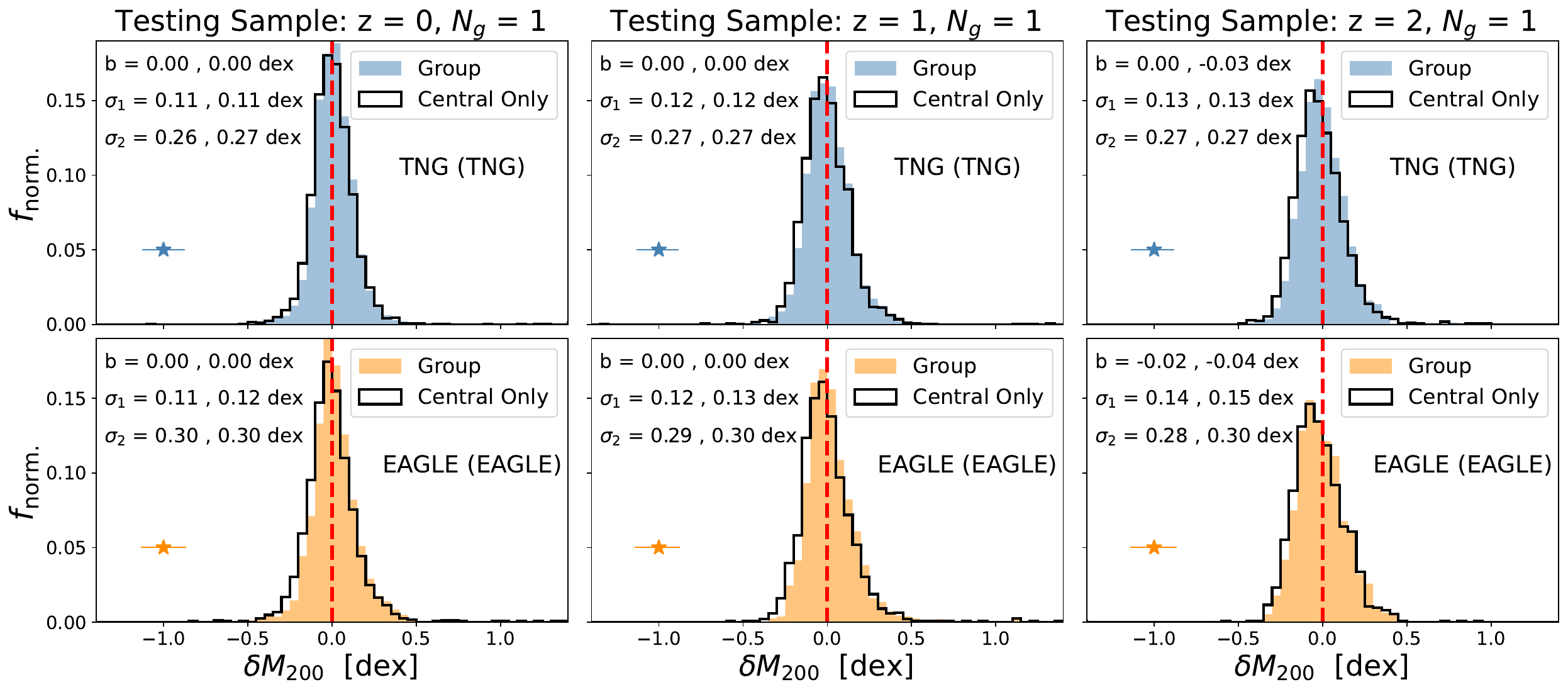}
\includegraphics[width=0.95\textwidth]{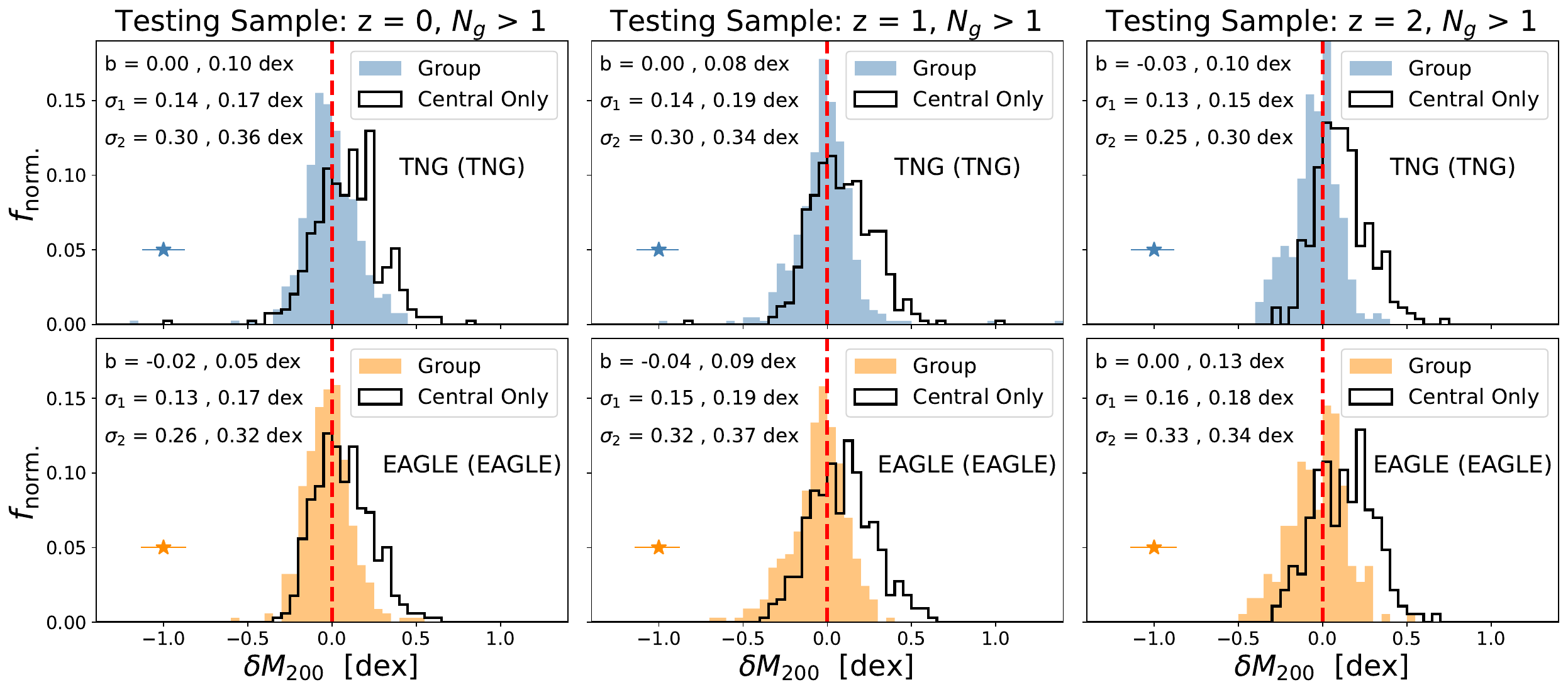}
\caption{Halo mass recovery performance with DfL IIb: Distributions in the offsets from true-to-predicted $M_{200}$ values ($\delta M_{200}$), shown here for single galaxy groups (top panels) and multi-galaxy groups (bottom panels). On each panel the distributions in halo mass offsets are shown for the full group RF model (shaded histograms) and compared to the central-only RF model (open black histograms), both of which are outputs from DfL. Statistics are presented on each panel for the group model, followed by the central only model. In the case of single galaxy groups, the two distributions are very similar, as expected. However, for multi-galaxy groups, the two halo mass offset distributions are significantly different. The central-only RF models yield halo masses which are biased to higher values of $\delta M_{200}$ (lower values of $M_{200}$ pred.) and have a higher dispersion relative to the group results. Moreover, the full group RF models have significantly better agreement with the true halo masses from each simulation. This demonstrates the critical importance of incorporating the group information into the {\small DfL} pipeline and hence justifies the need for the two stage approach, discussed in Section 3.}
\label{fig8}
\end{centering}
\end{figure*}

\begin{figure*}
\begin{centering}
\includegraphics[width=0.95\textwidth]{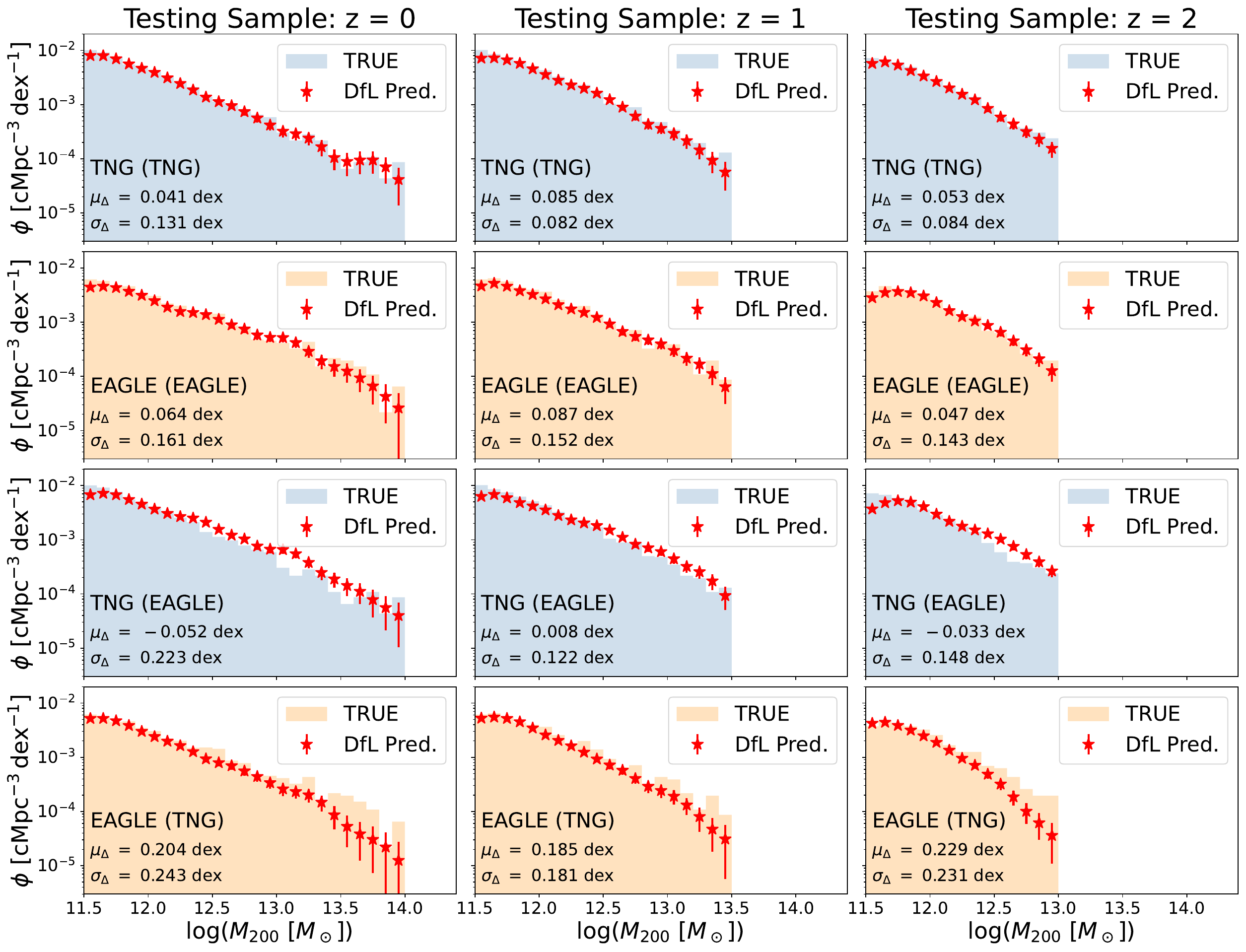}
\caption{Halo mass recovery performance with DfL III: Halo mass functions. This figure is structured as in Figs. 6 \& 7. On each panel the actual halo mass function for the simulation at each epoch (shaded regions) and the estimated halo mass function from the {\small DfL} pipeline (data points with errors) are compared. In the case of applying the same model and data pairing (albeit trained and tested on different subsamples), there is excellent agreement between the two distributions. In the case of applying different model and data pairings, there is more noticeable disagreement, originating from the differences in the true halo mass functions between simulations. Therefore, one can recover the halo mass function as it would be in either simulation from observational-like data to high fidelity, but the output halo mass function will be biased by the model choice. Additionally, on each panel we show the mean offset and standard deviation in offsets between the mass functions in order to quantify the statistical difference, and to facilitate comparison to the FOF-AM approach (see Section 4.4).}
\label{fig9}
\end{centering}
\end{figure*}

\begin{figure*}
\begin{centering}
\includegraphics[width=0.95\textwidth]{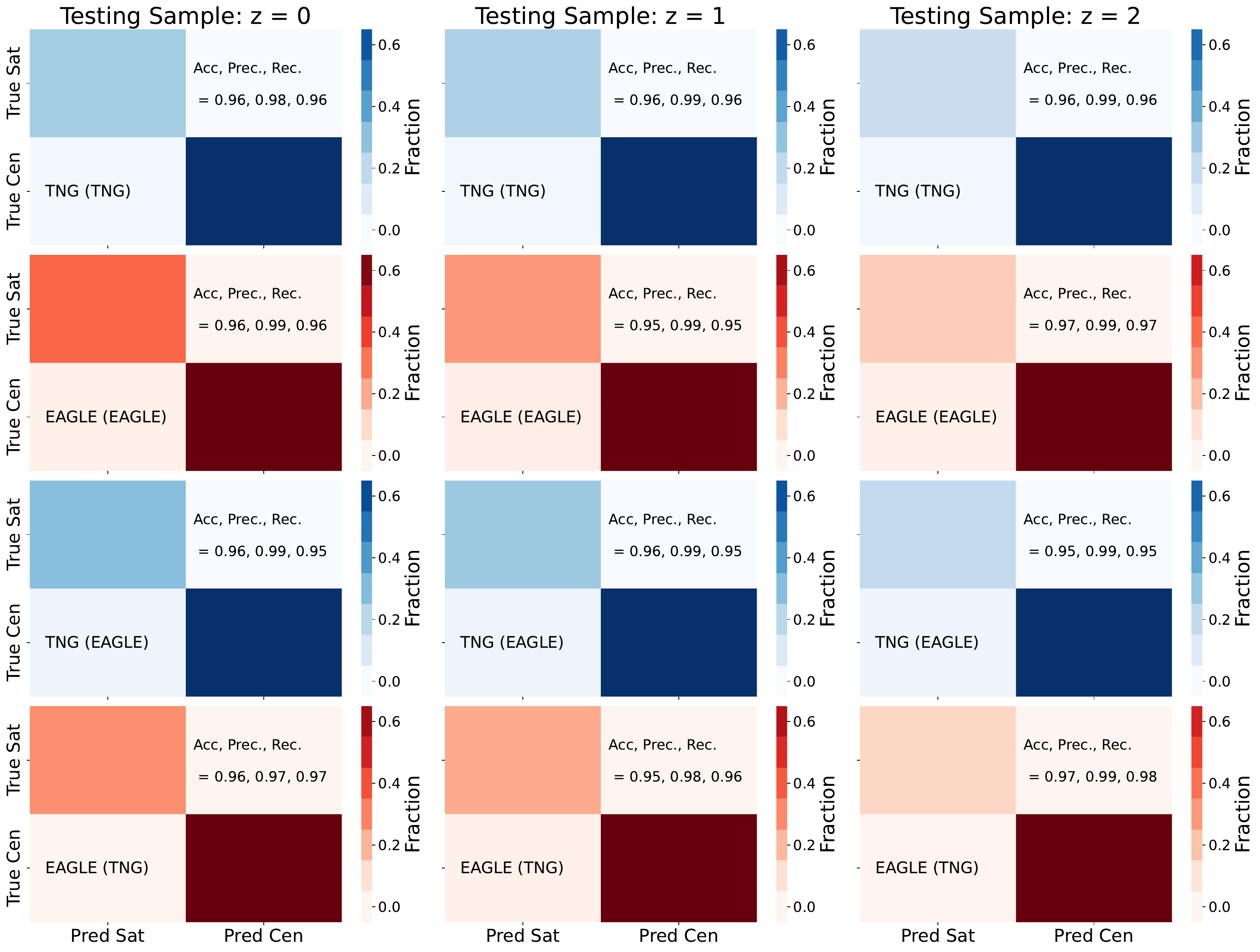}
\caption{Central - (core) satellite classification results from the {\small DfL} pipeline. The structure of this figure is identical to Figs. 6, 7 \& 9. Each panel displays the confusion matrix, which plots true centrals and satellites against predicted centrals and satellites, with the color of each rectangular region indicating the fraction of data in each quadrant (as labeled by the color bars). The sum of all quadrants equals unity. A perfect classification would have all data on the diagonal in each panel, or equivalently dark colored rectangles on the diagonal and white space in the off-diagonal regions. Three statistics are indicated on each panel (see eqs. 28 - 30 for definitions): the accuracy in separation between central and satellite galaxies (Acc.), the precision in identifying centrals (Prec.), and the recall (Rec.), which quantifies the completeness of the recovered central population. All results are shown here for centrals compared to the core satellite population. The average accuracy across all epochs is 96\%, and this is largely unaffected by model choice.}
\label{fig10}
\end{centering}
\end{figure*}

\noindent In Fig. 6, we present the recovery of halo masses from the DfL halo finder pipeline method. Density contours of true $M_{200}$ values, from the simulations, are plotted against the estimated $M_{200}$ values output from the pipeline. Due to the steepness of the halo mass function, the density contours are dominated by lower-mass haloes. To combat this issue, we additionally plot as star-shaped points (with error bars) the location of high mass groups/ clusters (explicitly those with $N_g > 5$). 

The top row of Fig. 6 shows results for TNG (using the appropriate TNG models) and the second row shows the results for EAGLE (using the appropriate EAGLE models). Note, however, that different data is used in training/ validation of the RF models and in the testing of DfL, shown here (as explained in Section 3.4). The lower two rows show cross-validation between models, which is discussed later in Section 4.2. In that section, completely different simulations (in addition to samples) are used in training the RF models and in testing the performance of DfL. Results are shown separately for $z = 0, 1, 2$ (ordered by column). Dashed red lines indicate the one-to-one relationship, with dotted red lines indicating $\pm$1 dex from the one-to-one relation, to help indicate scale.

Visually, the recovery of halo masses is very good in the top two rows. The contours are well centered on the one-to-one relationship and clearly reasonably tight. We quantify the strength of the relationship with the Spearman rank correlation ($\rho$), displayed on each panel. The average correlation strength is very high, at $\langle \rho \rangle$ = 0.91, indicating an excellent recovery of halo masses from the pipeline on average. Additionally, we show the fraction of catastrophic outliers ($f_{\rm out}$), cases where the discrepancy is an order of magnitude or more. These are very rare occurrences ($\langle f_{\rm out} \rangle < 0.01$), but are important to keep track of in the overall performance. We also show the correlation strengths in parenthesis for the $\sim$ 99\% of the data which is not an outlier. Here the correlation strength rises to $\langle \rho \rangle$ = 0.94.

In Fig. 7, we show the distributions in the the offsets in halo masses, defined explicitly as:

\begin{equation}
\delta M_{\rm Halo} \equiv  \,\, M_{200} \, ({\rm True}) - M_{200} \, ({\rm Predicted}).
\end{equation}

\noindent On each panel we present the bias ($b$, computed as the median value of $\delta M_{\rm Halo}$), the 1$\sigma$ uncertainty ($\sigma_1$, spanning the 16th - 84th percentile of the PDF), the 2$\sigma$ uncertainty ($\sigma_2$, spanning the 2.5th - 97.5th percentile of the PDF), and the catastrophic outlier fraction ($f_{\rm out}$, defined above). 

Considering just the top two rows, where the same model is used for each data set, we see that the halo finder recovers halo masses with essentially no bias ($|b| \leq 0.02$ dex) and an average uncertainty across redshifts of $\langle \sigma_1 \rangle \,\, (\langle \sigma_2 \rangle) = 0.12 \,\, (0.28)$ dex. This is an extremely good performance, which substantially improves on the current status quo in halo mass estimation for wide-field galaxy surveys (e.g., $\sigma_1 \sim$ 0.2 - 0.5 dex in \citealt{Yang2007, Yang2009, Woo2013}).

In Fig. 8, we show distributions of the offsets in estimated halo masses ($\delta M_{200}$) for the full group model from DfL (shaded histograms) and the central-only model (open black histograms). The top panel shows results for single galaxy `groups'. The bottom panel shows results for multi-galaxy groups (also including clusters). In both panels the TNG data is shown as the top row and the EAGLE data is shown as the bottom row.

For single galaxy groups, the group and the central models yield qualitatively very similar results, as expected. However, for the multi-parameter groups, the central galaxy model is significantly biased to lower halo mass predictions (higher values of $\delta M_{200}$) and has increased uncertainty, relative to the group model. This highlights the clear value in incorporating group information into the estimation of the halo masses within the DfL pipeline. Typically, the halo masses are biased by $\sim 0.1$ dex and there is an increase in uncertainty of 20 - 35\%.

In Fig. 9, we present the halo mass functions for the actual simulations (shown as shaded regions) compared to those output from the DfL halo finder pipeline (shown as red stars with error bars). This figure is structured as in Figs 6 \& 7, whereby different redshifts are shown along each column and different data - model pairings are shown along each row. Here we discuss the first two rows, which show correctly paired models and data.

Visually, there is excellent agreement between the actual and predicted halo mass functions for each snapshot and in both simulations. We quantify the similarity between the halo mass functions with the mean offset in number densities and the standard deviation in number density offsets (both presented on each panel). 

Finally, in Fig. 10 we present the confusion matrices for central - satellite classification from the halo finder pipeline. The confusion matrix plots true centrals and satellites against predicted centrals and satellites, resulting in four quadrants (from top left to bottom right): (i) true predicted satellites; (ii) false predicted centrals; (iii) false predicted satellites; and (iv) true predicted centrals. Hence, a perfect classification would have all data on the diagonal. The fraction of data in each quadrant is indicated by the darkness of each rectangular region, as indicated by the color bar. Additionally, we present the accuracy, precision, and recall of central - (core) satellite classification on each panel\footnote{For results incorporating the buffer zone of extended `satellites' see Fig. B6, where they are discussed in comparison to the FOF-AM technique.}.

These statistics are defined as:

\begin{equation}
{\rm Accuracy \,\, (Acc.) = \frac{N_{True \, Centrals} + N_{True \, Satellites}}{N_{Total \, Classified}}} \, ,
\end{equation}

\begin{equation}
{\rm Precision \,\, (Prec.) = \frac{N_{True \, Centrals}}{N_{True \, Centrals} + N_{False \,\, Centrals}}} \, ,
\end{equation}

\begin{equation}
{\rm Recall \,\, (Rec.) = \frac{N_{True \, Centrals}}{N_{True \, Centrals} + N_{False \,\, Satellites}}} \, .
\end{equation}

\noindent The accuracy is intuitive, but the precision and recall can be a little confusing. Essentially, the recall measures the completeness of the output sample and the precision measures its purity. Optimizing for precision often worsens the recall, whereas optimizing for recall often worsens precision. As such, our methodology is to optimize the overall accuracy, which is the best compromise between recall and precision.

\noindent Excluding the extended satellite region, and focusing on the matched models and data, the group finder pipeline achieves an excellent accuracy of $\langle {\rm Acc} \rangle = 96\%$. This indicates that just $4\%$ of galaxies are misclassified as centrals or satellites in the pipeline within one virial radius of each identified central galaxy. Including the extended satellite region (i.e, galaxies within $1 < D_c/R_{200} < 2$ and $1 < \Delta V_{\rm los} / V_{200} < 2$ of their central) the accuracy is significantly lowered to $\langle {\rm Acc} \rangle = 88\%$ (see Fig. B6). But we emphasize that this region is flagged and hence may be excluded from any given analysis.

The extended satellite region is a highly complex part of the parameter space, where true satellites and true centrals mingle. This region contains splash-back satellites, which have reached beyond the virial radius having previously been inside it (see, e.g., \citealt{Adhikari2014, More2015}), satellites falling into the group/ cluster for the first time (which can have $|\Delta V_{\rm los} / V_{200}| \leq \sqrt{2}$), and genuine central galaxies which are independent from the group, though close to it. This is further exacerbated by this extended phase space cut being more prone to chance alignments of galaxies which are at large radial distances from the central, but at similar line-of-sight radial velocities. As such, this region of parameter space is inevitably going to lead to more uncertainty in central - satellite classification. 

We have tested treating this region as all centrals or as all satellites, and the overall accuracy is essentially identical in both cases. We have also tried reducing, and increasing, the buffer zone region and here again the overall accuracy of central - satellite classification is not significantly impacted. Hence, we conclude that this is an unavoidable uncertainty resulting from restricting the true 6D phase space information (in simulations) to observational 2D+$V_{\rm los}$ space. In Section 4.3 we compare to another approach (abundance matching applied to FoF groups) and find similar issues.

As a result of the above tests, we recommend users of the halo finder pipeline to treat the extended satellite region with caution. For applications where purity in central - satellite classification is more important than completeness, we recommend to exclude the extended satellite region from the analysis. This will yield a very high central - satellite classification accuracy. Conversely, for applications where completeness is more important than purity, we recommend to include the extended satellite population, but to appreciate that this results in a $\sim 8\%$ reduction in overall accuracy. In total $\sim 20\%$ of the full sample (and $\sim 50\%$ of all satellites) are in the extended satellite region.

\subsection{Cross validation between models}

\noindent In the previous sub-section we considered the performance of the DfL halo finder pipeline in the case where the random forest regression models are trained on the same simulation as used for the test data (although, crucially, not the same data set or data format; see Section 3.4). This would be adequate if a given model was known to be correct in all aspects. However, in reality the true underlying galaxy formation model of the Universe is unknown. As such, it is essential to attempt to quantify the uncertainty on halo mass recovery induced by the choice of model. To achieve this, we apply the EAGLE model to TNG data, and vice versa. This enables us to determine a preliminary estimate of the impact of the model choice on the final halo mass estimates, and group membership.

We present the cross-validation of mismatched model and data pairings as the bottom two rows in Figs. 6, 7, 9 \& 10. The third row indicates the EAGLE model applied to TNG observational-like data, and the bottom row indicates the TNG model applied to EAGLE observational-like data.

Unlike in the case of matched models and data, in Fig. 6 we see that the contours (and data points) are slightly offset from the one-to-one relationship. The EAGLE model applied to TNG data systematically over-estimates the true values of halo masses, whereas the TNG model applied to EAGLE data systematically under-estimates the true values of halo masses. Nonetheless, the correlation strengths remain extremely high and the fraction of catastrophic outliers remains very low (see values on lower panels in Fig. 6).

More quantitatively, in the lower half of Fig. 7 we display the bias and error from applying a mismatched model to the simulated observational-like data. From applying the EAGLE model to TNG data, we find an average bias of $\langle b \rangle = -0.10$ dex. Conversely, for the TNG model applied to EAGLE data we find an average bias of $\langle b \rangle = +0.09$ dex. This implies that the choice of model results in a systematic uncertainty of $\sim \pm0.1$ dex on the halo mass estimates. The random error (computed as the $\sigma_1$ statistic) is essentially unchanged by applying mismatched models ($\langle \sigma \rangle =$ 0.13 dex). This indicates that the model uncertainty results in a systematic bias, but very little increase in random uncertainty after this is taken into account.

\begin{table*}
\begin{center}
\caption{Performance Comparison between {\small DfL} and {\small FOF-AM}.} 
\label{tab2}
\begin{tabular}{ | l | c | c | c |} 
 \hline 
Method/ Sim.  & z = 0  &  z = 1 & z = 2    \\
  \hline\hline & & &  \\
  
DfL: & & &    \\
TNG (TNG)       & (0.01, 0.11, 0.27, 0.01)  & (0.00, 0.12, 0.28, 0.00)   & (-0.01, 0.13, 0.27, 0.00)    \\ 
EAGLE (EAGLE)   & (0.00, 0.12, 0.30, 0.01)  & (-0.01, 0.12, 0.30, 0.01)   & (-0.02, 0.14, 0.28, 0.00)    \\
TNG (EAGLE)     & (-0.10, 0.12, 0.27, 0.00) & (-0.08, 0.13, 0.29, 0.00)  & (-0.13, 0.13, 0.28, 0.00)   \\
EAGLE (TNG)     & (0.11, 0.13, 0.44, 0.02)  & (0.07, 0.14, 0.32, 0.01)   & (0.10, 0.15, 0.30, 0.00)     \\
& & & \\

FOF-AM: & & &  \\
TNG (TNG)       & (0.01, 0.13, 1.13, 0.06)  & (-0.01, 0.15, 0.83, 0.04)  & (-0.01, 0.15, 0.55, 0.02)      \\ 
EAGLE (EAGLE)   & (0.00, 0.15, 1.06, 0.06)  & (-0.02, 0.15, 0.71, 0.03)  & (-0.01, 0.16, 0.39, 0.01)      \\
TNG (EAGLE)     & (-0.10, 0.15, 1.15, 0.06)  & (-0.08, 0.17, 0.86, 0.04, )  & (-0.13, 0.16, 0.58, 0.01)   \\
EAGLE (TNG)     & (0.12, 0.15, 1.14, 0.07)   & (0.05, 0.15, 0.70, 0.03)  & (0.10, 0.15, 0.39, 0.01)     \\
& & & \\
 \hline
\end{tabular}
\end{center}
Notes: For each cell the performance metrics are given as: ($b \, , \, \sigma_1, \, \sigma_2, \, f_{\rm out}$). The biases and dispersions are all given in units of [dex], whilst the catastrophic outlier fractions are unitless.
\end{table*}

Of course, exploration of further galaxy formation models may result in different biases in principle. Indeed, it is our intent to add to the usefulness of DfL as a tool by incorporating more simulations into later versions, enabling a greater sampling of the contemporary model parameter space. However, for now, we note that EAGLE and TNG are very different models in terms of the baryonic physics applied, especially in terms of feedback (see Section 2). Hence, EAGLE and TNG provide a good baseline comparison for exploring how significant variation in the baryonic sub-grid physics of feedback is on the estimation of halo masses from luminous tracers.

In the lower half of Fig. 9, we compare the estimated halo mass functions from the DfL halo finder pipeline (trained with mismatched models) to the true halo mass functions from the simulations. Here the results are quite instructive. Whilst the halo mass function recovery is certainly not unreasonable, it is noticeably worse than in the case of matched model and data pairings. The EAGLE model tends to over-estimate the number densities of high mass haloes relative to TNG, whereas the TNG model tends to under-estimate the number densities of high mass haloes.

Finally, in the lower half of Fig. 10, we show the performance of the halo finder on central - (core) satellite classification for the case of mismatched models and data. The accuracy in central - satellite classification, both including and excluding the extended satellite population, is extremely similar in the mismatched case to the matched case. This implies that the identification of galaxy classes is not strongly impacted by the galaxy formation model choice. Ultimately, this is logical because the impact of the model will only (slightly) bias the halo mass estimates, and not impact directly the relative location of centrals and satellites, or their stellar masses. It is the latter which is most critical for determining group membership in most cases.

\subsection{Benchmarking: performance comparison with FOF-AM}

\noindent In this sub-section we compare the performance of DfL to a more established set of techniques. More explicitly, we combine friends-of-friends group finding (FOF, see e.g., \citealt{Davis1985, Springel2001, Springel2005, More2011, Duarte2014, Tempel2016}) with abundance matching (AM; see e.g., \citealt{Vale2004, Moster2010, Behroozi2010, Hearin2013}) in order to make halo mass predictions on the basis of total group stellar mass. This approach has much precedence in observational applications (see especially \citealt{Yang2007, Yang2009}).

In order for a fair comparison we must conduct abundance matching with total group stellar mass, not central stellar mass. This is because we have already established that for multi-galaxy haloes information on the group stellar mass yields superior results to the central galaxy alone (see back to Section 4.1). To achieve this one must already have groups defined. Indeed, even to reliably run on central galaxies one must already have a reasonable estimate of the group structure of the survey. As such, for this comparison, we first run FOF on the observational-like testing data. Using the outputs of the FOF group-finding algorithm, we next use AM from the total stellar mass of each group compared to the halo mass, as constrained in TNG and EAGLE (as with DfL). We refer to this combined approach as: FOF-AM.

Full details on our application of FOF-AM in TNG and EAGLE are provided in Appendix B, including information on optimizing the linking-lengths used on sky and in redshift space (see Fig. B1), and a demonstration of the abundance matching procedure with the simulations (see Fig. B2). We also show in Appendix B reproductions of the main performance plots for FOF-AM (see Figs. B3 - B6). For the sake of brevity, in this sub-section we concentrate on the bottom line - a direct comparison of performance statistics.

In Table 2 we show a side-by-side comparison between DfL and FOF-AM for the various training samples and model pairings at $z = 0, 1, 2$. Within each cell of Table 2 we show four performance statistics: \\

\begin{enumerate}[i]

\item $b$: the median value of $\delta M_{200}$, \\
\item $\sigma_1$: the range in $\delta M_{200}$ containing 68\% of the full sample, \\
\item $\sigma_2$: the range in $\delta M_{200}$ containing 95\% of the full sample, \\
\item $f_{\rm out}$: the fraction of data with $|\delta M_{200}| > $ 1 dex.

\end{enumerate}

In both DfL and FOF-AM, halo masses are recovered without any significant bias when comparing runs with the same testing data and model pairings. In the case of mismatched model and data pairings, both DfL and FOF-AM experience very similar (almost identical) biases. This indicates that the biases learned from model choices will equally impact these different methodologies to infer halo masses. This is not surprising given that both approaches require a training set for the halo masses, and hence will learn from that a structure which will not in general be identical to other simulations and models.

In terms of $\sigma_1$, the performance of DfL and FOF-AM is quite similar, but DfL is marginally superior across all of the different testing modes and redshift snapshots. However, when viewing $\sigma_2$ and $f_{\rm out}$, we see that DfL performs with much greater accuracy than the more conventional FOF-AM approach. This indicates that DfL is much less likely to erroneously misplace galaxies into groups than the more standard FOF-AM approach, and hence is less likely to lead to catastrophic outliers and long tails in the $\delta M_{200}$ distributions.

By comparing Fig. 7 to Fig. B4, one can immediately see that both FOF-AM and DfL recover narrow Gaussian-like distributions centered on $\delta M_{200}$, but that there is more of a power-law tail in the former than the latter. This is further evidenced by the contour relationships (compare Fig. 6 to Fig. B3), which show significantly reduced correlation strengths in FOF-AM compared to DfL, except when excluding catastrophic outliers. Additionally, the recovery of the halo mass functions is visibly less accurate in FOF-AM compared to DfL (compare Fig. 9 to Fig. B5).

\begin{figure*}
\begin{centering}
\includegraphics[width=0.48\textwidth]{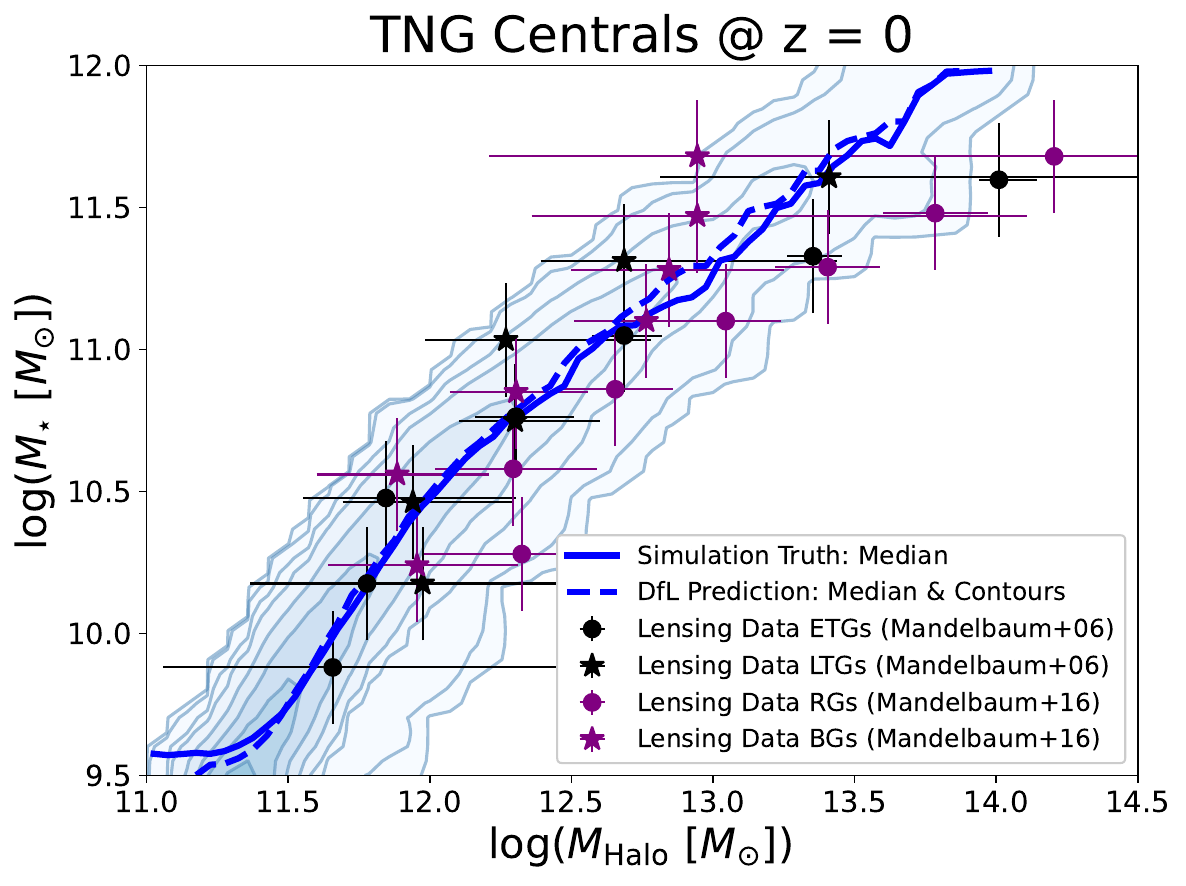}
\includegraphics[width=0.48\textwidth]{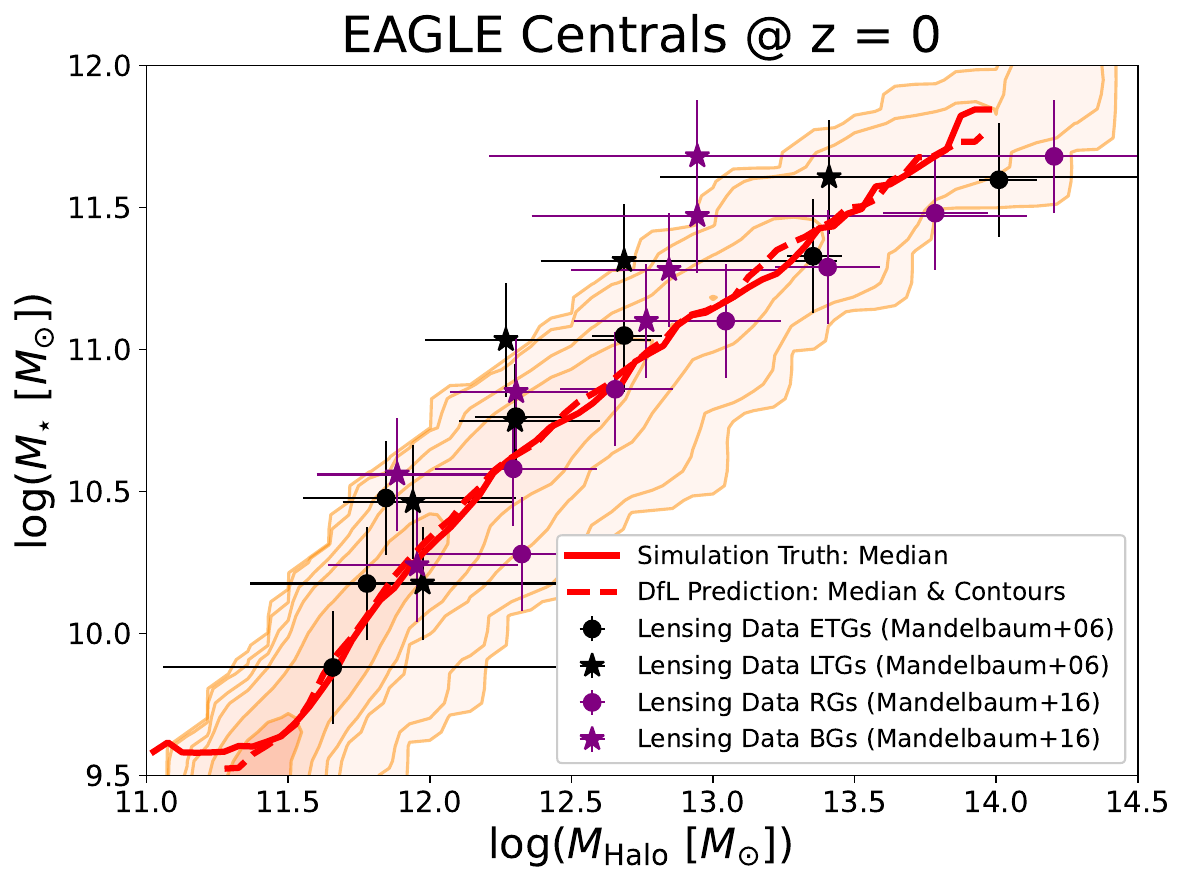}
\caption{A test of {\small DfL} with observational data at $z \sim 0$: The stellar mass - halo mass relation for central galaxies in TNG (left-hand panel) and EAGLE (right-hand panel), compared to observations from strong gravitational lensing. On both panels the solid colored line indicates the true stellar mass - halo mass relation from each simulation, with the dashed colored line indicating this relationship as recovered from observational-like data with the {\small DfL} pipeline. Clearly, there is excellent agreement, demonstrating that the {\small DfL} pipeline can reproduce the true stellar mass - halo mass relationship from each simulation to high fidelity. The contours on each panel show the full range in the stellar mass - halo mass relationship as ascertained through the {\small DfL} pipeline. Additionally, observational constraints from gravitational lensing are shown on each panel, taken from \cite{Mandelbaum2006, Mandelbaum2016}. In both cases, {\small DfL} does a good job at reproducing the observations (both the general trend and the observed scatter). However, the TNG simulation appears visually to be slightly more accurate compared to EAGLE at $z = 0$. }
\label{fig11}
\end{centering}
\end{figure*}

Taken in concert, the results outlined in Table 2 indicate that DfL performs at least as well as FOF-AM in all metrics, and clearly outperforms FOF-AM in the wings of the $\delta M_{200}$ distributions. Ultimately, this is achieved by DfL accounting for the virial structure of the haloes by incorporating physically motivated group sizes (in both physical and velocity space) into the group finding algorithm (see Section 3). Additionally, the buffer zone incorporated into the DfL group structure helps to insulate groups from the uncertain region on their edges, preventing the over- or under- assignment of haloes to galaxies. This is a key aspect of why there are significantly fewer catastrophic outliers in DfL compared to FOF-AM.

In principle one could incorporate a similar outlier region in FOF-AM. However, it is important to appreciate that this is not normally done and, moreover, that this is not a natural way to implement a linking-length algorithm. In more detail, FOF operates by assigning group membership based on the distance to {\it any} group member in comparison to a universal threshold, whereas DfL operates by assigning group membership based on the distance to the {\it central galaxy} compared to its estimated virial properties. In order to define the outlier region one needs to establish the distance from the central to each galaxy in addition to the virial parameters, and hence one ends up with an approach which is essentially a hybrid of DfL and FOF-AM. This is not suitable for a clear comparison of methodologies.

One final thing to note is that DfL provides a single use algorithm to infer the group structure, central - satellite classification, and estimation of halo masses (plus virial parameters) in observational data. In comparison, FOF-AM requires a three stage approach -- (i) optimizing the group finder to the observational data set, (ii) running the group finder, and (iii) running abundance matching on the group outputs in comparison to a given simulation or model. Hence, in addition to outperforming FOF-AM in a number of performance metrics, DfL is also more efficient and faster to run on observational data.

\subsection{Comparison to observations}

\noindent So far in the Results section we have assessed the performance of DfL on various simulated data sets, and in comparison to a more established technique (FOF-AM). In this sub-section we compare the outputs of DfL applied to TNG and EAGLE to observational constraints at $z \sim 0$ from gravitational lensing.

In Fig. 11 we present the stellar mass - halo mass ($M_* - M_{\rm Halo}$) relationship for TNG (left panel) and EAGLE (right panel), both at $z = 0$. In both panels the estimates derived from DfL are shown as contours, with the median relationship shown as a dashed solid line. We compare this with the simulation truth (shown as a solid line of the same color). Clearly, DfL does an excellent job of recovering the true $M_* - M_{\rm Halo}$ relationship from unseen, observational-like input data in both TNG and EAGLE. This is very reassuring as it demonstrates that one can reconstruct this key scaling relationship in observer (2D+$z$) space to high fidelity using the DfL methodology.

Furthermore, on both panels in Fig. 11 we show observational constraints on the $M_* - M_{\rm Halo}$ relationship from strong gravitational lensing. Data points are taken from \cite{Mandelbaum2006, Mandelbaum2016}. Observational measurements are shown separately for early type galaxies (ETGs), late type galaxies (LTGs), red galaxies (RGs), and blue galaxies (BGs). As such, these data span a high diversity of galaxy types and a large range in both stellar and halo masses. Uncertainties on the halo masses are taken from \cite{Mandelbaum2006, Mandelbaum2016} directly. However, no uncertainties on stellar masses are provided in those publications. As such, we add a blanket stellar mass uncertainty of 0.2 dex to all data points as the Y-axis error bars, which is well established as a reasonable baseline (see, e.g., \citealt{Brinchmann2004, Peng2010, Mendel2014}).

The simulated $M_* - M_{\rm Halo}$ relationship (and most importantly, the recovered DfL version) are in very good accord with the observational constraints. In the case of TNG, the DfL output contours are almost perfectly spanned by the observational constraints. In the case of EAGLE, there is a slight tendency for the observational constraints to be higher in stellar mass at a fixed halo mass (i.e., more data points lying above the DfL median relationship than below). Nonetheless, the observational data is still reasonably well supported by the DfL recovered simulated extent of this relationship.

Taken in concert, Fig 11 demonstrates the potential of DfL to recover the true underlying $M_* - M_{\rm Halo}$ relationship from simulations under observational limitations. Moreover, Fig. 11 further establishes that this approach leads to a very reasonable estimate of the $M_* - M_{\rm Halo}$ relationship as constrained by strong gravitational lensing observations at $z \sim 0$. This is very encouraging for the application of DfL at higher redshifts, where the observational constraints from gravitational lensing are much more sparse, and biased to the highest masses.

\subsection{The impact of stellar mass uncertainty \& incompleteness on halo mass recovery}

\noindent Up to this point, we have analyzed the performance of DfL with perfect input parameters, albeit limited to observational-like formats. In this sub-section we relax that assumption and systematically assess the impact of observational limitations on the performance of DfL under various conditions.  

In observational data, modern astrometry combined with precision constraints on cosmological parameters is sufficient to yield physical distance accuracy to at least $\sim$ kpc scales at the highest redshifts probed in this work (\citealt{Grogin2011, Cirasuolo2014, PLANCK2020}). This is at least two orders of magnitude smaller than the typical virial radii of groups (0.1 - 1\,Mpc). Additionally, through application of spectroscopic redshifts, line-of-sight velocities are obtainable to $\sim$25-75\,km/s accuracy at cosmic noon (e.g., \citealt{Steidel2014, Cirasuolo2014, Kriek2015, Maiolino2020}). This is significantly smaller than typical group virial velocities at these redshifts ($\sim$ 100 - 1000 km/s). Hence, we do not consider uncertainties in either coordinates or redshifts here as significant limitations to DfL performance in observational data.

On the other hand, stellar mass estimates in observational data are notoriously imprecise, even when using spectroscopic redshifts or high sampled multi-band photometry (see, e.g., \citealt{Mendel2014, Sanchez2016, Dimauro2018, Mucesh2021}). This is in large part due to uncertainty in stellar population synthesis models arising from both stellar population library uncertainty and incompleteness, and from inherent degeneracies between stellar population age and metallicity (among several other parameters). Accounting for all of this uncertainty, most modern stellar mass estimates applied with known (spectroscopic) redshifts are $\sigma (M_*) \sim 0.2 - 0.3$ dex (e.g., \citealt{Mendel2014, Dimauro2018, Sanchez2016, Sanchez2020}). Hence, the impact of stellar mass uncertainty in halo mass recovery with DfL is an essential issue to investigate.

Additionally, incompleteness in observational data due to both spectroscopic fiber collisions and photometric magnitude limits may bias the results obtained through the halo finder pipeline (e.g., \citealt{Cirasuolo2020}). Here we consider the case where the survey is stellar mass complete at $M_*~>~10^{9.5} M_{\odot}$ (the approximate mass completeness limit of VLT-MOONRISE, see \citealt{Maiolino2020}), but incurs a variable (mass-independent) incompleteness, due to fiber collisions and survey planning limitations.

To simulate stellar mass uncertainty we construct a new stellar masses from the original data set by taking a random draw from a Gaussian distribution centered on the true stellar mass, with a standard deviation set equal to $\sigma(M_*)$ = 0 - 0.5 dex (in steps of 0.1 dex). That is, before sending to the pipeline in each run, we set:

\begin{equation}
M_* \,\, \rightarrow \,\, \mathcal{R} \big(\,\mathcal{N}\,[\mu = M_* \,;\, \sigma = \sigma(M_*)] \, \big)
\end{equation}

\noindent for each individual galaxy in the observational-like simulated galaxy survey. This is a rather simplistic modeling of stellar mass uncertainty. However, this approach is appropriate for modeling stellar mass uncertainties in spectroscopic data where the full PDFs are typically uni-modal, due to the accurate constraints on redshifts. Moreover, the use of more complex PDF shapes is not strongly motivated here. That said, DfL enables users to input any known PDF format for stellar masses into the pipeline. This is then fully propagated through the pipeline to yield the impact on the final halo mass PDF (see Section 3). Here our intent is simply to assess the impact of the width of the stellar mass PDF on the resultant errors incurred on halo masses through the DfL pipeline. This is most straightforwardly achieved with a normal distribution.

To simulate incompleteness we randomly remove 0 - 50\% of the data from the input sample (in steps of 10\%) prior to running the halo finder pipeline. Of course, in any real galaxy survey the loss in galaxies will not be strictly random. However, until a specific observing strategy is available this is the most logical manner in which to assess the impact of incompleteness. Obviously, this results in cases where there is no halo estimate at all. However, our interest lies in the case where there is an estimate (i.e., a predicted central galaxy exists in the survey data). Hence, we aim to quantify how the incompleteness of other galaxies impacts the halo mass estimate of surviving galaxies.

\begin{figure*}[h!]
\begin{centering}
\includegraphics[width=0.45\textwidth]{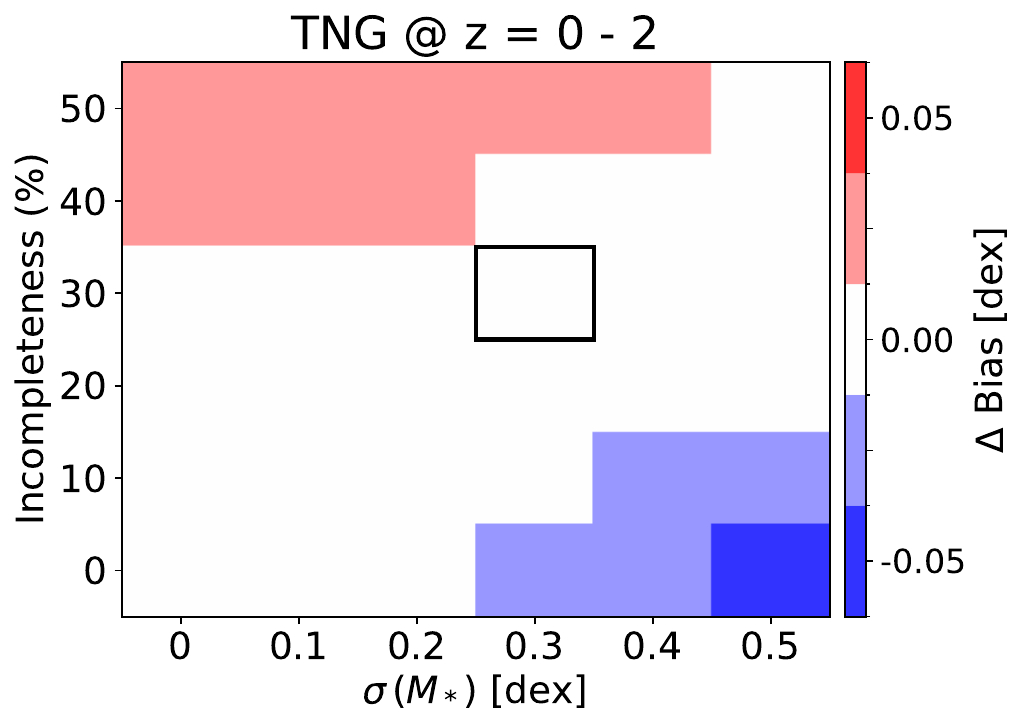}
\includegraphics[width=0.45\textwidth]{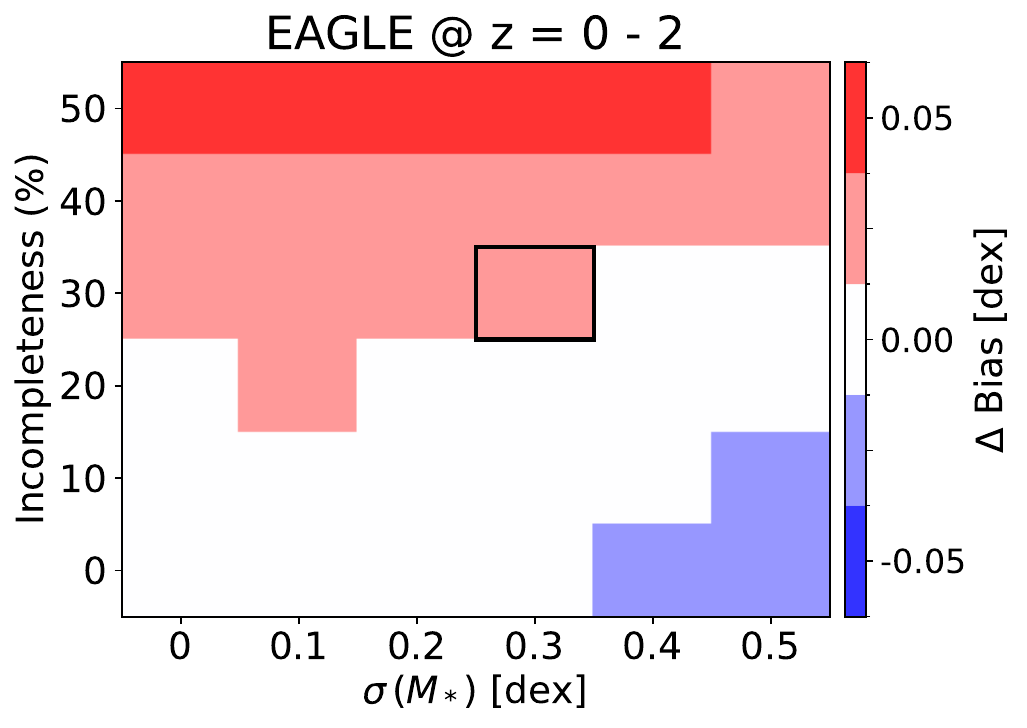}
\includegraphics[width=0.45\textwidth]{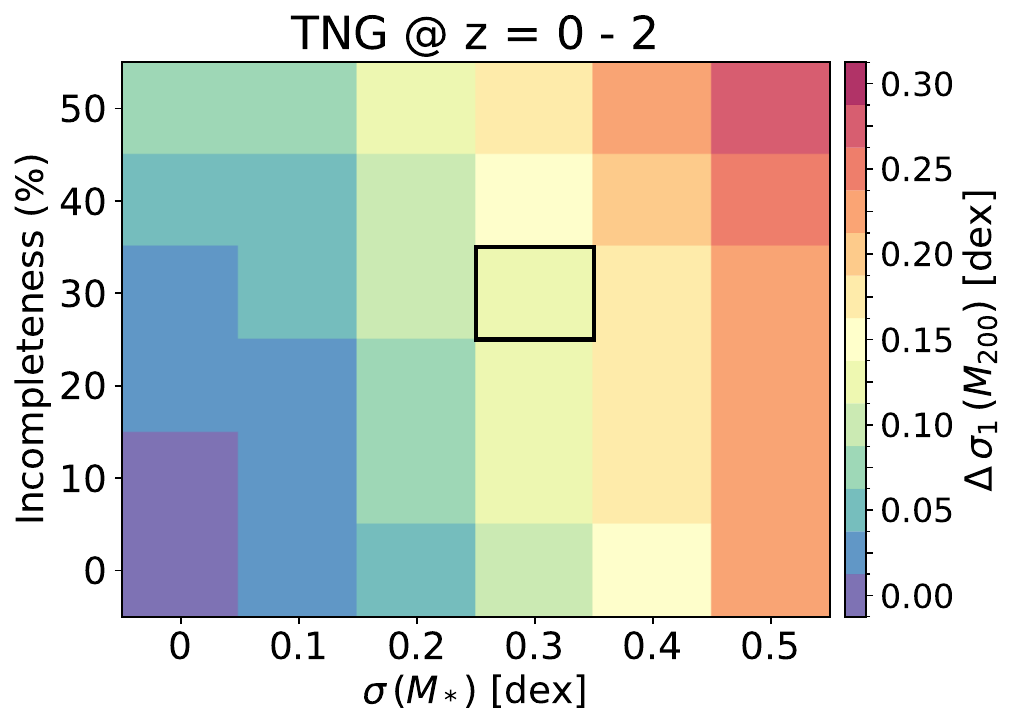}
\includegraphics[width=0.45\textwidth]{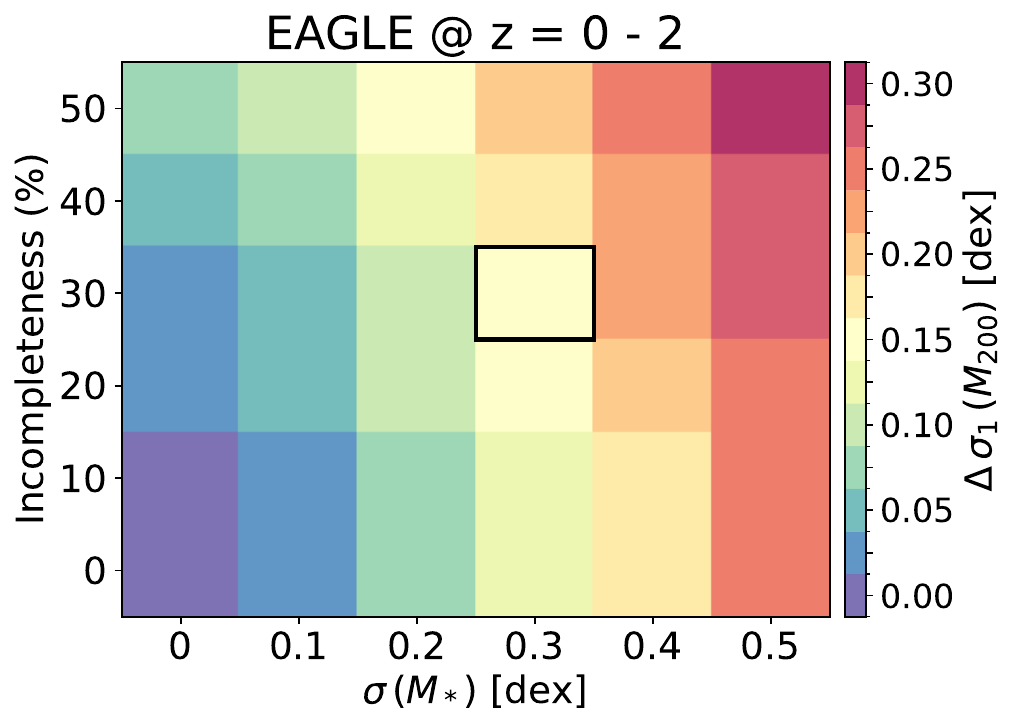}
\includegraphics[width=0.45\textwidth]{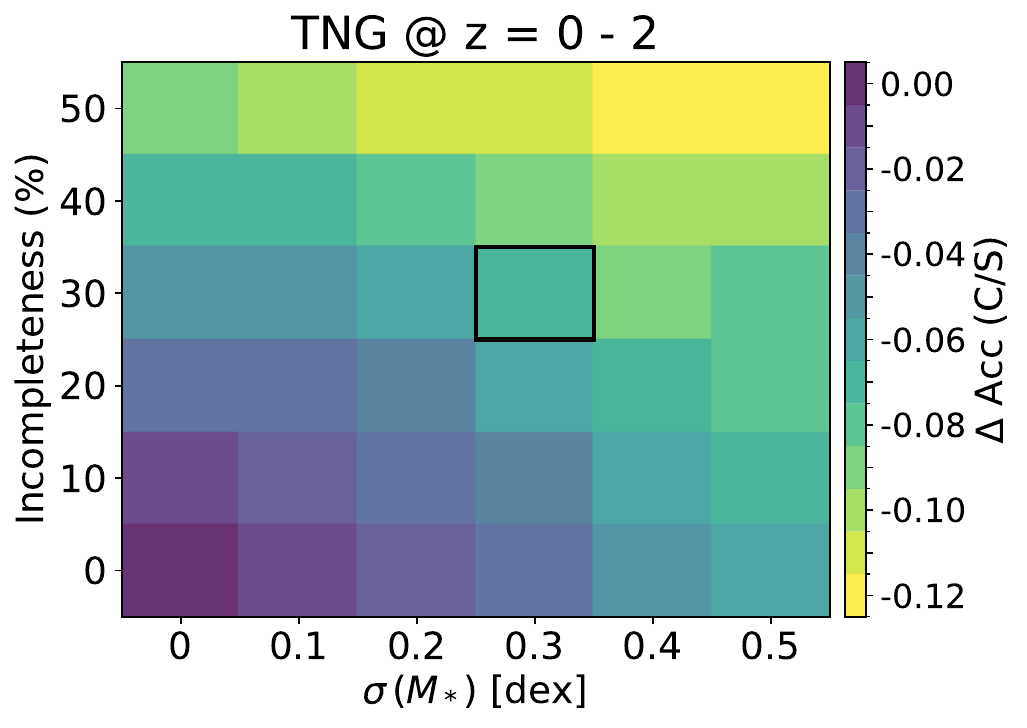}
\includegraphics[width=0.45\textwidth]{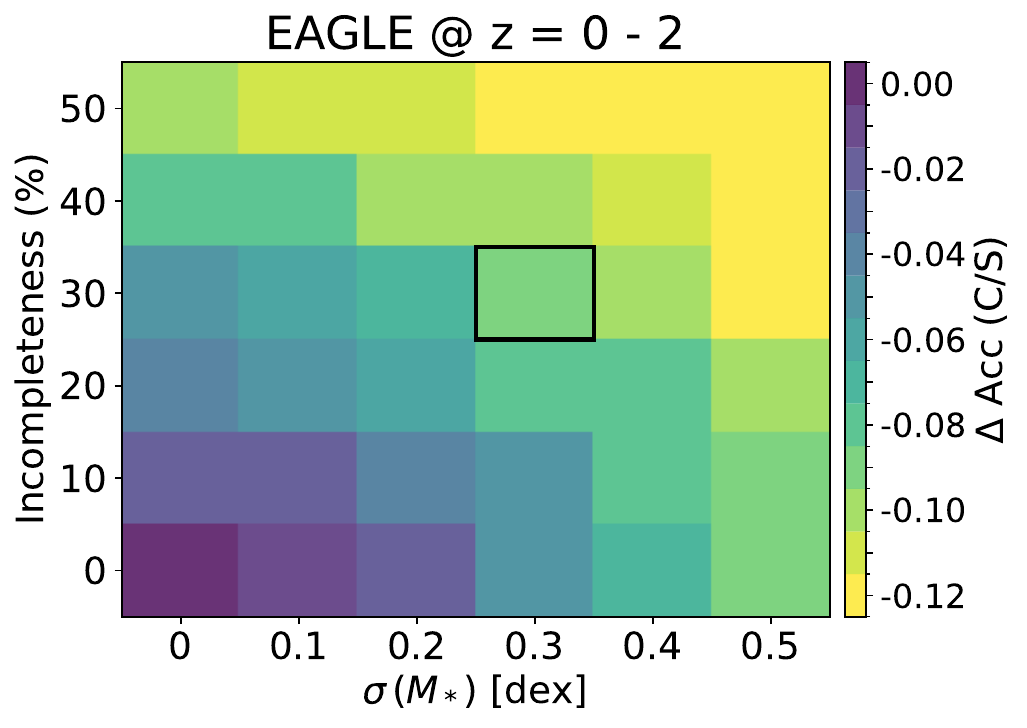}
\caption{Halo recovery performance from {\small DfL} as a function of spectroscopic incompleteness and stellar mass uncertainty. The left-hand column shows results for the TNG simulation and the right-hand column shows results for the EAGLE simulation. The top row shows the systematic shift in bias of halo mass recovery ($\Delta$\,Bias), the middle row shows the systematic shift in statistical uncertainty of halo mass recovery ($\Delta \sigma_1 (M_{200})$), and the bottom row shows the systematic shift in central - satellite classification accuracy ($\Delta$\,Acc(C/S)). These statistics are all calculated as a function of increasing stellar mass uncertainty and survey incompleteness, relative to the case of a complete sample with zero stellar mass uncertainty. Hence, this figure enables one to ascertain how these two critical observational limitations will impact the output halo properties from {\small DfL} under realistic observational conditions. The expected survey incompleteness and stellar mass uncertainty of MOONRISE is indicated by a black box on each grid. We expect DfL run on MOONRISE not to incur increased bias, to have increased statistical uncertainty of $\sim$0.15\,dex, and to have decreased accuracy in central satellite classification of 5 - 8\% (dependent on simulation). These values are measured relative to the case of running on a complete observational spectroscopic sample, with perfect knowledge of stellar masses. }
\label{fig12}
\end{centering}
\end{figure*}

In Fig. 12 we systematically explore the impact of stellar mass uncertainty and survey incompleteness on the recovery of halo mass and central - satellite classification with DfL. The left column shows results for the TNG simulation, with the right column showing results for the EAGLE simulation. The top row shows the impact on the bias ($b$), the second row shows the impact on the dispersion ($\sigma_1$), and the bottom row shows the impact on central - satellite classification accuracy (Acc). On each panel, data is used from all redshift ranges from $z = 0 - 2$, and the statistics are measured as offsets relative to the performance with $\sigma(M_*) = 0$\,dex and incompleteness (INC) = 0\%.

In terms of bias (top row of Fig. 12), we note a weak systematic bias to overestimate halo masses with increasing stellar mass uncertainty, and a weak systematic bias to underestimate halo masses with increasing incompleteness. Both of these trends are expected. In the case of missing galaxies, in general the stellar mass of groups will be lower leading to underestimates in halo masses. Furthermore, due to the steepness of the stellar mass function, increasing stellar mass uncertainty leads to more low mas galaxies moving higher in mass estimates (beyond the mass limit) than the other way around. This in turn leads to a small systematic bias to overestimate the halo masses of groups. That said, the biases incurred by increasing stellar mass uncertainty and survey incompleteness are generally quite small. For example, in VLT-MOONRISE, with an estimated $\sigma(M_*) = 0.3$\,dex and INC = 30\% (see \citealt{Maiolino2020}), we would expect to engender no significant additional bias in the recovery of halo masses. This region is highlighted with a solid box in Fig. 12 on all panels for reference. Note that the bias incurred due to model choice will still be an important factor (see Section 4.2).

In terms of dispersion (center row of Fig. 12), we note a systematic increase in the uncertainty on halo mass recovery with both increasing stellar mass uncertainty and survey incompleteness. Again, this is as expected. Increasing the uncertainty on the stellar mass inputs will lead to more uncertainty on the halo mass outputs. Additionally, increasing incompleteness will add more noise into the group membership, leading to more uncertainty in halo mass recovery. However, the dependence is not symmetric. The impact of stellar mass uncertainty is greater than the impact of survey incompleteness. For the VLT-MOONISE specifications (mentioned above), we anticipate increasing the uncertainty on the halo mass estimates over their optimal values (see Fig. 7) by $\sim$0.15\,dex. This leads to a total uncertainty of $\sim 0.25 - 0.3$\, dex expected in this upcoming survey. Note that this is a significant improvement on halo mass recovery with stellar mass uncertainty at $z \sim 0$ with abundance matching techniques ($\sim 0.3 - 0.5$\,dex, see, e.g., \citealt{Yang2007, Woo2013}).

Finally, in the bottom panels we explore the impact of stellar mass uncertainty and incompleteness on the classification of central and satellite galaxies in DfL. We find that the classification accuracy systematically decreases with both increasing stellar mass uncertainty and increasing incompleteness in the survey. Here, however, the impact of incompleteness is similarly important to that of stellar mass uncertainty (unlike in the case of dispersion, above). Again this is as expected. Higher uncertainty on stellar masses will lead to a greater chance that a satellite overtakes the true central in measured stellar mass, leading to a misclassification. Additionally, absent data may on occasion remove a central from a group, which will lead to a satellite taking its place (as the most massive galaxy), again leading to misclassification. For VLT-MOONRISE specifications, we anticipate a reduction in central - satellite classification accuracy of 5 - 8\%. This would lead to a final central - (core) satellite classification accuracy of 88 - 91\%.

Taken in concert, Fig. 12 can be used to infer how survey limitations (explicitly due to stellar mass uncertainty and incompleteness) will impact the performance of DfL in recovering halo masses and in central - satellite classification. We note that these effects are not strongly redshift dependent, and hence we present in Fig. 12 the results for the entire redshift range for the sake of brevity. However, redshift specific performance matrices are available from the corresponding author upon request.

\section{Discussion: applications of DfL in wide-field spectroscopic galaxy surveys}
\label{sec5}

\noindent One of the immediate intended applications of the DfL halo finder pipeline presented in this work is to the VLT-MOONRISE wide-field spectroscopic galaxy survey targeting cosmic noon, explicitly $z = 0.8 - 2.0$ (see \citealt{Cirasuolo2020, Maiolino2020}). MOONRISE will obtain rest-frame optical spectra for $\sim$500k galaxies across three wide-field regions within (i)~COSMOS; (ii) EGS; and (iii) VIDEO, each with $\geq 1 {\rm deg}^2$ area on sky. As such, MOONRISE will provide near optimal input data for DfL, accessing a unique cosmological parameter space hitherto unexplored with dense spectroscopic sampling across a large volume. Initial data from this unprecedented galaxy survey will become available to the community by the end of 2026, with the final survey data publicly available by the end of 2031\footnote{Dates may be subject to change due to technical and survey planning issues.}. 

It is anticipated that MOONRISE will achieve approximate stellar mass completeness above $M_* > 10^{9.5} M_{\odot}$, and spectroscopic sampling completeness of $\sim$70\% for a large fraction of the survey (\citealt{Maiolino2020}). Additionally, we estimate that stellar masses will be recovered using extant multi-band photometry plus new spectroscopic redshifts with an accuracy of $\sigma(M_*) \leq 0.3$\,dex. This will likely be improved upon by the subset of data for which full continuum fitting is feasible (anticipated to be $\sim$50\% of galaxies), and for galaxies with extensive JWST multi-band follow-up. The latter will improve stellar mass estimates by constraining the redder part of the spectrum, enabling more stringent constraints on dust content.

The performance expectations for DfL under the planned limitations of MOONRISE are indicated by bold rectangular regions in Fig. 12 (as discussed in the previous sub-section). We anticipate to be able to achieve halo masses from $M_{\rm Halo} = 10^{11} - 10^{14} \, M_{\odot}$ with uncertainties of $\sigma(M_{\rm halo}) \leq 0.3$\,dex and a model dependent bias of $\pm 0.1$\,dex. Moreover, we will provide full PDFs for all halo mass estimates, enabling rigorous accounting for uncertainty in various scientific applications. Furthermore, we also anticipate to achieve a central - (core) satellite classification accuracy of $\sim$ 90\%. In this subsection we list several potential scientific applications of the MOONRISE halo mass and group catalog from DfL. Many, if not all, of these applications will also be feasible with other wide-field galaxy surveys moving forwards.

(i) Halo mass functions \& the growth of large scale structure: Perhaps the most direct application of the halo finder pipeline to MOONRISE will be the formation of halo mass functions from $z = 0.8 - 2.0$. This will be the first time that accurate halo mass functions can be empirically determined at these relatively early cosmic times across a large dynamic range in halo mass (up to three orders of magnitude). This will directly probe the growth of large scale structure as a function of cosmic time. Moreover, these halo mass functions can be directly compared to measured halo mass functions at lower redshifts, enabling strong empirical constraints across $\sim$10\,Gyrs of cosmic history. 

Furthermore, one can also use these constraints on the growth of large scale structure to test and constrain cosmological models. In the first instance, one can compare to dark mater only simulations to gauge the accuracy of the underlying cosmological model and hierarchical assembly physics from gravitation alone. Interestingly, one can even use each hydrodynamical simulation in question to predict the best light-to-dark mapping. Differences in the halo mass functions which persist will point to statistically meaningful differences between the galaxy formation model and the real Universe, which can be used to constrain the feedback mechanisms used in future generations of cosmological simulations. In the most extreme case, a fundamental mismatch between predicted and inferred halo mass distributions may point towards new fundamental physics (see also point (v)).

(ii) The role of cosmic environment in galaxy formation: In the hierarchical theory of structure formation, galaxies form within dark matter haloes and grow as a result of both accretion of dark matter and baryons and through halo (and galaxy) mergers (e.g., \citealt{Cole2000, Bower2006, Vogelsberger2014a, Schaye2015, Dave2019}). Using the DfL halo finder presented in this work applied to MOONRISE, in conjunction with computational morphology on extant imaging and close pair statistics in MOONRISE fields, one may investigate how the merger rate of galaxies varies as a function of halo mass and galaxy class (central vs. satellite) across cosmic time. This will reveal where and when in the history of the Universe galaxy formation occurs and is most efficient. Furthermore, through combining this data with star formation rate tracers (e.g., extinction corrected $H$$\alpha$ luminosity, see \citealt{Kennicutt1998}), one can also quantify whether in situ star formation or mergers dominate the growth of stellar mass in galaxies as a function of location within the cosmic web.

Ultimately, probing the relationship between halo mass and stellar mass as a function of cosmic time will provide precision new constraints on the formation of galaxies within their dark matter haloes. Comparison with simulations will be particularly constraining of feedback mechanisms, both stellar and AGN. Note that even though stellar mass is used as an input to the halo finder, the number of galaxies within a group and the total group stellar mass significantly ameliorate the issue of comparing the same parameter to itself. For instance, using extant halo catalogs based on stellar mass inputs at $z \sim 0$, important differences in the dependencies of galaxy properties on stellar and halo mass have been found (e.g., \citealt{Yang2009, Woo2013, Bluck2014, Bluck2016}).

(iii) The role of cosmic environment in galaxy quenching: One of the two primary routes to quench a galaxy is through environment, the other being via AGN feedback (see, e.g., \citealt{Peng2010, Peng2012, Bluck2020b, Bluck2020a, Bluck2022, Piotrowska2022, Goubert2024}). Using the group catalogs output by the halo finder, in addition to the halo mass estimates, one can probe how galactic star formation and its cessation through quenching is connected with cosmic environment as a function of cosmic time. 

We will be able to reveal where and when in the history of the Universe dense environments are conducive to star formation and when they become hostile to future star formation, resulting in galaxy quenching. Moreover, we will also be able to investigate the relationship of galactic star formation and quenching to both the mass of groups and clusters and to the location of satellites within their groups. This is unprecedented science at cosmic noon, due largely to the prior lack of wide-field spectroscopic galaxy surveys targeting this epoch. This is essential for accurate group identification on large scales and for the estimation of the dark sector through DfL.

(iv) The role of cosmic environment in triggering active galactic nuclei (AGN): Using the DfL halo mass catalogs for MOONRISE in addition to optical AGN identification through emission lines (e.g., \citealt{Baldwin1981, Kewley2001, Kauffmann2003, Kewley2006}), one will be able to search for possible links between AGN and cosmic environment. Questions we can answer with DfL applied to MOONRISE include: are the gas rich cores of high mass groups and proto-clusters the origin of the most powerful quasars in the Universe?; how and when do these environments become stabilized through feedback processes?; is cosmic environment the fundamental driver of supermassive black hole growth?; and, furthermore, which grows first, the dark matter halo, the galaxy, or the supermassive black hole? 

Supermassive black hole masses may be inferred in MOONRISE for broad-line AGN utilizing emission line calibrations (e.g., \citealt{Maiolino2024b, Maiolino2024a}). Additionally, black hole masses may be statistically constrained on a population basis through calibrations with photometric (e.g., the stellar potential, $\phi_* \sim M_*/R_e$) and kinematic (e.g., the central velocity dispersion, $\sigma_*$) measurements (see \citealt{Bluck2022, Bluck2023, Bluck2024}). Hence, another powerful science application will be to explore how the halo mass - black hole mass relation varies with cosmic time, and location of galaxies within the cosmic web. This will reveal key new insights on whether supermassive black holes grow early or late in the history of the Universe, ultimately answering whether they are the end product of galaxy formation, or, perhaps, a trigger for it.

(v) Direct tests of $\Lambda CDM$ \& alternative cosmologies: Perhaps the most exciting potential application of the halo finder pipeline applied to VLT-MOONRISE data is the potential to constrain the underlying background cosmology. The evolution in the halo mass function at high-mass scales is largely unaffected by baryonic physics (e.g., \citealt{Vogelsberger2014a, Vogelsberger2014b, Schaye2015, Henriques2015, Henriques2019}). Hence, variations in the observed halo mass functions at large scales, when compared to $\Lambda$CDM models, may indicate the need for new physics, if significant tensions emerge. Given the precise knowledge we have gained in this work on both the systematic and random uncertainties of the recovered halo masses, we can rigorously quantify any high-mass deviations from the expected halo mass functions, potentially revealing the need for modifications to $\Lambda$CDM and/or new physics. 

Yet another powerful cosmological constraint can be obtained through the halo merger rate, which is strongly dependent upon cosmological parameters (e.g., \citealt{Conselice2014}). Through kinematics of central - central close pairs, using the classifications from the halo finder, we will be able to make the first direct measurements of halo assembly at cosmic noon. The dependence of halo merger rates is especially sensitive to the dark energy equation of state. Therefore, by comparing the cosmological dark matter halo merger rate at high redshifts to the local Universe, we can obtain independent constraints on the transition from matter to dark energy domination in the Universe, and potentially find evidence for, or against, quintessence (e.g., \citealt{Steinhardt1999, Copeland2006}).

(vi) Environmental physics beyond the virial radius: During the course of this paper the extended satellite population has been more of a challenge than an advantage. However, identifying the region on the edge of galaxy groups and clusters has many potential science benefits. For instance, looking for correlations between star formation enhancements and/or suppression as a function of physical distance from the group center will yield new constraints on the splashback radius (e.g., \citealt{Adhikari2014, Baxter2017}), and the fraction of non-virialized high mass objects at cosmic noon (e.g., \citealt{Chiang2013, Overzier2016}). Ultimately, studies of the extended satellite population (despite its relative impurity) will yield new insights on how environment shapes galaxy evolution beyond the virial radius.

(vii) Cross validation of halo properties with alternative methods: In addition to the many scientific applications outlined above there are also important technical applications. Researchers may also infer halo masses from clustering statistics (e.g., \citealt{Berlind2002, Zheng2007}), sub-halo and/or conditional abundance matching (e.g., \citealt{Hearin2013, Hearin2014}), in addition to the machine learning approach presented in this work (in comparison to the more standard FOF-AM technique, see \citealt{Vale2004, Conroy2006}). Cross validation through comparison of different approaches will be highly beneficial in exposing advantages and disadvantages between methods and establishing whether or not key insights are method dependent. 

There are, of course, many more scientific applications which will become tractable once a halo catalog with well-defined errors is available at high redshifts through DfL. Nonetheless, we hope the above provides a strong motivation for the approach outlined in this paper, and provides some useful ideas for applications within the extragalactic community in the coming years.

In future work we plan to add more simulations to the DfL library, especially SIMBA (\citealt{Dave2019}), Horizon-AGN (\citealt{Dubois2014}), MassiveBlack-II (\citealt{Khandai2015}), and Romulus (\citealt{Tremmel2017}), among several others. Moreover, we also plan to add greater flexibility to DfL by adding the option to include more galaxy parameters in halo estimation, and by enabling an ANN (in addition to RF) module for mapping between light and dark sectors. Furthermore, we also plan to make further benchmarking comparisons to a host of other methodologies from CAM to HOD, in addition to FOF-AM (considered in Appendix B). Finally, we anticipate adding a magnitude limit option to the DfL halo finder, enabling straightforward applications to galaxy surveys beyond MOONRISE. As such, DfL is intended as living code repository. What is presented in this work is essentially version-1 (see Appendix A for access), with proof of concept, initial testing, and a comprehensive discussion of the methodology. Much more is planned for the coming years.

\section{Summary}
\label{sec6}

\noindent In this paper we present DfL: a novel method to infer the dark sector from luminous tracers at all epochs from the present to cosmic noon. DfL leverages a machine learning based approach utilizing random forest regression, trained on cosmological simulations (specifically, EAGLE and IllustrisTNG). The essential idea behind DfL is to learn an effective mapping from observational-like input data to dark matter halo properties, including masses, group membership, and other derived properties (e.g., virial radii, velocities, and temperatures).

In Section 3 we present a detailed overview of the DfL halo finder pipeline. Briefly, two regression models are prepared for each simulation, spanning $z = 0 - 3$ in snapshots. The first traces a direct mapping from stellar to halo mass for central galaxies (like in abundance matching), but with redshift information added. The second incorporates the total stellar mass of the group and the number of galaxies in the group, in addition to the stellar mass and redshift of the central, in order to yield a more accurate halo mass estimate (see Fig. 2). 

The first regression model is used to estimate the physical and phase-space size of the group, from just the central galaxy stellar mass and redshift. Central galaxies are identified initially as the most massive galaxy in the survey and then sequentially as the most massive galaxy remaining in the survey after each group is removed in turn. Satellites are identified as galaxies within one virial radius and one virial velocity of the central galaxy, with an extended satellite population counted out to two virial radii and two virial velocities. This information is then used to infer the inputs for the second stage from the group membership. 

The above described approach results in two estimates of halo mass - from the central and from the group as a whole. If they are similar enough ($\delta M_{\rm Halo} < $ lim.), the halo mass and group membership is set, and the group is removed from the pipeline. If they are not sufficiently similar the pipeline iterates, allowing the group to grow or shrink in phase-space based on the new group membership. This process continues until convergence is reached, or else a hard cutoff is exceeded. 

Crucially, DfL requires only coordinates, spectroscopic redshifts, and a stellar mass estimate for every galaxy in the survey in order to run. This makes it extremely convenient for use on wide-field spectroscopic galaxy surveys. Additionally, stellar mass uncertainties may be specified as either the width of a Gaussian PDF, or else as a full non-parametric PDF. DfL propagates the uncertainty on stellar masses through the pipeline, outputting full PDFs for the halo mass estimates, which also take account of the distribution of individual tree predictions for each group from the RF regression analysis.

In Section 4 we test the performance of the halo finder on observational-like inputs drawn from the two simulations, which are unseen by the RF models in either training or validation. This enables us to compare the halo mass estimates to the actual halo masses in each simulation, testing the performance of DfL under observational conditions. In the case of applying the same model and data pairing, we recover a bias-free estimate of halo mass with a random uncertainty of $\langle \sigma \rangle = 0.12$\,dex. Furthermore, we also obtain a 96\% accuracy in central - (core) satellite classification, which is lowered to 88\% including the extended satellite population.

We perform cross-validation tests of the pipeline, where we mismatch the model and data pairing to test the impact of the galaxy formation model choice on the recovery of halo properties (see Section 4.2). We find that the random error is largely unaffected by this, but that the bias rises to $\langle b \rangle = \pm 0.1$\,dex. This is taken as a preliminary estimate of the systematic uncertainty induced by model choice, resulting from the fundamental uncertainty of the true underlying galaxy formation model of the Universe. We plan to investigate this further in future work by comparing training and testing of DfL on other cosmological simulations.

We compare the performance of DfL to the more standard FOF-AM technique (abundance matching applied to the total group stellar mass, inferred through a friends-of-friends group finder; see Appendix B). DfL and FOF-AM perform similarly in terms of bias and the central dispersion of the offset in halo masses ($\delta M_{200}$). However, DfL yields significantly fewer catastrophic outliers and hence leads to a much improved recovery of halo masses in the outskirts of the distribution (see Section 4.3). Additionally, we compare the output $M_* - M_{\rm Halo}$ relationships from DfL to observational constraints from strong gravitational lensing in Section 4.4. The recovered relationship is in very good accord with these observational constraints at low redshifts.

We systematically explore how stellar mass uncertainty and spectroscopic incompleteness impact the recovery of halo masses in DfL (see Section 4.5). For VLT-MOONRISE, we anticipate a reduction in accuracy of $\sim 0.15$\,dex due to the combination of these effects, but little impact on the overall bias in halo mass recovery. Additionally, we anticipate a reduction in central - satellite accuracy of $\sim 5 - 8$\%.

Finally, in Section 5 we present some potential scientific applications of the DfL halo finder applied to VLT-MOONRISE, ranging from cosmology to galaxy formation and evolution. In the appendix we provide information on how to access and use the halo finder pipeline (Appendix A), details on the FOF-AM method and results, which are used to compare to DfL (Appendix~B), and provide a list of hyper-parameters used in the RF and ANN approaches for reproducibility (Appendix C). 

In conclusion, DfL leverages machine learning trained on cosmological hydrodynamical simulations to provide a fast, straightforward, and accurate route to infer the dark sector from luminous tracers at all epochs up to cosmic noon. This software is publicly available for use in research (see below for access).

\section*{Data availability}

\noindent All of the data used in this paper is publicly available, or else currently in preparation for public release.\\

\noindent DfL Halo Finder\footnote{\url{https://github.com/abluck/Dark-from-Light}}:\\
See Appendix A for full details.\\

\noindent Simulations:\\
Eagle data access\footnote{\url{http://icc.dur.ac.uk/Eagle/}}\\
IllustrisTNG data access\footnote{\url{www.tng-project.org/}} \\
Docker\footnote{\url{https://hub.docker.com/u/jpiotrowska}}\\

\noindent {\it Analysis}:\\
All analyses in this work were performed using {\small PYTHON-3}\footnote{\url{https://www.python.org/}}, including {\small NUMPY, ASTROPY, SCIPY, PANDAS, SEABORN, MATPLOTLIB}. All machine learning analyses were performed using {\small SCIKIT-LEARN}\footnote{\url{https://scikit-learn.org}}. \\

\begin{acknowledgements}

\noindent AFLB gratefully acknowledges support from an NSF research grant: NSF-AST 2408009, in addition to an ORAU Junior Faculty Enhancement Award in Physical Sciences, and research start-up funds from FIU. PG also acknowledges support from NSF-AST 2408009. RM acknowledges support from the Science and Technology Facilities Council (STFC), ERC Advanced Grant 695671 “QUENCH", and by the UKRI Frontier Research grant RISEandFALL. RM also acknowledges funding from a research professorship from the Royal Society.

\end{acknowledgements}

\bibliographystyle{aa}
\bibliography{aa54702-25}

\begin{thebibliography}{172}
\expandafter\ifx\csname natexlab\endcsname\relax\def\natexlab#1{#1}\fi

\bibitem[{{Abbott} {et~al.}(2016){Abbott}, {Abbott}, {Abbott}, {Abernathy},
  {Acernese}, {Ackley}, {Adams}, {Adams}, {Addesso}, {Adhikari}, {Adya},
  {Affeldt}, {Agathos}, {Agatsuma}, {Aggarwal}, {Aguiar}, {Aiello}, {Ain},
  {Ajith}, {Allen}, {Allocca}, {Altin}, {Anderson}, {Anderson}, {Arai},
  {Arain}, {Araya}, {Arceneaux}, {Areeda}, {Arnaud}, {Arun}, {Ascenzi},
  {Ashton}, {Ast}, {Aston}, {Astone}, {Aufmuth}, {Aulbert}, {Babak}, {Bacon},
  {Bader}, {Baker}, {Baldaccini}, {Ballardin}, {Ballmer}, {Barayoga},
  {Barclay}, {Barish}, {Barker}, {Barone}, {Barr}, {Barsotti}, {Barsuglia},
  {Barta}, {Bartlett}, {Barton}, {Bartos}, {Bassiri}, {Basti}, {Batch},
  {Baune}, {Bavigadda}, {Bazzan}, {Behnke}, {Bejger}, {Belczynski}, {Bell},
  {Bell}, {Berger}, {Bergman}, {Bergmann}, {Berry}, {Bersanetti}, {Bertolini},
  {Betzwieser}, {Bhagwat}, {Bhandare}, {Bilenko}, {Billingsley}, {Birch},
  {Birney}, {Birnholtz}, {Biscans}, {Bisht}, {Bitossi}, {Biwer}, {Bizouard},
  {Blackburn}, {Blair}, {Blair}, {Blair}, {Bloemen}, {Bock}, {Bodiya}, {Boer},
  {Bogaert}, {Bogan}, {Bohe}, {Bojtos}, {Bond}, {Bondu}, {Bonnand}, {Boom},
  {Bork}, {Boschi}, {Bose}, {Bouffanais}, {Bozzi}, {Bradaschia}, {Brady},
  {Braginsky}, {Branchesi}, {Brau}, {Briant}, {Brillet}, {Brinkmann},
  {Brisson}, {Brockill}, {Brooks}, {Brown}, {Brown}, {Brown}, {Buchanan},
  {Buikema}, {Bulik}, {Bulten}, {Buonanno}, {Buskulic}, {Buy}, {Byer},
  {Cabero}, {Cadonati}, {Cagnoli}, {Cahillane}, {Bustillo}, {Callister},
  {Calloni}, {Camp}, {Cannon}, {Cao}, {Capano}, {Capocasa}, {Carbognani},
  {Caride}, {Diaz}, {Casentini}, {Caudill}, {Cavagli{\`a}}, {Cavalier},
  {Cavalieri}, {Cella}, {Cepeda}, {Baiardi}, {Cerretani}, {Cesarini},
  {Chakraborty}, {Chalermsongsak}, {Chamberlin}, {Chan}, {Chao}, {Charlton},
  {Chassande-Mottin}, {Chen}, {Chen}, {Cheng}, {Chincarini}, {Chiummo}, {Cho},
  {Cho}, {Chow}, {Christensen}, {Chu}, {Chua}, {Chung}, {Ciani}, {Clara},
  {Clark}, {Cleva}, {Coccia}, {Cohadon}, {Colla}, {Collette}, {Cominsky},
  {Constancio}, {Conte}, {Conti}, {Cook}, {Corbitt}, {Cornish}, {Corsi},
  {Cortese}, {Costa}, {Coughlin}, {Coughlin}, {Coulon}, {Countryman},
  {Couvares}, {Cowan}, {Coward}, \& {Cowart}}]{Abbott2016}
{Abbott}, B.~P., {Abbott}, R., {Abbott}, T.~D., {et~al.} 2016, \prl, 116,
  061102

\bibitem[{{Abbott} {et~al.}(2017){Abbott}, {Abbott}, {Abbott}, {Acernese},
  {Ackley}, {Adams}, {Adams}, {Addesso}, {Adhikari}, {Adya}, {Affeldt},
  {Afrough}, {Agarwal}, {Agathos}, {Agatsuma}, {Aggarwal}, {Aguiar}, {Aiello},
  {Ain}, {Ajith}, {Allen}, {Allen}, {Allocca}, {Altin}, {Amato}, {Ananyeva},
  {Anderson}, {Anderson}, {Angelova}, {Antier}, {Appert}, {Arai}, {Araya},
  {Areeda}, {Arnaud}, {Arun}, {Ascenzi}, {Ashton}, {Ast}, {Aston}, {Astone},
  {Atallah}, {Aufmuth}, {Aulbert}, {AultONeal}, {Austin}, {Avila-Alvarez},
  {Babak}, {Bacon}, {Bader}, {Bae}, {Bailes}, {Baker}, {Baldaccini},
  {Ballardin}, {Ballmer}, {Banagiri}, {Barayoga}, {Barclay}, {Barish},
  {Barker}, {Barkett}, {Barone}, {Barr}, {Barsotti}, {Barsuglia}, {Barta},
  {Barthelmy}, {Bartlett}, {Bartos}, {Bassiri}, {Basti}, {Batch}, {Bawaj},
  {Bayley}, {Bazzan}, {B{\'e}csy}, {Beer}, {Bejger}, {Belahcene}, {Bell},
  {Berger}, {Bergmann}, {Bernuzzi}, {Bero}, {Berry}, {Bersanetti}, {Bertolini},
  {Betzwieser}, {Bhagwat}, {Bhandare}, {Bilenko}, {Billingsley}, {Billman},
  {Birch}, {Birney}, {Birnholtz}, {Biscans}, {Biscoveanu}, {Bisht}, {Bitossi},
  {Biwer}, {Bizouard}, {Blackburn}, {Blackman}, {Blair}, {Blair}, {Blair},
  {Bloemen}, {Bock}, {Bode}, {Boer}, {Bogaert}, {Bohe}, {Bondu}, {Bonilla},
  {Bonnand}, {Boom}, {Bork}, {Boschi}, {Bose}, {Bossie}, {Bouffanais}, {Bozzi},
  {Bradaschia}, {Brady}, {Branchesi}, {Brau}, {Briant}, {Brillet}, {Brinkmann},
  {Brisson}, {Brockill}, {Broida}, {Brooks}, {Brown}, {Brown}, {Brunett},
  {Buchanan}, {Buikema}, {Bulik}, {Bulten}, {Buonanno}, {Buskulic}, {Buy},
  {Byer}, {Cabero}, {Cadonati}, {Cagnoli}, {Cahillane}, {Calder{\'o}n
  Bustillo}, {Callister}, {Calloni}, {Camp}, {Canepa}, {Canizares}, {Cannon},
  {Cao}, {Cao}, {Capano}, {Capocasa}, {Carbognani}, {Caride}, {Carney},
  {Carullo}, {Casanueva Diaz}, {Casentini}, {Caudill}, {Cavagli{\`a}},
  {Cavalier}, {Cavalieri}, {Cella}, {Cepeda}, {Cerd{\'a}-Dur{\'a}n},
  {Cerretani}, {Cesarini}, {Chamberlin}, {Chan}, {Chao}, {Charlton}, {Chase},
  {Chassande-Mottin}, {Chatterjee}, {Chatziioannou}, {Cheeseboro}, {Chen},
  {Chen}, {Chen}, {Cheng}, {Chia}, {Chincarini}, {Chiummo}, {Chmiel}, {Cho},
  {Cho}, {Chow}, {Christensen}, {Chu}, {Chua}, \& {Chua}}]{Abbott2017}
{Abbott}, B.~P., {Abbott}, R., {Abbott}, T.~D., {et~al.} 2017, \prl, 119,
  161101

\bibitem[{{Adhikari} {et~al.}(2014){Adhikari}, {Dalal}, \&
  {Chamberlain}}]{Adhikari2014}
{Adhikari}, S., {Dalal}, N., \& {Chamberlain}, R.~T. 2014, \jcap, 2014, 019

\bibitem[{{Ansoldi} {et~al.}(2018){Ansoldi}, {Antonelli}, {Arcaro}, {Baack},
  {Babi{\'c}}, {Banerjee}, {Bangale}, {Barres de Almeida}, {Barrio}, {Becerra
  Gonz{\'a}lez}, {Bednarek}, {Bernardini}, {Berse}, {Berti}, {Besenrieder},
  {Bhattacharyya}, {Bigongiari}, {Biland}, {Blanch}, {Bonnoli}, {Carosi},
  {Ceribella}, {Chatterjee}, {Colak}, {Colin}, {Colombo}, {Contreras},
  {Cortina}, {Covino}, {Cumani}, {D'Elia}, {Da Vela}, {Dazzi}, {De Angelis},
  {De Lotto}, {Delfino}, {Delgado}, {Di Pierro}, {Dom{\'\i}nguez}, {Dominis
  Prester}, {Dorner}, {Doro}, {Einecke}, {Elsaesser}, {Fallah Ramazani},
  {Fattorini}, {Fern{\'a}ndez-Barral}, {Ferrara}, {Fidalgo}, {Foffano},
  {Fonseca}, {Font}, {Fruck}, {Gallozzi}, {Garc{\'\i}a L{\'o}pez},
  {Garczarczyk}, {Gaug}, {Giammaria}, {Godinovi{\'c}}, {Guberman}, {Hadasch},
  {Hahn}, {Hassan}, {Hayashida}, {Herrera}, {Hoang}, {Hrupec}, {Inoue},
  {Ishio}, {Iwamura}, {Konno}, {Kubo}, {Kushida}, {Lamastra}, {Lelas}, {Leone},
  {Lindfors}, {Lombardi}, {Longo}, {L{\'o}pez}, {Maggio}, {Majumdar},
  {Makariev}, {Maneva}, {Manganaro}, {Mannheim}, {Maraschi}, {Mariotti},
  {Mart{\'\i}nez}, {Masuda}, {Mazin}, {Mielke}, {Minev}, {Miranda}, {Mirzoyan},
  {Moralejo}, {Moreno}, {Moretti}, {Neustroev}, {Niedzwiecki}, {Nievas
  Rosillo}, {Nigro}, {Nilsson}, {Ninci}, {Nishijima}, {Noda}, {Nogu{\'e}s},
  {Paiano}, {Palacio}, {Paneque}, {Paoletti}, {Paredes}, {Pedaletti},
  {Pe{\~n}il}, {Peresano}, {Persic}, {Pfrang}, {Prada Moroni}, {Prandini},
  {Puljak}, {Garcia}, {Rhode}, {Rib{\'o}}, {Rico}, {Righi}, {Rugliancich},
  {Saha}, {Saito}, {Satalecka}, {Schweizer}, {Sitarek}, {{\v{S}}nidari{\'c}},
  {Sobczynska}, {Stamerra}, {Strzys}, {Suri{\'c}}, {Tavecchio}, {Temnikov},
  {Terzi{\'c}}, {Teshima}, {Torres-Alb{\'a}}, {Tsujimoto}, {Vanzo}, {Vazquez
  Acosta}, {Vovk}, {Ward}, {Will}, {Zari{\'c}}, \& {Cerruti}}]{Ansoldi2018}
{Ansoldi}, S., {Antonelli}, L.~A., {Arcaro}, C., {et~al.} 2018, \apjl, 863, L10

\bibitem[{{Artale} {et~al.}(2018){Artale}, {Zehavi}, {Contreras}, \&
  {Norberg}}]{Artale2018}
{Artale}, M.~C., {Zehavi}, I., {Contreras}, S., \& {Norberg}, P. 2018, \mnras,
  480, 3978

\bibitem[{{Atek} {et~al.}(2018){Atek}, {Richard}, {Kneib}, \&
  {Schaerer}}]{Atek2018}
{Atek}, H., {Richard}, J., {Kneib}, J.-P., \& {Schaerer}, D. 2018, \mnras, 479,
  5184

\bibitem[{{Auger} {et~al.}(2010){Auger}, {Treu}, {Gavazzi}, {Bolton},
  {Koopmans}, \& {Marshall}}]{Auger2010}
{Auger}, M.~W., {Treu}, T., {Gavazzi}, R., {et~al.} 2010, \apjl, 721, L163

\bibitem[{{Baldwin} {et~al.}(1981){Baldwin}, {Phillips}, \&
  {Terlevich}}]{Baldwin1981}
{Baldwin}, J.~A., {Phillips}, M.~M., \& {Terlevich}, R. 1981, \pasp, 93, 5

\bibitem[{{Banerji} {et~al.}(2010){Banerji}, {Lahav}, {Lintott}, {Abdalla},
  {Schawinski}, {Bamford}, {Andreescu}, {Murray}, {Raddick}, {Slosar},
  {Szalay}, {Thomas}, \& {Vandenberg}}]{Banerji2010}
{Banerji}, M., {Lahav}, O., {Lintott}, C.~J., {et~al.} 2010, \mnras, 406, 342

\bibitem[{{Baron}(2019)}]{Baron2019}
{Baron}, D. 2019, arXiv e-prints, arXiv:1904.07248

\bibitem[{{Baxter} {et~al.}(2017){Baxter}, {Chang}, {Jain}, {Adhikari},
  {Dalal}, {Kravtsov}, {More}, {Rozo}, {Rykoff}, \& {Sheth}}]{Baxter2017}
{Baxter}, E., {Chang}, C., {Jain}, B., {et~al.} 2017, \apj, 841, 18

\bibitem[{{Begeman}(1989)}]{Begeman1989}
{Begeman}, K.~G. 1989, \aap, 223, 47

\bibitem[{{Behroozi} {et~al.}(2010){Behroozi}, {Conroy}, \&
  {Wechsler}}]{Behroozi2010}
{Behroozi}, P.~S., {Conroy}, C., \& {Wechsler}, R.~H. 2010, \apj, 717, 379

\bibitem[{{Benson} {et~al.}(2000){Benson}, {Cole}, {Frenk}, {Baugh}, \&
  {Lacey}}]{Benson2000}
{Benson}, A.~J., {Cole}, S., {Frenk}, C.~S., {Baugh}, C.~M., \& {Lacey}, C.~G.
  2000, \mnras, 311, 793

\bibitem[{{Berlind} \& {Weinberg}(2002)}]{Berlind2002}
{Berlind}, A.~A. \& {Weinberg}, D.~H. 2002, \apj, 575, 587

\bibitem[{{Bertotti} {et~al.}(2003){Bertotti}, {Iess}, \&
  {Tortora}}]{Bertotti2003}
{Bertotti}, B., {Iess}, L., \& {Tortora}, P. 2003, \nat, 425, 374

\bibitem[{{Bluck} {et~al.}(2024){Bluck}, {Conselice}, {Ormerod}, {Piotrowska},
  {Adams}, {Austin}, {Caruana}, {Duncan}, {Ferreira}, {Goubert}, {Harvey},
  {Trussler}, \& {Maiolino}}]{Bluck2024}
{Bluck}, A. F.~L., {Conselice}, C.~J., {Ormerod}, K., {et~al.} 2024, \apj, 961,
  163

\bibitem[{{Bluck} {et~al.}(2022){Bluck}, {Maiolino}, {Brownson}, {Conselice},
  {Ellison}, {Piotrowska}, \& {Thorp}}]{Bluck2022}
{Bluck}, A. F.~L., {Maiolino}, R., {Brownson}, S., {et~al.} 2022, \aap, 659,
  A160

\bibitem[{{Bluck} {et~al.}(2020{\natexlab{a}}){Bluck}, {Maiolino},
  {Piotrowska}, {Trussler}, {Ellison}, {S{\'a}nchez}, {Thorp}, {Teimoorinia},
  {Moreno}, \& {Conselice}}]{Bluck2020b}
{Bluck}, A. F.~L., {Maiolino}, R., {Piotrowska}, J.~M., {et~al.}
  2020{\natexlab{a}}, \mnras, 499, 230

\bibitem[{{Bluck} {et~al.}(2020{\natexlab{b}}){Bluck}, {Maiolino},
  {S{\'a}nchez}, {Ellison}, {Thorp}, {Piotrowska}, {Teimoorinia}, \&
  {Bundy}}]{Bluck2020a}
{Bluck}, A. F.~L., {Maiolino}, R., {S{\'a}nchez}, S.~F., {et~al.}
  2020{\natexlab{b}}, \mnras, 492, 96

\bibitem[{{Bluck} {et~al.}(2014){Bluck}, {Mendel}, {Ellison}, {Moreno},
  {Simard}, {Patton}, \& {Starkenburg}}]{Bluck2014}
{Bluck}, A. F.~L., {Mendel}, J.~T., {Ellison}, S.~L., {et~al.} 2014, \mnras,
  441, 599

\bibitem[{{Bluck} {et~al.}(2016){Bluck}, {Mendel}, {Ellison}, {Patton},
  {Simard}, {Henriques}, {Torrey}, {Teimoorinia}, {Moreno}, \&
  {Starkenburg}}]{Bluck2016}
{Bluck}, A. F.~L., {Mendel}, J.~T., {Ellison}, S.~L., {et~al.} 2016, \mnras,
  462, 2559

\bibitem[{{Bluck} {et~al.}(2023){Bluck}, {Piotrowska}, \&
  {Maiolino}}]{Bluck2023}
{Bluck}, A. F.~L., {Piotrowska}, J.~M., \& {Maiolino}, R. 2023, \apj, 944, 108

\bibitem[{{Bocquet} {et~al.}(2019){Bocquet}, {Dietrich}, {Schrabback}, {Bleem},
  {Klein}, {Allen}, {Applegate}, {Ashby}, {Bautz}, {Bayliss}, {Benson},
  {Brodwin}, {Bulbul}, {Canning}, {Capasso}, {Carlstrom}, {Chang}, {Chiu},
  {Cho}, {Clocchiatti}, {Crawford}, {Crites}, {de Haan}, {Desai}, {Dobbs},
  {Foley}, {Forman}, {Garmire}, {George}, {Gladders}, {Gonzalez}, {Grandis},
  {Gupta}, {Halverson}, {Hlavacek-Larrondo}, {Hoekstra}, {Holder}, {Holzapfel},
  {Hou}, {Hrubes}, {Huang}, {Jones}, {Khullar}, {Knox}, {Kraft}, {Lee}, {von
  der Linden}, {Luong-Van}, {Mantz}, {Marrone}, {McDonald}, {McMahon}, {Meyer},
  {Mocanu}, {Mohr}, {Morris}, {Padin}, {Patil}, {Pryke}, {Rapetti},
  {Reichardt}, {Rest}, {Ruhl}, {Saliwanchik}, {Saro}, {Sayre}, {Schaffer},
  {Shirokoff}, {Stalder}, {Stanford}, {Staniszewski}, {Stark}, {Story},
  {Strazzullo}, {Stubbs}, {Vanderlinde}, {Vieira}, {Vikhlinin}, {Williamson},
  \& {Zenteno}}]{Bocquet2019}
{Bocquet}, S., {Dietrich}, J.~P., {Schrabback}, T., {et~al.} 2019, \apj, 878,
  55

\bibitem[{{Bolton} {et~al.}(2006){Bolton}, {Burles}, {Koopmans}, {Treu}, \&
  {Moustakas}}]{Bolton2006}
{Bolton}, A.~S., {Burles}, S., {Koopmans}, L. V.~E., {Treu}, T., \&
  {Moustakas}, L.~A. 2006, \apj, 638, 703

\bibitem[{{Bondi} \& {Hoyle}(1944)}]{Bondi1944}
{Bondi}, H. \& {Hoyle}, F. 1944, \mnras, 104, 273

\bibitem[{Bower {et~al.}(2006)Bower, Benson, Malbon, {et~al.}}]{Bower2006}
Bower, R.~G., Benson, A.~J., Malbon, R., {et~al.} 2006, Monthly Notices of the
  Royal Astronomical Society, 370, 645

\bibitem[{Bower {et~al.}(2008)Bower, McCarthy, \& Benson}]{Bower2008}
Bower, R.~G., McCarthy, I.~G., \& Benson, A.~J. 2008, Monthly Notices of the
  Royal Astronomical Society, 390, 1399

\bibitem[{{Breiman}(2001)}]{Breiman2001}
{Breiman}, L. 2001, Machine Learning, 45, 5

\bibitem[{{Brinchmann} {et~al.}(2004){Brinchmann}, {Charlot}, {White},
  {Tremonti}, {Kauffmann}, {Heckman}, \& {Brinkmann}}]{Brinchmann2004}
{Brinchmann}, J., {Charlot}, S., {White}, S.~D.~M., {et~al.} 2004, \mnras, 351,
  1151

\bibitem[{{Brink} {et~al.}(2013){Brink}, {Richards}, {Poznanski}, {Bloom},
  {Rice}, {Negahban}, \& {Wainwright}}]{Brink2013}
{Brink}, H., {Richards}, J.~W., {Poznanski}, D., {et~al.} 2013, \mnras, 435,
  1047

\bibitem[{{Campbell} {et~al.}(2018){Campbell}, {van den Bosch}, {Padmanabhan},
  {Mao}, {Zentner}, {Lange}, {Jiang}, \& {Villarreal}}]{Campbell2018}
{Campbell}, D., {van den Bosch}, F.~C., {Padmanabhan}, N., {et~al.} 2018,
  \mnras, 477, 359

\bibitem[{{Cappellari} {et~al.}(2006){Cappellari}, {Bacon}, {Bureau}, {Damen},
  {Davies}, {de Zeeuw}, {Emsellem}, {Falc{\'o}n-Barroso}, {Krajnovi{\'c}},
  {Kuntschner}, {McDermid}, {Peletier}, {Sarzi}, {van den Bosch}, \& {van de
  Ven}}]{Cappellari2006}
{Cappellari}, M., {Bacon}, R., {Bureau}, M., {et~al.} 2006, \mnras, 366, 1126

\bibitem[{{Cappellari} {et~al.}(2013){Cappellari}, {McDermid}, {Alatalo},
  {Blitz}, {Bois}, {Bournaud}, {Bureau}, {Crocker}, {Davies}, {Davis}, {de
  Zeeuw}, {Duc}, {Emsellem}, {Khochfar}, {Krajnovi{\'c}}, {Kuntschner},
  {Morganti}, {Naab}, {Oosterloo}, {Sarzi}, {Scott}, {Serra}, {Weijmans}, \&
  {Young}}]{Cappellari2013}
{Cappellari}, M., {McDermid}, R.~M., {Alatalo}, K., {et~al.} 2013, \mnras, 432,
  1862

\bibitem[{{Carlberg} {et~al.}(1996){Carlberg}, {Yee}, {Ellingson}, {Abraham},
  {Gravel}, {Morris}, \& {Pritchet}}]{Carlberg1996}
{Carlberg}, R.~G., {Yee}, H.~K.~C., {Ellingson}, E., {et~al.} 1996, \apj, 462,
  32

\bibitem[{{Carlstrom} {et~al.}(2002){Carlstrom}, {Holder}, \&
  {Reese}}]{Carlstrom2002}
{Carlstrom}, J.~E., {Holder}, G.~P., \& {Reese}, E.~D. 2002, \araa, 40, 643

\bibitem[{{Chabrier}(2003)}]{Chabrier2003}
{Chabrier}, G. 2003, \pasp, 115, 763

\bibitem[{{Chaves-Montero} {et~al.}(2016){Chaves-Montero}, {Angulo}, {Schaye},
  {Schaller}, {Crain}, {Furlong}, \& {Theuns}}]{Chaves2016}
{Chaves-Montero}, J., {Angulo}, R.~E., {Schaye}, J., {et~al.} 2016, \mnras,
  460, 3100

\bibitem[{{Chiang} {et~al.}(2013){Chiang}, {Overzier}, \&
  {Gebhardt}}]{Chiang2013}
{Chiang}, Y.-K., {Overzier}, R., \& {Gebhardt}, K. 2013, \apj, 779, 127

\bibitem[{{Chua} \& {Vallisneri}(2020)}]{Chua2020}
{Chua}, A. J.~K. \& {Vallisneri}, M. 2020, \prl, 124, 041102

\bibitem[{{Cirasuolo} {et~al.}(2014){Cirasuolo}, {Afonso}, {Carollo}, {Flores},
  {Maiolino}, {Oliva}, {Paltani}, {Vanzi}, {Evans}, {Abreu}, {Atkinson},
  {Babusiaux}, {Beard}, {Bauer}, {Bellazzini}, {Bender}, {Best}, {Bezawada},
  {Bonifacio}, {Bragaglia}, {Bryson}, {Busher}, {Cabral}, {Caputi}, {Centrone},
  {Chemla}, {Cimatti}, {Cioni}, {Clementini}, {Coelho}, {Crnojevic}, {Daddi},
  {Dunlop}, {Eales}, {Feltzing}, {Ferguson}, {Fisher}, {Fontana}, {Fynbo},
  {Garilli}, {Gilmore}, {Glauser}, {Guinouard}, {Hammer}, {Hastings}, {Hess},
  {Ivison}, {Jagourel}, {Jarvis}, {Kaper}, {Kauffman}, {Kitching}, {Lawrence},
  {Lee}, {Lemasle}, {Licausi}, {Lilly}, {Lorenzetti}, {Lunney}, {Maiolino},
  {Mannucci}, {McLure}, {Minniti}, {Montgomery}, {Muschielok}, {Nandra},
  {Navarro}, {Norberg}, {Oliver}, {Origlia}, {Padilla}, {Peacock}, {Pedichini},
  {Peng}, {Pentericci}, {Pragt}, {Puech}, {Randich}, {Rees}, {Renzini}, {Ryde},
  {Rodrigues}, {Roseboom}, {Royer}, {Saglia}, {Sanchez}, {Schiavon},
  {Schnetler}, {Sobral}, {Speziali}, {Sun}, {Stuik}, {Taylor}, {Taylor},
  {Todd}, {Tolstoy}, {Torres}, {Tosi}, {Vanzella}, {Venema}, {Vitali},
  {Wegner}, {Wells}, {Wild}, {Wright}, {Zamorani}, \&
  {Zoccali}}]{Cirasuolo2014}
{Cirasuolo}, M., {Afonso}, J., {Carollo}, M., {et~al.} 2014, in Society of
  Photo-Optical Instrumentation Engineers (SPIE) Conference Series, Vol. 9147,
  Ground-based and Airborne Instrumentation for Astronomy V, ed. S.~K.
  {Ramsay}, I.~S. {McLean}, \& H.~{Takami}, 91470N

\bibitem[{{Cirasuolo} {et~al.}(2020){Cirasuolo}, {Fairley}, {Rees}, {Gonzalez},
  {Taylor}, {Maiolino}, {Afonso}, {Evans}, {Flores}, {Lilly}, {Oliva},
  {Paltani}, {Vanzi}, {Abreu}, {Accardo}, {Adams}, {{\'A}lvarez M{\'e}ndez},
  {Amans}, {Amarantidis}, {Atek}, {Atkinson}, {Banerji}, {Barrett},
  {Barrientos}, {Bauer}, {Beard}, {B{\'e}chet}, {Belfiore}, {Bellazzini},
  {Benoist}, {Best}, {Biazzo}, {Black}, {Boettger}, {Bonifacio}, {Bowler},
  {Bragaglia}, {Brierley}, {Brinchmann}, {Brinkmann}, {Buat}, {Buitrago},
  {Burgarella}, {Burningham}, {Buscher}, {Cabral}, {Caffau}, {Cardoso},
  {Carnall}, {Carollo}, {Castillo}, {Castignani}, {Catelan}, {Cicone},
  {Cimatti}, {Cioni}, {Clementini}, {Cochrane}, {Coelho}, {Colling}, {Contini},
  {Contreras}, {Conzelmann}, {Cresci}, {Cropper}, {Cucciati}, {Cullen},
  {Cumani}, {Curti}, {Da Silva}, {Daddi}, {Dalessandro}, {Dalessio}, {Dauvin},
  {Davidson}, {de Laverny}, {Delplancke-Str{\"o}bele}, {De Lucia}, {Del
  Vecchio}, {Dessauges-Zavadsky}, {Di Matteo}, {Dole}, {Drass}, {Dunlop},
  {D{\"u}nner}, {Eales}, {Ellis}, {Enriques}, {Fasola}, {Ferguson}, {Ferruzzi},
  {Fisher}, {Flores}, {Fontana}, {Forchi}, {Francois}, {Franzetti}, {Gargiulo},
  {Garilli}, {Gaudemard}, {Gieles}, {Gilmore}, {Ginolfi}, {Gomes}, {Guinouard},
  {Gutierrez}, {Haigron}, {Hammer}, {Hammersley}, {Haniff}, {Harrison},
  {Haywood}, {Hill}, {Hubin}, {Humphrey}, {Ibata}, {Infante}, {Ives}, {Ivison},
  {Iwert}, {Jablonka}, {Jakob}, {Jarvis}, {King}, {Kneib}, {Laporte},
  {Lawrence}, {Lee}, {Li Causi}, {Lorenzoni}, {Lucatello}, {Luco}, {Macleod},
  {Magliocchetti}, {Magrini}, {Mainieri}, {Maire}, {Mannucci}, {Martin},
  {Matute}, {Maurogordato}, {McGee}, {Mcleod}, {McLure}, {McMahon}, {Melse},
  {Messias}, {Mucciarelli}, {Nisini}, {Nix}, {Norberg}, {Oesch}, {Oliveira},
  {Origlia}, {Padilla}, {Palsa}, {Pancino}, {Papaderos}, {Pappalardo}, {Parry},
  {Pasquini}, {Peacock}, {Pedichini}, {Pello}, {Peng}, {Pentericci}, {Pfuhl},
  {Piazzesi}, {Popovic}, {Pozzetti}, {Puech}, {Puzia}, {Raichoor}, {Randich},
  {Recio-Blanco}, {Reis}, {Reix}, {Renzini}, {Rodrigues}, {Rojas},
  {Rojas-Arriagada}, {Rota}, {Royer}, {Sacco}, {Sanchez-Janssen}, {Sanna},
  {Santos}, {Sarzi}, {Schaerer}, {Schiavon}, {Schnell}, {Schultheis},
  {Scodeggio}, {Serjeant}, {Shen}, {Simmonds}, {Smoker}, {Sobral}, {Sordet}, \&
  {Sp{\'e}rone}}]{Cirasuolo2020}
{Cirasuolo}, M., {Fairley}, A., {Rees}, P., {et~al.} 2020, The Messenger, 180,
  10

\bibitem[{{Cochrane} {et~al.}(2018){Cochrane}, {Best}, {Sobral}, {Smail},
  {Geach}, {Stott}, \& {Wake}}]{Cochrane2018}
{Cochrane}, R.~K., {Best}, P.~N., {Sobral}, D., {et~al.} 2018, \mnras, 475,
  3730

\bibitem[{{Cole} {et~al.}(2000){Cole}, {Lacey}, {Baugh}, \& {Frenk}}]{Cole2000}
{Cole}, S., {Lacey}, C.~G., {Baugh}, C.~M., \& {Frenk}, C.~S. 2000, \mnras,
  319, 168

\bibitem[{{Conroy} {et~al.}(2006){Conroy}, {Wechsler}, \&
  {Kravtsov}}]{Conroy2006}
{Conroy}, C., {Wechsler}, R.~H., \& {Kravtsov}, A.~V. 2006, \apj, 647, 201

\bibitem[{{Conselice} {et~al.}(2014){Conselice}, {Bluck}, {Mortlock},
  {Palamara}, \& {Benson}}]{Conselice2014}
{Conselice}, C.~J., {Bluck}, A. F.~L., {Mortlock}, A., {Palamara}, D., \&
  {Benson}, A.~J. 2014, \mnras, 444, 1125

\bibitem[{{Copeland} {et~al.}(2006){Copeland}, {Sami}, \&
  {Tsujikawa}}]{Copeland2006}
{Copeland}, E.~J., {Sami}, M., \& {Tsujikawa}, S. 2006, International Journal
  of Modern Physics D, 15, 1753

\bibitem[{{Crain} {et~al.}(2015){Crain}, {Schaye}, {Bower}, {Furlong},
  {Schaller}, {Theuns}, {Dalla Vecchia}, {Frenk}, {McCarthy}, {Helly},
  {Jenkins}, {Rosas-Guevara}, {White}, \& {Trayford}}]{Crain2015}
{Crain}, R.~A., {Schaye}, J., {Bower}, R.~G., {et~al.} 2015, \mnras, 450, 1937

\bibitem[{{Croton} {et~al.}(2007){Croton}, {Gao}, \& {White}}]{Croton2007}
{Croton}, D.~J., {Gao}, L., \& {White}, S. D.~M. 2007, \mnras, 374, 1303

\bibitem[{{Dav{\'e}} {et~al.}(2019){Dav{\'e}}, {Angl{\'e}s-Alc{\'a}zar},
  {Narayanan}, {Li}, {Rafieferantsoa}, \& {Appleby}}]{Dave2019}
{Dav{\'e}}, R., {Angl{\'e}s-Alc{\'a}zar}, D., {Narayanan}, D., {et~al.} 2019,
  \mnras, 486, 2827

\bibitem[{{Davis} {et~al.}(1985){Davis}, {Efstathiou}, {Frenk}, \&
  {White}}]{Davis1985}
{Davis}, M., {Efstathiou}, G., {Frenk}, C.~S., \& {White}, S.~D.~M. 1985, \apj,
  292, 371

\bibitem[{{de Blok} {et~al.}(2008){de Blok}, {Walter}, {Brinks},
  {Trachternach}, {Oh}, \& {Kennicutt}}]{deBlok2008}
{de Blok}, W.~J.~G., {Walter}, F., {Brinks}, E., {et~al.} 2008, \aj, 136, 2648

\bibitem[{{Dieleman} {et~al.}(2015){Dieleman}, {Willett}, \&
  {Dambre}}]{Dieleman2015}
{Dieleman}, S., {Willett}, K.~W., \& {Dambre}, J. 2015, \mnras, 450, 1441

\bibitem[{{Dimauro} {et~al.}(2018){Dimauro}, {Huertas-Company}, {Daddi},
  {P{\'e}rez-Gonz{\'a}lez}, {Bernardi}, {Barro}, {Buitrago}, {Caro},
  {Cattaneo}, {Dominguez-S{\'a}nchez}, {Faber}, {H{\"a}u{\ss}ler}, {Kocevski},
  {Koekemoer}, {Koo}, {Lee}, {Mei}, {Margalef-Bentabol}, {Primack},
  {Rodriguez-Puebla}, {Salvato}, {Shankar}, \& {Tuccillo}}]{Dimauro2018}
{Dimauro}, P., {Huertas-Company}, M., {Daddi}, E., {et~al.} 2018, \mnras, 478,
  5410

\bibitem[{{Dolag} {et~al.}(2009){Dolag}, {Borgani}, {Murante}, \&
  {Springel}}]{Dolag2009}
{Dolag}, K., {Borgani}, S., {Murante}, G., \& {Springel}, V. 2009, \mnras, 399,
  497

\bibitem[{{Duarte} \& {Mamon}(2014)}]{Duarte2014}
{Duarte}, M. \& {Mamon}, G.~A. 2014, \mnras, 440, 1763

\bibitem[{{Dubois} {et~al.}(2014){Dubois}, {Pichon}, {Welker}, {Le Borgne},
  {Devriendt}, {Laigle}, {Codis}, {Pogosyan}, {Arnouts}, {Benabed}, {Bertin},
  {Blaizot}, {Bouchet}, {Cardoso}, {Colombi}, {de Lapparent}, {Desjacques},
  {Gavazzi}, {Kassin}, {Kimm}, {McCracken}, {Milliard}, {Peirani}, {Prunet},
  {Rouberol}, {Silk}, {Slyz}, {Sousbie}, {Teyssier}, {Tresse}, {Treyer},
  {Vibert}, \& {Volonteri}}]{Dubois2014}
{Dubois}, Y., {Pichon}, C., {Welker}, C., {et~al.} 2014, \mnras, 444, 1453

\bibitem[{{Efstathiou} {et~al.}(1990){Efstathiou}, {Sutherland}, \&
  {Maddox}}]{Efstathiou1990}
{Efstathiou}, G., {Sutherland}, W.~J., \& {Maddox}, S.~J. 1990, \nat, 348, 705

\bibitem[{{Erb} {et~al.}(2006){Erb}, {Steidel}, {Shapley}, {Pettini}, {Reddy},
  \& {Adelberger}}]{Erb2006}
{Erb}, D.~K., {Steidel}, C.~C., {Shapley}, A.~E., {et~al.} 2006, \apj, 646, 107

\bibitem[{{Everitt} {et~al.}(2011){Everitt}, {Debra}, {Parkinson}, {Turneaure},
  {Conklin}, {Heifetz}, {Keiser}, {Silbergleit}, {Holmes}, {Kolodziejczak},
  {Al-Meshari}, {Mester}, {Muhlfelder}, {Solomonik}, {Stahl}, {Worden},
  {Bencze}, {Buchman}, {Clarke}, {Al-Jadaan}, {Al-Jibreen}, {Li}, {Lipa},
  {Lockhart}, {Al-Suwaidan}, {Taber}, \& {Wang}}]{Everitt2011}
{Everitt}, C.~W.~F., {Debra}, D.~B., {Parkinson}, B.~W., {et~al.} 2011, \prl,
  106, 221101

\bibitem[{{Evrard} {et~al.}(1996){Evrard}, {Metzler}, \&
  {Navarro}}]{Evrard1996}
{Evrard}, A.~E., {Metzler}, C.~A., \& {Navarro}, J.~F. 1996, \apj, 469, 494

\bibitem[{{Frenk} {et~al.}(1988){Frenk}, {White}, {Davis}, \&
  {Efstathiou}}]{Frenk1988}
{Frenk}, C.~S., {White}, S. D.~M., {Davis}, M., \& {Efstathiou}, G. 1988, \apj,
  327, 507

\bibitem[{{Gerke} {et~al.}(2005){Gerke}, {Newman}, {Davis}, {Marinoni}, {Yan},
  {Coil}, {Conroy}, {Cooper}, {Faber}, {Finkbeiner}, {Guhathakurta}, {Kaiser},
  {Koo}, {Phillips}, {Weiner}, \& {Willmer}}]{Gerke2005}
{Gerke}, B.~F., {Newman}, J.~A., {Davis}, M., {et~al.} 2005, \apj, 625, 6

\bibitem[{{Giodini} {et~al.}(2009){Giodini}, {Pierini}, {Finoguenov}, {Pratt},
  {Boehringer}, {Leauthaud}, {Guzzo}, {Aussel}, {Bolzonella}, {Capak}, {Elvis},
  {Hasinger}, {Ilbert}, {Kartaltepe}, {Koekemoer}, {Lilly}, {Massey},
  {McCracken}, {Rhodes}, {Salvato}, {Sanders}, {Scoville}, {Sasaki}, {Smolcic},
  {Taniguchi}, {Thompson}, \& {COSMOS Collaboration}}]{Giodini2009}
{Giodini}, S., {Pierini}, D., {Finoguenov}, A., {et~al.} 2009, \apj, 703, 982

\bibitem[{{Goubert} {et~al.}(2024){Goubert}, {Bluck}, {Piotrowska}, \&
  {Maiolino}}]{Goubert2024}
{Goubert}, P.~H., {Bluck}, A. F.~L., {Piotrowska}, J.~M., \& {Maiolino}, R.
  2024, \mnras, 528, 4891

\bibitem[{{Grogin} {et~al.}(2011){Grogin}, {Kocevski}, {Faber}, {Ferguson},
  {Koekemoer}, {Riess}, {Acquaviva}, {Alexander}, {Almaini}, {Ashby}, {Barden},
  {Bell}, {Bournaud}, {Brown}, {Caputi}, {Casertano}, {Cassata}, {Castellano},
  {Challis}, {Chary}, {Cheung}, {Cirasuolo}, {Conselice}, {Roshan Cooray},
  {Croton}, {Daddi}, {Dahlen}, {Dav{\'e}}, {de Mello}, {Dekel}, {Dickinson},
  {Dolch}, {Donley}, {Dunlop}, {Dutton}, {Elbaz}, {Fazio}, {Filippenko},
  {Finkelstein}, {Fontana}, {Gardner}, {Garnavich}, {Gawiser}, {Giavalisco},
  {Grazian}, {Guo}, {Hathi}, {H{\"a}ussler}, {Hopkins}, {Huang}, {Huang},
  {Jha}, {Kartaltepe}, {Kirshner}, {Koo}, {Lai}, {Lee}, {Li}, {Lotz}, {Lucas},
  {Madau}, {McCarthy}, {McGrath}, {McIntosh}, {McLure}, {Mobasher},
  {Moustakas}, {Mozena}, {Nandra}, {Newman}, {Niemi}, {Noeske}, {Papovich},
  {Pentericci}, {Pope}, {Primack}, {Rajan}, {Ravindranath}, {Reddy}, {Renzini},
  {Rix}, {Robaina}, {Rodney}, {Rosario}, {Rosati}, {Salimbeni}, {Scarlata},
  {Siana}, {Simard}, {Smidt}, {Somerville}, {Spinrad}, {Straughn}, {Strolger},
  {Telford}, {Teplitz}, {Trump}, {van der Wel}, {Villforth}, {Wechsler},
  {Weiner}, {Wiklind}, {Wild}, {Wilson}, {Wuyts}, {Yan}, \& {Yun}}]{Grogin2011}
{Grogin}, N.~A., {Kocevski}, D.~D., {Faber}, S.~M., {et~al.} 2011, \apjs, 197,
  35

\bibitem[{{Guo} {et~al.}(2010){Guo}, {White}, {Li}, \&
  {Boylan-Kolchin}}]{Guo2010}
{Guo}, Q., {White}, S., {Li}, C., \& {Boylan-Kolchin}, M. 2010, \mnras, 404,
  1111

\bibitem[{{Hadzhiyska} {et~al.}(2021){Hadzhiyska}, {Bose}, {Eisenstein}, \&
  {Hernquist}}]{Hadzhiyska2021}
{Hadzhiyska}, B., {Bose}, S., {Eisenstein}, D., \& {Hernquist}, L. 2021,
  \mnras, 501, 1603

\bibitem[{{Hadzhiyska} {et~al.}(2020){Hadzhiyska}, {Bose}, {Eisenstein},
  {Hernquist}, \& {Spergel}}]{Hadzhiyska2020}
{Hadzhiyska}, B., {Bose}, S., {Eisenstein}, D., {Hernquist}, L., \& {Spergel},
  D.~N. 2020, \mnras, 493, 5506

\bibitem[{{Hadzhiyska} {et~al.}(2022){Hadzhiyska}, {Eisenstein}, {Bose},
  {Garrison}, \& {Maksimova}}]{Hadzhiyska2022}
{Hadzhiyska}, B., {Eisenstein}, D., {Bose}, S., {Garrison}, L.~H., \&
  {Maksimova}, N. 2022, \mnras, 509, 501

\bibitem[{{Hadzhiyska} {et~al.}(2023{\natexlab{a}}){Hadzhiyska}, {Eisenstein},
  {Hernquist}, {Pakmor}, {Bose}, {Delgado}, {Contreras}, {Kannan}, {White},
  {Springel}, {Frenk}, {Hern{\'a}ndez-Aguayo}, {Barrera}, \&
  {Monica}}]{Hadzhiyska2023a}
{Hadzhiyska}, B., {Eisenstein}, D., {Hernquist}, L., {et~al.}
  2023{\natexlab{a}}, \mnras, 524, 2507

\bibitem[{{Hadzhiyska} {et~al.}(2023{\natexlab{b}}){Hadzhiyska}, {Hernquist},
  {Eisenstein}, {Delgado}, {Bose}, {Kannan}, {Pakmor}, {Springel}, {Contreras},
  {Barrera}, {Ferlito}, {Hern{\'a}ndez-Aguayo}, {White}, \&
  {Frenk}}]{Hadzhiyska2023b}
{Hadzhiyska}, B., {Hernquist}, L., {Eisenstein}, D., {et~al.}
  2023{\natexlab{b}}, \mnras, 524, 2524

\bibitem[{{Hahn} {et~al.}(2024){Hahn}, {Bottrell}, \& {Lee}}]{Hahn2024}
{Hahn}, C., {Bottrell}, C., \& {Lee}, K.-G. 2024, \apj, 968, 90

\bibitem[{{Hearin} {et~al.}(2014){Hearin}, {Watson}, {Becker}, {Reyes},
  {Berlind}, \& {Zentner}}]{Hearin2014}
{Hearin}, A.~P., {Watson}, D.~F., {Becker}, M.~R., {et~al.} 2014, \mnras, 444,
  729

\bibitem[{{Hearin} {et~al.}(2013){Hearin}, {Zentner}, {Berlind}, \&
  {Newman}}]{Hearin2013}
{Hearin}, A.~P., {Zentner}, A.~R., {Berlind}, A.~A., \& {Newman}, J.~A. 2013,
  \mnras, 433, 659

\bibitem[{{Henriques} {et~al.}(2019){Henriques}, {White}, {Lilly}, {Bell},
  {Bluck}, \& {Terrazas}}]{Henriques2019}
{Henriques}, B. M.~B., {White}, S. D.~M., {Lilly}, S.~J., {et~al.} 2019,
  \mnras, 485, 3446

\bibitem[{{Henriques} {et~al.}(2015){Henriques}, {White}, {Thomas}, {Angulo},
  {Guo}, {Lemson}, {Springel}, \& {Overzier}}]{Henriques2015}
{Henriques}, B. M.~B., {White}, S. D.~M., {Thomas}, P.~A., {et~al.} 2015,
  \mnras, 451, 2663

\bibitem[{{Hickson} {et~al.}(1992){Hickson}, {Mendes de Oliveira}, {Huchra}, \&
  {Palumbo}}]{Hickson1992}
{Hickson}, P., {Mendes de Oliveira}, C., {Huchra}, J.~P., \& {Palumbo}, G.~G.
  1992, \apj, 399, 353

\bibitem[{{Ho} {et~al.}(2024){Ho}, {Bartlett}, {Chartier}, {Cuesta-Lazaro},
  {Ding}, {Lapel}, {Lemos}, {Lovell}, {Makinen}, {Modi}, {Pandya}, {Pandey},
  {Perez}, {Wandelt}, \& {Bryan}}]{Ho2024}
{Ho}, M., {Bartlett}, D.~J., {Chartier}, N., {et~al.} 2024, The Open Journal of
  Astrophysics, 7, 54

\bibitem[{{Hoekstra} {et~al.}(2013){Hoekstra}, {Bartelmann}, {Dahle}, {Israel},
  {Limousin}, \& {Meneghetti}}]{Hoekstra2013}
{Hoekstra}, H., {Bartelmann}, M., {Dahle}, H., {et~al.} 2013, \ssr, 177, 75

\bibitem[{{Hort{\'u}a} {et~al.}(2023){Hort{\'u}a}, {Garc{\'\i}a}, \&
  {Casta{\~n}eda C.}}]{Hortua2023}
{Hort{\'u}a}, H.~J., {Garc{\'\i}a}, L.~{\'A}., \& {Casta{\~n}eda C.}, L. 2023,
  Frontiers in Astronomy and Space Sciences, 10, 1139120

\bibitem[{{Hoyle} \& {Lyttleton}(1939)}]{Hoyle1939}
{Hoyle}, F. \& {Lyttleton}, R.~A. 1939, Proceedings of the Cambridge
  Philosophical Society, 35, 405

\bibitem[{{Hu} \& {Sugiyama}(1995)}]{Hu1995}
{Hu}, W. \& {Sugiyama}, N. 1995, \apj, 444, 489

\bibitem[{{Huertas-Company} {et~al.}(2024){Huertas-Company}, {Iyer},
  {Angeloudi}, {Bagley}, {Finkelstein}, {Kartaltepe}, {McGrath}, {Sarmiento},
  {Vega-Ferrero}, {Arrabal Haro}, {Behroozi}, {Buitrago}, {Cheng}, {Costantin},
  {Dekel}, {Dickinson}, {Elbaz}, {Grogin}, {Hathi}, {Holwerda}, {Koekemoer},
  {Lucas}, {Papovich}, {P{\'e}rez-Gonz{\'a}lez}, {Pirzkal}, {Seill{\'e}}, {de
  la Vega}, {Wuyts}, {Yang}, \& {Yung}}]{Huertas-Company2024}
{Huertas-Company}, M., {Iyer}, K.~G., {Angeloudi}, E., {et~al.} 2024, \aap,
  685, A48

\bibitem[{{Ilbert} {et~al.}(2006){Ilbert}, {Arnouts}, {McCracken},
  {Bolzonella}, {Bertin}, {Le F{\`e}vre}, {Mellier}, {Zamorani}, {Pell{\`o}},
  {Iovino}, {Tresse}, {Le Brun}, {Bottini}, {Garilli}, {Maccagni}, {Picat},
  {Scaramella}, {Scodeggio}, {Vettolani}, {Zanichelli}, {Adami}, {Bardelli},
  {Cappi}, {Charlot}, {Ciliegi}, {Contini}, {Cucciati}, {Foucaud}, {Franzetti},
  {Gavignaud}, {Guzzo}, {Marano}, {Marinoni}, {Mazure}, {Meneux}, {Merighi},
  {Paltani}, {Pollo}, {Pozzetti}, {Radovich}, {Zucca}, {Bondi}, {Bongiorno},
  {Busarello}, {de La Torre}, {Gregorini}, {Lamareille}, {Mathez}, {Merluzzi},
  {Ripepi}, {Rizzo}, \& {Vergani}}]{Ilbert2006}
{Ilbert}, O., {Arnouts}, S., {McCracken}, H.~J., {et~al.} 2006, \aap, 457, 841

\bibitem[{{Jones} {et~al.}(2024){Jones}, {Do}, {Boscoe}, {Singal}, {Wan}, \&
  {Nguyen}}]{Jones2024}
{Jones}, E., {Do}, T., {Boscoe}, B., {et~al.} 2024, \apj, 964, 130

\bibitem[{{Kauffmann} {et~al.}(2003){Kauffmann}, {Heckman}, {Tremonti},
  {Brinchmann}, {Charlot}, {White}, {Ridgway}, {Brinkmann}, {Fukugita}, {Hall},
  {Ivezi{\'c}}, {Richards}, \& {Schneider}}]{Kauffmann2003}
{Kauffmann}, G., {Heckman}, T.~M., {Tremonti}, C., {et~al.} 2003, \mnras, 346,
  1055

\bibitem[{{Kennicutt}(1998)}]{Kennicutt1998}
{Kennicutt}, Jr., R.~C. 1998, \araa, 36, 189

\bibitem[{{Kewley} {et~al.}(2001){Kewley}, {Dopita}, {Sutherland}, {Heisler},
  \& {Trevena}}]{Kewley2001}
{Kewley}, L.~J., {Dopita}, M.~A., {Sutherland}, R.~S., {Heisler}, C.~A., \&
  {Trevena}, J. 2001, \apj, 556, 121

\bibitem[{{Kewley} {et~al.}(2006){Kewley}, {Groves}, {Kauffmann}, \&
  {Heckman}}]{Kewley2006}
{Kewley}, L.~J., {Groves}, B., {Kauffmann}, G., \& {Heckman}, T. 2006, \mnras,
  372, 961

\bibitem[{{Khandai} {et~al.}(2015){Khandai}, {Di Matteo}, {Croft}, {Wilkins},
  {Feng}, {Tucker}, {DeGraf}, \& {Liu}}]{Khandai2015}
{Khandai}, N., {Di Matteo}, T., {Croft}, R., {et~al.} 2015, \mnras, 450, 1349

\bibitem[{{Klypin} {et~al.}(2015){Klypin}, {Prada}, {Yepes}, {He{\ss}}, \&
  {Gottl{\"o}ber}}]{Klypin2015}
{Klypin}, A., {Prada}, F., {Yepes}, G., {He{\ss}}, S., \& {Gottl{\"o}ber}, S.
  2015, \mnras, 447, 3693

\bibitem[{{Klypin} {et~al.}(2011){Klypin}, {Trujillo-Gomez}, \&
  {Primack}}]{Klypin2011}
{Klypin}, A.~A., {Trujillo-Gomez}, S., \& {Primack}, J. 2011, \apj, 740, 102

\bibitem[{{Knebe} {et~al.}(2011){Knebe}, {Knollmann}, {Muldrew}, {Pearce},
  {Aragon-Calvo}, {Ascasibar}, {Behroozi}, {Ceverino}, {Colombi}, {Diemand},
  {Dolag}, {Falck}, {Fasel}, {Gardner}, {Gottl{\"o}ber}, {Hsu}, {Iannuzzi},
  {Klypin}, {Luki{\'c}}, {Maciejewski}, {McBride}, {Neyrinck}, {Planelles},
  {Potter}, {Quilis}, {Rasera}, {Read}, {Ricker}, {Roy}, {Springel}, {Stadel},
  {Stinson}, {Sutter}, {Turchaninov}, {Tweed}, {Yepes}, \& {Zemp}}]{Knabe2011}
{Knebe}, A., {Knollmann}, S.~R., {Muldrew}, S.~I., {et~al.} 2011, \mnras, 415,
  2293

\bibitem[{{Knobel} {et~al.}(2009){Knobel}, {Lilly}, {Iovino}, {Porciani},
  {Kova{\v{c}}}, {Cucciati}, {Finoguenov}, {Kitzbichler}, {Carollo}, {Contini},
  {Kneib}, {Le F{\`e}vre}, {Mainieri}, {Renzini}, {Scodeggio}, {Zamorani},
  {Bardelli}, {Bolzonella}, {Bongiorno}, {Caputi}, {Coppa}, {de la Torre}, {de
  Ravel}, {Franzetti}, {Garilli}, {Kampczyk}, {Lamareille}, {Le Borgne}, {Le
  Brun}, {Maier}, {Mignoli}, {Pello}, {Peng}, {Perez Montero}, {Ricciardelli},
  {Silverman}, {Tanaka}, {Tasca}, {Tresse}, {Vergani}, {Zucca}, {Abbas},
  {Bottini}, {Cappi}, {Cassata}, {Cimatti}, {Fumana}, {Guzzo}, {Koekemoer},
  {Leauthaud}, {Maccagni}, {Marinoni}, {McCracken}, {Memeo}, {Meneux}, {Oesch},
  {Pozzetti}, \& {Scaramella}}]{Knobel2009}
{Knobel}, C., {Lilly}, S.~J., {Iovino}, A., {et~al.} 2009, \apj, 697, 1842

\bibitem[{{Komatsu} {et~al.}(2011){Komatsu}, {Smith}, {Dunkley}, {Bennett},
  {Gold}, {Hinshaw}, {Jarosik}, {Larson}, {Nolta}, {Page}, {Spergel},
  {Halpern}, {Hill}, {Kogut}, {Limon}, {Meyer}, {Odegard}, {Tucker}, {Weiland},
  {Wollack}, \& {Wright}}]{Komatsu2011}
{Komatsu}, E., {Smith}, K.~M., {Dunkley}, J., {et~al.} 2011, \apjs, 192, 18

\bibitem[{{Kravtsov} {et~al.}(2004){Kravtsov}, {Berlind}, {Wechsler}, {Klypin},
  {Gottl{\"o}ber}, {Allgood}, \& {Primack}}]{Kravtsov2004}
{Kravtsov}, A.~V., {Berlind}, A.~A., {Wechsler}, R.~H., {et~al.} 2004, \apj,
  609, 35

\bibitem[{{Kriek} {et~al.}(2015){Kriek}, {Shapley}, {Reddy}, {Siana}, {Coil},
  {Mobasher}, {Freeman}, {de Groot}, {Price}, {Sanders}, {Shivaei}, {Brammer},
  {Momcheva}, {Skelton}, {van Dokkum}, {Whitaker}, {Aird}, {Azadi}, {Kassis},
  {Bullock}, {Conroy}, {Dav{\'e}}, {Kere{\v{s}}}, \& {Krumholz}}]{Kriek2015}
{Kriek}, M., {Shapley}, A.~E., {Reddy}, N.~A., {et~al.} 2015, \apjs, 218, 15

\bibitem[{{Lehmann} {et~al.}(2017){Lehmann}, {Mao}, {Becker}, {Skillman}, \&
  {Wechsler}}]{Lehmann2017}
{Lehmann}, B.~V., {Mao}, Y.-Y., {Becker}, M.~R., {Skillman}, S.~W., \&
  {Wechsler}, R.~H. 2017, \apj, 834, 37

\bibitem[{{Looser} {et~al.}(2021){Looser}, {Lilly}, {Sin}, {Henriques},
  {Maiolino}, \& {Cirasuolo}}]{Looser2021}
{Looser}, T.~J., {Lilly}, S.~J., {Sin}, L. P.~T., {et~al.} 2021, \mnras, 504,
  3029

\bibitem[{{Maiolino} {et~al.}(2020){Maiolino}, {Cirasuolo}, {Afonso}, {Bauer},
  {Bowler}, {Cucciati}, {Daddi}, {De Lucia}, {Evans}, {Flores}, {Gargiulo},
  {Garilli}, {Jablonka}, {Jarvis}, {Kneib}, {Lilly}, {Looser}, {Magliocchetti},
  {Man}, {Mannucci}, {Maurogordato}, {McLure}, {Norberg}, {Oesch}, {Oliva},
  {Paltani}, {Pappalardo}, {Peng}, {Pentericci}, {Pozzetti}, {Renzini},
  {Rodrigues}, {Royer}, {Serjeant}, {Vanzi}, {Wild}, \&
  {Zamorani}}]{Maiolino2020}
{Maiolino}, R., {Cirasuolo}, M., {Afonso}, J., {et~al.} 2020, The Messenger,
  180, 24

\bibitem[{{Maiolino} {et~al.}(2024{\natexlab{a}}){Maiolino}, {Scholtz},
  {Curtis-Lake}, {Carniani}, {Baker}, {de Graaff}, {Tacchella}, {{\"U}bler},
  {D'Eugenio}, {Witstok}, {Curti}, {Arribas}, {Bunker}, {Charlot},
  {Chevallard}, {Eisenstein}, {Egami}, {Ji}, {Jones}, {Lyu}, {Rawle},
  {Robertson}, {Rujopakarn}, {Perna}, {Sun}, {Venturi}, {Williams}, \&
  {Willott}}]{Maiolino2024b}
{Maiolino}, R., {Scholtz}, J., {Curtis-Lake}, E., {et~al.} 2024{\natexlab{a}},
  \aap, 691, A145

\bibitem[{{Maiolino} {et~al.}(2024{\natexlab{b}}){Maiolino}, {Scholtz},
  {Witstok}, {Carniani}, {D'Eugenio}, {de Graaff}, {{\"U}bler}, {Tacchella},
  {Curtis-Lake}, {Arribas}, {Bunker}, {Charlot}, {Chevallard}, {Curti},
  {Looser}, {Maseda}, {Rawle}, {Rodr{\'\i}guez del Pino}, {Willott}, {Egami},
  {Eisenstein}, {Hainline}, {Robertson}, {Williams}, {Willmer}, {Baker},
  {Boyett}, {DeCoursey}, {Fabian}, {Helton}, {Ji}, {Jones}, {Kumari},
  {Laporte}, {Nelson}, {Perna}, {Sandles}, {Shivaei}, \& {Sun}}]{Maiolino2024a}
{Maiolino}, R., {Scholtz}, J., {Witstok}, J., {et~al.} 2024{\natexlab{b}},
  \nat, 627, 59

\bibitem[{{Mandelbaum}(2018)}]{Mandelbaum2018}
{Mandelbaum}, R. 2018, \araa, 56, 393

\bibitem[{{Mandelbaum} {et~al.}(2006){Mandelbaum}, {Seljak}, {Kauffmann},
  {Hirata}, \& {Brinkmann}}]{Mandelbaum2006}
{Mandelbaum}, R., {Seljak}, U., {Kauffmann}, G., {Hirata}, C.~M., \&
  {Brinkmann}, J. 2006, \mnras, 368, 715

\bibitem[{{Mandelbaum} {et~al.}(2016){Mandelbaum}, {Wang}, {Zu}, {White},
  {Henriques}, \& {More}}]{Mandelbaum2016}
{Mandelbaum}, R., {Wang}, W., {Zu}, Y., {et~al.} 2016, \mnras, 457, 3200

\bibitem[{{Mannheim} {et~al.}(2000){Mannheim}, {Protheroe}, \&
  {Rachen}}]{Mannheim2000}
{Mannheim}, K., {Protheroe}, R.~J., \& {Rachen}, J.~P. 2000, \prd, 63, 023003

\bibitem[{{Marinacci} {et~al.}(2018){Marinacci}, {Vogelsberger}, {Pakmor},
  {Torrey}, {Springel}, {Hernquist}, {Nelson}, {Weinberger}, {Pillepich},
  {Naiman}, \& {Genel}}]{Marinacci2018}
{Marinacci}, F., {Vogelsberger}, M., {Pakmor}, R., {et~al.} 2018, \mnras, 480,
  5113

\bibitem[{{McAlpine} {et~al.}(2016){McAlpine}, {Helly}, {Schaller}, {Trayford},
  {Qu}, {Furlong}, {Bower}, {Crain}, {Schaye}, {Theuns}, {Dalla Vecchia},
  {Frenk}, {McCarthy}, {Jenkins}, {Rosas-Guevara}, {White}, {Baes}, {Camps}, \&
  {Lemson}}]{McAlpine2016}
{McAlpine}, S., {Helly}, J.~C., {Schaller}, M., {et~al.} 2016, Astronomy and
  Computing, 15, 72

\bibitem[{{Mendel} {et~al.}(2014){Mendel}, {Simard}, {Palmer}, {Ellison}, \&
  {Patton}}]{Mendel2014}
{Mendel}, J.~T., {Simard}, L., {Palmer}, M., {Ellison}, S.~L., \& {Patton},
  D.~R. 2014, \apjs, 210, 3

\bibitem[{{Mo} {et~al.}(2010){Mo}, {van den Bosch}, \& {White}}]{Mo2010}
{Mo}, H., {van den Bosch}, F.~C., \& {White}, S. 2010, {Galaxy Formation and
  Evolution} (Cambridge)

\bibitem[{{More} {et~al.}(2015){More}, {Diemer}, \& {Kravtsov}}]{More2015}
{More}, S., {Diemer}, B., \& {Kravtsov}, A.~V. 2015, \apj, 810, 36

\bibitem[{{More} {et~al.}(2011){More}, {Kravtsov}, {Dalal}, \&
  {Gottl{\"o}ber}}]{More2011}
{More}, S., {Kravtsov}, A.~V., {Dalal}, N., \& {Gottl{\"o}ber}, S. 2011, \apjs,
  195, 4

\bibitem[{{More} {et~al.}(2016){More}, {Miyatake}, {Takada}, {Diemer},
  {Kravtsov}, {Dalal}, {More}, {Murata}, {Mandelbaum}, {Rozo}, {Rykoff},
  {Oguri}, \& {Spergel}}]{More2016}
{More}, S., {Miyatake}, H., {Takada}, M., {et~al.} 2016, \apj, 825, 39

\bibitem[{{Moster} {et~al.}(2010){Moster}, {Somerville}, {Maulbetsch}, {van den
  Bosch}, {Macci{\`o}}, {Naab}, \& {Oser}}]{Moster2010}
{Moster}, B.~P., {Somerville}, R.~S., {Maulbetsch}, C., {et~al.} 2010, \apj,
  710, 903

\bibitem[{{Mucesh} {et~al.}(2021){Mucesh}, {Hartley}, {Palmese}, {Lahav},
  {Whiteway}, {Bluck}, {Alarcon}, {Amon}, {Bechtol}, {Bernstein}, {Carnero
  Rosell}, {Carrasco Kind}, {Choi}, {Eckert}, {Everett}, {Gruen}, {Gruendl},
  {Harrison}, {Huff}, {Kuropatkin}, {Sevilla-Noarbe}, {Sheldon}, {Yanny},
  {Aguena}, {Allam}, {Bacon}, {Bertin}, {Bhargava}, {Brooks}, {Carretero},
  {Castander}, {Conselice}, {Costanzi}, {Crocce}, {da Costa}, {Pereira}, {De
  Vicente}, {Desai}, {Diehl}, {Drlica-Wagner}, {Evrard}, {Ferrero}, {Flaugher},
  {Fosalba}, {Frieman}, {Garc{\'\i}a-Bellido}, {Gaztanaga}, {Gerdes},
  {Gschwend}, {Gutierrez}, {Hinton}, {Hollowood}, {Honscheid}, {James},
  {Kuehn}, {Lima}, {Lin}, {Maia}, {Melchior}, {Menanteau}, {Miquel}, {Morgan},
  {Paz-Chinch{\'o}n}, {Plazas}, {Sanchez}, {Scarpine}, {Schubnell}, {Serrano},
  {Smith}, {Suchyta}, {Tarle}, {Thomas}, {To}, {Varga}, {Wilkinson}, \& {DES
  Collaboration}}]{Mucesh2021}
{Mucesh}, S., {Hartley}, W.~G., {Palmese}, A., {et~al.} 2021, \mnras, 502, 2770

\bibitem[{{Naiman} {et~al.}(2018){Naiman}, {Pillepich}, {Springel},
  {Ramirez-Ruiz}, {Torrey}, {Vogelsberger}, {Pakmor}, {Nelson}, {Marinacci},
  {Hernquist}, {Weinberger}, \& {Genel}}]{Naiman2018}
{Naiman}, J.~P., {Pillepich}, A., {Springel}, V., {et~al.} 2018, \mnras, 477,
  1206

\bibitem[{{Navarro} {et~al.}(1996){Navarro}, {Frenk}, \& {White}}]{Navarro1996}
{Navarro}, J.~F., {Frenk}, C.~S., \& {White}, S. D.~M. 1996, \apj, 462, 563

\bibitem[{{Nelson} {et~al.}(2018){Nelson}, {Pillepich}, {Springel},
  {Weinberger}, {Hernquist}, {Pakmor}, {Genel}, {Torrey}, {Vogelsberger},
  {Kauffmann}, {Marinacci}, \& {Naiman}}]{Nelson2018}
{Nelson}, D., {Pillepich}, A., {Springel}, V., {et~al.} 2018, \mnras, 475, 624

\bibitem[{{Overzier}(2016)}]{Overzier2016}
{Overzier}, R.~A. 2016, \aapr, 24, 14

\bibitem[{{Peacock} \& {Smith}(2000)}]{Peacock2000}
{Peacock}, J.~A. \& {Smith}, R.~E. 2000, \mnras, 318, 1144

\bibitem[{{Pedregosa} {et~al.}(2011){Pedregosa}, {Varoquaux}, {Gramfort},
  {Michel}, {Thirion}, {Grisel}, {Blondel}, {M{\"u}ller}, {Nothman}, {Louppe},
  {Prettenhofer}, {Weiss}, {Dubourg}, {Vanderplas}, {Passos}, {Cournapeau},
  {Brucher}, {Perrot}, \& {Duchesnay}}]{Pedregosa2011}
{Pedregosa}, F., {Varoquaux}, G., {Gramfort}, A., {et~al.} 2011, Journal of
  Machine Learning Research, 12, 2825

\bibitem[{{Peng} {et~al.}(2010){Peng}, {Lilly}, {Kova{\v{c}}}, {Bolzonella},
  {Pozzetti}, {Renzini}, {Zamorani}, {Ilbert}, {Knobel}, {Iovino}, {Maier},
  {Cucciati}, {Tasca}, {Carollo}, {Silverman}, {Kampczyk}, {de Ravel},
  {Sanders}, {Scoville}, {Contini}, {Mainieri}, {Scodeggio}, {Kneib}, {Le
  F{\`e}vre}, {Bardelli}, {Bongiorno}, {Caputi}, {Coppa}, {de la Torre},
  {Franzetti}, {Garilli}, {Lamareille}, {Le Borgne}, {Le Brun}, {Mignoli},
  {Perez Montero}, {Pello}, {Ricciardelli}, {Tanaka}, {Tresse}, {Vergani},
  {Welikala}, {Zucca}, {Oesch}, {Abbas}, {Barnes}, {Bordoloi}, {Bottini},
  {Cappi}, {Cassata}, {Cimatti}, {Fumana}, {Hasinger}, {Koekemoer},
  {Leauthaud}, {Maccagni}, {Marinoni}, {McCracken}, {Memeo}, {Meneux}, {Nair},
  {Porciani}, {Presotto}, \& {Scaramella}}]{Peng2010}
{Peng}, Y.-j., {Lilly}, S.~J., {Kova{\v{c}}}, K., {et~al.} 2010, \apj, 721, 193

\bibitem[{{Peng} {et~al.}(2012){Peng}, {Lilly}, {Renzini}, \&
  {Carollo}}]{Peng2012}
{Peng}, Y.-j., {Lilly}, S.~J., {Renzini}, A., \& {Carollo}, M. 2012, \apj, 757,
  4

\bibitem[{{Persic} {et~al.}(1996){Persic}, {Salucci}, \& {Stel}}]{Persic1996}
{Persic}, M., {Salucci}, P., \& {Stel}, F. 1996, \mnras, 281, 27

\bibitem[{{Pillepich} {et~al.}(2018){Pillepich}, {Springel}, {Nelson}, {Genel},
  {Naiman}, {Pakmor}, {Hernquist}, {Torrey}, {Vogelsberger}, {Weinberger}, \&
  {Marinacci}}]{Pillepich2018}
{Pillepich}, A., {Springel}, V., {Nelson}, D., {et~al.} 2018, \mnras, 473, 4077

\bibitem[{{Piotrowska} {et~al.}(2022){Piotrowska}, {Bluck}, {Maiolino}, \&
  {Peng}}]{Piotrowska2022}
{Piotrowska}, J.~M., {Bluck}, A. F.~L., {Maiolino}, R., \& {Peng}, Y. 2022,
  \mnras, 512, 1052

\bibitem[{{Planck Collaboration} {et~al.}(2014){Planck Collaboration}, {Ade},
  {Aghanim}, {Armitage-Caplan}, {Arnaud}, {Ashdown}, {Atrio-Barandela},
  {Aumont}, {Baccigalupi}, {Banday}, {Barreiro}, {Bartlett}, {Battaner},
  {Benabed}, {Beno{\^\i}t}, {Benoit-L{\'e}vy}, {Bernard}, {Bersanelli},
  {Bielewicz}, {Bobin}, {Bock}, {Bonaldi}, {Bond}, {Borrill}, {Bouchet},
  {Bridges}, {Bucher}, {Burigana}, {Butler}, {Calabrese}, {Cappellini},
  {Cardoso}, {Catalano}, {Challinor}, {Chamballu}, {Chary}, {Chen}, {Chiang},
  {Chiang}, {Christensen}, {Church}, {Clements}, {Colombi}, {Colombo},
  {Couchot}, {Coulais}, {Crill}, {Curto}, {Cuttaia}, {Danese}, {Davies},
  {Davis}, {de Bernardis}, {de Rosa}, {de Zotti}, {Delabrouille}, {Delouis},
  {D{\'e}sert}, {Dickinson}, {Diego}, {Dolag}, {Dole}, {Donzelli}, {Dor{\'e}},
  {Douspis}, {Dunkley}, {Dupac}, {Efstathiou}, {Elsner}, {En{\ss}lin},
  {Eriksen}, {Finelli}, {Forni}, {Frailis}, {Fraisse}, {Franceschi}, {Gaier},
  {Galeotta}, {Galli}, {Ganga}, {Giard}, {Giardino}, {Giraud-H{\'e}raud},
  {Gjerl{\o}w}, {Gonz{\'a}lez-Nuevo}, {G{\'o}rski}, {Gratton}, {Gregorio},
  {Gruppuso}, {Gudmundsson}, {Haissinski}, {Hamann}, {Hansen}, {Hanson},
  {Harrison}, {Henrot-Versill{\'e}}, {Hern{\'a}ndez-Monteagudo}, {Herranz},
  {Hildebrandt}, {Hivon}, {Hobson}, {Holmes}, {Hornstrup}, {Hou}, {Hovest},
  {Huffenberger}, {Jaffe}, {Jaffe}, {Jewell}, {Jones}, {Juvela},
  {Keih{\"a}nen}, {Keskitalo}, {Kisner}, {Kneissl}, {Knoche}, {Knox}, {Kunz},
  {Kurki-Suonio}, {Lagache}, {L{\"a}hteenm{\"a}ki}, {Lamarre}, {Lasenby},
  {Lattanzi}, {Laureijs}, {Lawrence}, {Leach}, {Leahy}, {Leonardi},
  {Le{\'o}n-Tavares}, {Lesgourgues}, {Lewis}, {Liguori}, {Lilje},
  {Linden-V{\o}rnle}, {L{\'o}pez-Caniego}, {Lubin}, {Mac{\'\i}as-P{\'e}rez},
  {Maffei}, {Maino}, {Mandolesi}, {Maris}, {Marshall}, {Martin},
  {Mart{\'\i}nez-Gonz{\'a}lez}, {Masi}, {Massardi}, {Matarrese}, {Matthai},
  {Mazzotta}, {Meinhold}, {Melchiorri}, {Melin}, {Mendes}, {Menegoni},
  {Mennella}, {Migliaccio}, {Millea}, {Mitra}, {Miville-Desch{\^e}nes},
  {Moneti}, {Montier}, {Morgante}, {Mortlock}, {Moss}, {Munshi}, {Murphy},
  {Naselsky}, {Nati}, {Natoli}, {Netterfield}, {N{\o}rgaard-Nielsen},
  {Noviello}, {Novikov}, {Novikov}, {O'Dwyer}, {Osborne}, {Oxborrow}, {Paci},
  {Pagano}, {Pajot}, {Paladini}, {Paoletti}, {Partridge}, {Pasian},
  {Patanchon}, {Pearson}, {Pearson}, {Peiris}, {Perdereau}, {Perotto},
  {Perrotta}, {Pettorino}, {Piacentini}, {Piat}, {Pierpaoli}, {Pietrobon},
  {Plaszczynski}, {Platania}, \& {Pointecouteau}}]{PLANCK2014}
{Planck Collaboration}, {Ade}, P.~A.~R., {Aghanim}, N., {et~al.} 2014, \aap,
  571, A16

\bibitem[{{Planck Collaboration} {et~al.}(2016){Planck Collaboration}, {Ade},
  {Aghanim}, {Arnaud}, {Ashdown}, {Aumont}, {Baccigalupi}, {Banday},
  {Barreiro}, {Bartlett}, {Bartolo}, {Battaner}, {Battye}, {Benabed},
  {Beno{\^\i}t}, {Benoit-L{\'e}vy}, {Bernard}, {Bersanelli}, {Bielewicz},
  {Bock}, {Bonaldi}, {Bonavera}, {Bond}, {Borrill}, {Bouchet}, {Boulanger},
  {Bucher}, {Burigana}, {Butler}, {Calabrese}, {Cardoso}, {Catalano},
  {Challinor}, {Chamballu}, {Chary}, {Chiang}, {Chluba}, {Christensen},
  {Church}, {Clements}, {Colombi}, {Colombo}, {Combet}, {Coulais}, {Crill},
  {Curto}, {Cuttaia}, {Danese}, {Davies}, {Davis}, {de Bernardis}, {de Rosa},
  {de Zotti}, {Delabrouille}, {D{\'e}sert}, {Di Valentino}, {Dickinson},
  {Diego}, {Dolag}, {Dole}, {Donzelli}, {Dor{\'e}}, {Douspis}, {Ducout},
  {Dunkley}, {Dupac}, {Efstathiou}, {Elsner}, {En{\ss}lin}, {Eriksen},
  {Farhang}, {Fergusson}, {Finelli}, {Forni}, {Frailis}, {Fraisse},
  {Franceschi}, {Frejsel}, {Galeotta}, {Galli}, {Ganga}, {Gauthier}, {Gerbino},
  {Ghosh}, {Giard}, {Giraud-H{\'e}raud}, {Giusarma}, {Gjerl{\o}w},
  {Gonz{\'a}lez-Nuevo}, {G{\'o}rski}, {Gratton}, {Gregorio}, {Gruppuso},
  {Gudmundsson}, {Hamann}, {Hansen}, {Hanson}, {Harrison}, {Helou},
  {Henrot-Versill{\'e}}, {Hern{\'a}ndez-Monteagudo}, {Herranz}, {Hildebrandt},
  {Hivon}, {Hobson}, {Holmes}, {Hornstrup}, {Hovest}, {Huang}, {Huffenberger},
  {Hurier}, {Jaffe}, {Jaffe}, {Jones}, {Juvela}, {Keih{\"a}nen}, {Keskitalo},
  {Kisner}, {Kneissl}, {Knoche}, {Knox}, {Kunz}, {Kurki-Suonio}, {Lagache},
  {L{\"a}hteenm{\"a}ki}, {Lamarre}, {Lasenby}, {Lattanzi}, {Lawrence}, {Leahy},
  {Leonardi}, {Lesgourgues}, {Levrier}, {Lewis}, {Liguori}, {Lilje},
  {Linden-V{\o}rnle}, {L{\'o}pez-Caniego}, {Lubin}, {Mac{\'\i}as-P{\'e}rez},
  {Maggio}, {Maino}, {Mandolesi}, {Mangilli}, {Marchini}, {Maris}, {Martin},
  {Martinelli}, {Mart{\'\i}nez-Gonz{\'a}lez}, {Masi}, {Matarrese}, {McGehee},
  {Meinhold}, {Melchiorri}, {Melin}, {Mendes}, {Mennella}, {Migliaccio},
  {Millea}, {Mitra}, {Miville-Desch{\^e}nes}, {Moneti}, {Montier}, {Morgante},
  {Mortlock}, {Moss}, {Munshi}, {Murphy}, {Naselsky}, {Nati}, {Natoli},
  {Netterfield}, {N{\o}rgaard-Nielsen}, {Noviello}, {Novikov}, {Novikov},
  {Oxborrow}, {Paci}, {Pagano}, {Pajot}, {Paladini}, {Paoletti}, {Partridge},
  {Pasian}, {Patanchon}, {Pearson}, {Perdereau}, {Perotto}, {Perrotta},
  {Pettorino}, {Piacentini}, {Piat}, {Pierpaoli}, {Pietrobon}, {Plaszczynski},
  {Pointecouteau}, {Polenta}, {Popa}, {Pratt}, \& {Pr{\'e}zeau}}]{PLANCK2016}
{Planck Collaboration}, {Ade}, P.~A.~R., {Aghanim}, N., {et~al.} 2016, \aap,
  594, A13

\bibitem[{{Planck Collaboration} {et~al.}(2020){Planck Collaboration},
  {Aghanim}, {Akrami}, {Ashdown}, {Aumont}, {Baccigalupi}, {Ballardini},
  {Banday}, {Barreiro}, {Bartolo}, {Basak}, {Battye}, {Benabed}, {Bernard},
  {Bersanelli}, {Bielewicz}, {Bock}, {Bond}, {Borrill}, {Bouchet}, {Boulanger},
  {Bucher}, {Burigana}, {Butler}, {Calabrese}, {Cardoso}, {Carron},
  {Challinor}, {Chiang}, {Chluba}, {Colombo}, {Combet}, {Contreras}, {Crill},
  {Cuttaia}, {de Bernardis}, {de Zotti}, {Delabrouille}, {Delouis}, {Di
  Valentino}, {Diego}, {Dor{\'e}}, {Douspis}, {Ducout}, {Dupac}, {Dusini},
  {Efstathiou}, {Elsner}, {En{\ss}lin}, {Eriksen}, {Fantaye}, {Farhang},
  {Fergusson}, {Fernandez-Cobos}, {Finelli}, {Forastieri}, {Frailis},
  {Fraisse}, {Franceschi}, {Frolov}, {Galeotta}, {Galli}, {Ganga},
  {G{\'e}nova-Santos}, {Gerbino}, {Ghosh}, {Gonz{\'a}lez-Nuevo}, {G{\'o}rski},
  {Gratton}, {Gruppuso}, {Gudmundsson}, {Hamann}, {Handley}, {Hansen},
  {Herranz}, {Hildebrandt}, {Hivon}, {Huang}, {Jaffe}, {Jones}, {Karakci},
  {Keih{\"a}nen}, {Keskitalo}, {Kiiveri}, {Kim}, {Kisner}, {Knox},
  {Krachmalnicoff}, {Kunz}, {Kurki-Suonio}, {Lagache}, {Lamarre}, {Lasenby},
  {Lattanzi}, {Lawrence}, {Le Jeune}, {Lemos}, {Lesgourgues}, {Levrier},
  {Lewis}, {Liguori}, {Lilje}, {Lilley}, {Lindholm}, {L{\'o}pez-Caniego},
  {Lubin}, {Ma}, {Mac{\'\i}as-P{\'e}rez}, {Maggio}, {Maino}, {Mandolesi},
  {Mangilli}, {Marcos-Caballero}, {Maris}, {Martin}, {Martinelli},
  {Mart{\'\i}nez-Gonz{\'a}lez}, {Matarrese}, {Mauri}, {McEwen}, {Meinhold},
  {Melchiorri}, {Mennella}, {Migliaccio}, {Millea}, {Mitra},
  {Miville-Desch{\^e}nes}, {Molinari}, {Montier}, {Morgante}, {Moss}, {Natoli},
  {N{\o}rgaard-Nielsen}, {Pagano}, {Paoletti}, {Partridge}, {Patanchon},
  {Peiris}, {Perrotta}, {Pettorino}, {Piacentini}, {Polastri}, {Polenta},
  {Puget}, {Rachen}, {Reinecke}, {Remazeilles}, {Renzi}, {Rocha}, {Rosset},
  {Roudier}, {Rubi{\~n}o-Mart{\'\i}n}, {Ruiz-Granados}, {Salvati}, {Sandri},
  {Savelainen}, {Scott}, {Shellard}, {Sirignano}, {Sirri}, {Spencer},
  {Sunyaev}, {Suur-Uski}, {Tauber}, {Tavagnacco}, {Tenti}, {Toffolatti},
  {Tomasi}, {Trombetti}, {Valenziano}, {Valiviita}, {Van Tent}, {Vibert},
  {Vielva}, {Villa}, {Vittorio}, {Wandelt}, {Wehus}, {White}, {White},
  {Zacchei}, \& {Zonca}}]{PLANCK2020}
{Planck Collaboration}, {Aghanim}, N., {Akrami}, Y., {et~al.} 2020, \aap, 641,
  A6

\bibitem[{{Pratt} {et~al.}(2010){Pratt}, {Arnaud}, {Piffaretti},
  {B{\"o}hringer}, {Ponman}, {Croston}, {Voit}, {Borgani}, \&
  {Bower}}]{Pratt2010}
{Pratt}, G.~W., {Arnaud}, M., {Piffaretti}, R., {et~al.} 2010, \aap, 511, A85

\bibitem[{{Press} \& {Davis}(1982)}]{Press1982}
{Press}, W.~H. \& {Davis}, M. 1982, \apj, 259, 449

\bibitem[{{Refsdal}(1964)}]{Refsdal1964}
{Refsdal}, S. 1964, \mnras, 128, 307

\bibitem[{{Remy} {et~al.}(2023){Remy}, {Lanusse}, {Jeffrey}, {Liu}, {Starck},
  {Osato}, \& {Schrabback}}]{Remy2023}
{Remy}, B., {Lanusse}, F., {Jeffrey}, N., {et~al.} 2023, \aap, 672, A51

\bibitem[{{Rubin} {et~al.}(1978){Rubin}, {Ford}, \& {Thonnard}}]{Rubin1978}
{Rubin}, V.~C., {Ford}, Jr., W.~K., \& {Thonnard}, N. 1978, \apjl, 225, L107

\bibitem[{{S{\'a}nchez}(2020)}]{Sanchez2020}
{S{\'a}nchez}, S.~F. 2020, \araa, 58, 99

\bibitem[{{S{\'a}nchez} {et~al.}(2016){S{\'a}nchez}, {P{\'e}rez},
  {S{\'a}nchez-Bl{\'a}zquez}, {Garc{\'\i}a-Benito}, {Ibarra-Mede},
  {Gonz{\'a}lez}, {Rosales-Ortega}, {S{\'a}nchez-Menguiano}, {Ascasibar},
  {Bitsakis}, {Law}, {Cano-D{\'\i}az}, {L{\'o}pez-Cob{\'a}}, {Marino}, {Gil de
  Paz}, {L{\'o}pez-S{\'a}nchez}, {Barrera-Ballesteros}, {Galbany}, {Mast},
  {Abril-Melgarejo}, \& {Roman-Lopes}}]{Sanchez2016}
{S{\'a}nchez}, S.~F., {P{\'e}rez}, E., {S{\'a}nchez-Bl{\'a}zquez}, P., {et~al.}
  2016, \rmxaa, 52, 171

\bibitem[{{Schaye} {et~al.}(2015){Schaye}, {Crain}, {Bower}, {Furlong},
  {Schaller}, {Theuns}, {Dalla Vecchia}, {Frenk}, {McCarthy}, {Helly},
  {Jenkins}, {Rosas-Guevara}, {White}, {Baes}, {Booth}, {Camps}, {Navarro},
  {Qu}, {Rahmati}, {Sawala}, {Thomas}, \& {Trayford}}]{Schaye2015}
{Schaye}, J., {Crain}, R.~A., {Bower}, R.~G., {et~al.} 2015, \mnras, 446, 521

\bibitem[{{Schneider} {et~al.}(1992){Schneider}, {Ehlers}, \&
  {Falco}}]{Schneider1992}
{Schneider}, P., {Ehlers}, J., \& {Falco}, E.~E. 1992, {Gravitational Lenses}
  (Springer)

\bibitem[{{Shapiro}(1964)}]{Shapiro1964}
{Shapiro}, I.~I. 1964, \prl, 13, 789

\bibitem[{{Somerville} \& {Dav{\'e}}(2015)}]{Somerville2015}
{Somerville}, R.~S. \& {Dav{\'e}}, R. 2015, \araa, 53, 51

\bibitem[{{Spergel} {et~al.}(2007){Spergel}, {Bean}, {Dor{\'e}}, {Nolta},
  {Bennett}, {Dunkley}, {Hinshaw}, {Jarosik}, {Komatsu}, {Page}, {Peiris},
  {Verde}, {Halpern}, {Hill}, {Kogut}, {Limon}, {Meyer}, {Odegard}, {Tucker},
  {Weiland}, {Wollack}, \& {Wright}}]{Spergel2007}
{Spergel}, D.~N., {Bean}, R., {Dor{\'e}}, O., {et~al.} 2007, \apjs, 170, 377

\bibitem[{{Spergel} {et~al.}(2003){Spergel}, {Verde}, {Peiris}, {Komatsu},
  {Nolta}, {Bennett}, {Halpern}, {Hinshaw}, {Jarosik}, {Kogut}, {Limon},
  {Meyer}, {Page}, {Tucker}, {Weiland}, {Wollack}, \& {Wright}}]{Spergel2003}
{Spergel}, D.~N., {Verde}, L., {Peiris}, H.~V., {et~al.} 2003, \apjs, 148, 175

\bibitem[{{Springel}(2010)}]{Springel2010}
{Springel}, V. 2010, \mnras, 401, 791

\bibitem[{{Springel} {et~al.}(2018){Springel}, {Pakmor}, {Pillepich},
  {Weinberger}, {Nelson}, {Hernquist}, {Vogelsberger}, {Genel}, {Torrey},
  {Marinacci}, \& {Naiman}}]{Springel2018}
{Springel}, V., {Pakmor}, R., {Pillepich}, A., {et~al.} 2018, \mnras, 475, 676

\bibitem[{{Springel} {et~al.}(2005){Springel}, {White}, {Jenkins}, {Frenk},
  {Yoshida}, {Gao}, {Navarro}, {Thacker}, {Croton}, {Helly}, {Peacock}, {Cole},
  {Thomas}, {Couchman}, {Evrard}, {Colberg}, \& {Pearce}}]{Springel2005}
{Springel}, V., {White}, S. D.~M., {Jenkins}, A., {et~al.} 2005, \nat, 435, 629

\bibitem[{{Springel} {et~al.}(2001{\natexlab{a}}){Springel}, {White}, {Tormen},
  \& {Kauffmann}}]{Springel2001}
{Springel}, V., {White}, S. D.~M., {Tormen}, G., \& {Kauffmann}, G.
  2001{\natexlab{a}}, \mnras, 328, 726

\bibitem[{{Springel} {et~al.}(2001{\natexlab{b}}){Springel}, {White}, {Tormen},
  \& {Kauffmann}}]{Springel2001a}
{Springel}, V., {White}, S. D.~M., {Tormen}, G., \& {Kauffmann}, G.
  2001{\natexlab{b}}, \mnras, 328, 726

\bibitem[{{Steidel} {et~al.}(2014){Steidel}, {Rudie}, {Strom}, {Pettini},
  {Reddy}, {Shapley}, {Trainor}, {Erb}, {Turner}, {Konidaris}, {Kulas}, {Mace},
  {Matthews}, \& {McLean}}]{Steidel2014}
{Steidel}, C.~C., {Rudie}, G.~C., {Strom}, A.~L., {et~al.} 2014, \apj, 795, 165

\bibitem[{{Steinhardt} {et~al.}(1999){Steinhardt}, {Wang}, \&
  {Zlatev}}]{Steinhardt1999}
{Steinhardt}, P.~J., {Wang}, L., \& {Zlatev}, I. 1999, \prd, 59, 123504

\bibitem[{{Sunyaev} \& {Zeldovich}(1972)}]{Sunyaev1972}
{Sunyaev}, R.~A. \& {Zeldovich}, Y.~B. 1972, Comments on Astrophysics and Space
  Physics, 4, 173

\bibitem[{{Teimoorinia} {et~al.}(2016){Teimoorinia}, {Bluck}, \&
  {Ellison}}]{Teimoorinia2016}
{Teimoorinia}, H., {Bluck}, A. F.~L., \& {Ellison}, S.~L. 2016, \mnras, 457,
  2086

\bibitem[{{Tempel} {et~al.}(2016){Tempel}, {Kipper}, {Tamm}, {Gramann},
  {Einasto}, {Sepp}, \& {Tuvikene}}]{Tempel2016}
{Tempel}, E., {Kipper}, R., {Tamm}, A., {et~al.} 2016, \aap, 588, A14

\bibitem[{{Tremmel} {et~al.}(2017){Tremmel}, {Karcher}, {Governato},
  {Volonteri}, {Quinn}, {Pontzen}, {Anderson}, \& {Bellovary}}]{Tremmel2017}
{Tremmel}, M., {Karcher}, M., {Governato}, F., {et~al.} 2017, \mnras, 470, 1121

\bibitem[{{Trujillo-Gomez} {et~al.}(2011){Trujillo-Gomez}, {Klypin}, {Primack},
  \& {Romanowsky}}]{Trujillo2011}
{Trujillo-Gomez}, S., {Klypin}, A., {Primack}, J., \& {Romanowsky}, A.~J. 2011,
  \apj, 742, 16

\bibitem[{{Turner} {et~al.}(1984){Turner}, {Ostriker}, \& {Gott}}]{Turner1984}
{Turner}, E.~L., {Ostriker}, J.~P., \& {Gott}, III, J.~R. 1984, \apj, 284, 1

\bibitem[{{Vale} \& {Ostriker}(2004)}]{Vale2004}
{Vale}, A. \& {Ostriker}, J.~P. 2004, \mnras, 353, 189

\bibitem[{{Vegetti} {et~al.}(2014){Vegetti}, {Koopmans}, {Auger}, {Treu}, \&
  {Bolton}}]{Vegetti2014}
{Vegetti}, S., {Koopmans}, L.~V.~E., {Auger}, M.~W., {Treu}, T., \& {Bolton},
  A.~S. 2014, \mnras, 442, 2017

\bibitem[{{Vogelsberger} {et~al.}(2014{\natexlab{a}}){Vogelsberger}, {Genel},
  {Springel}, {Torrey}, {Sijacki}, {Xu}, {Snyder}, {Bird}, {Nelson}, \&
  {Hernquist}}]{Vogelsberger2014b}
{Vogelsberger}, M., {Genel}, S., {Springel}, V., {et~al.} 2014{\natexlab{a}},
  \nat, 509, 177

\bibitem[{{Vogelsberger} {et~al.}(2014{\natexlab{b}}){Vogelsberger}, {Genel},
  {Springel}, {Torrey}, {Sijacki}, {Xu}, {Snyder}, {Nelson}, \&
  {Hernquist}}]{Vogelsberger2014a}
{Vogelsberger}, M., {Genel}, S., {Springel}, V., {et~al.} 2014{\natexlab{b}},
  \mnras, 444, 1518

\bibitem[{{Walsh}(1979)}]{Walsh1979}
{Walsh}, M.~J. 1979, AIAA Journal, 17, 783

\bibitem[{{Warren} {et~al.}(2006){Warren}, {Abazajian}, {Holz}, \&
  {Teodoro}}]{Warren2006}
{Warren}, M.~S., {Abazajian}, K., {Holz}, D.~E., \& {Teodoro}, L. 2006, \apj,
  646, 881

\bibitem[{{Weinberger} {et~al.}(2017){Weinberger}, {Springel}, {Hernquist},
  {Pillepich}, {Marinacci}, {Pakmor}, {Nelson}, {Genel}, {Vogelsberger},
  {Naiman}, \& {Torrey}}]{Weinberger2017}
{Weinberger}, R., {Springel}, V., {Hernquist}, L., {et~al.} 2017, \mnras, 465,
  3291

\bibitem[{{Weinberger} {et~al.}(2018){Weinberger}, {Springel}, {Pakmor},
  {Nelson}, {Genel}, {Pillepich}, {Vogelsberger}, {Marinacci}, {Naiman},
  {Torrey}, \& {Hernquist}}]{Weinberger2018}
{Weinberger}, R., {Springel}, V., {Pakmor}, R., {et~al.} 2018, \mnras, 479,
  4056

\bibitem[{{Wolf} {et~al.}(2010){Wolf}, {Martinez}, {Bullock}, {Kaplinghat},
  {Geha}, {Mu{\~n}oz}, {Simon}, \& {Avedo}}]{Wolf2010}
{Wolf}, J., {Martinez}, G.~D., {Bullock}, J.~S., {et~al.} 2010, \mnras, 406,
  1220

\bibitem[{{Woo} {et~al.}(2013){Woo}, {Dekel}, {Faber}, {Noeske}, {Koo},
  {Gerke}, {Cooper}, {Salim}, {Dutton}, {Newman}, {Weiner}, {Bundy}, {Willmer},
  {Davis}, \& {Yan}}]{Woo2013}
{Woo}, J., {Dekel}, A., {Faber}, S.~M., {et~al.} 2013, \mnras, 428, 3306

\bibitem[{{Xue} {et~al.}(2008){Xue}, {Rix}, {Zhao}, {Re Fiorentin}, {Naab},
  {Steinmetz}, {van den Bosch}, {Beers}, {Lee}, {Bell}, {Rockosi}, {Yanny},
  {Newberg}, {Wilhelm}, {Kang}, {Smith}, \& {Schneider}}]{Xue2008}
{Xue}, X.~X., {Rix}, H.~W., {Zhao}, G., {et~al.} 2008, \apj, 684, 1143

\bibitem[{{Yang} {et~al.}(2009){Yang}, {Mo}, \& {van den Bosch}}]{Yang2009}
{Yang}, X., {Mo}, H.~J., \& {van den Bosch}, F.~C. 2009, \apj, 695, 900

\bibitem[{{Yang} {et~al.}(2007){Yang}, {Mo}, {van den Bosch}, {Pasquali}, {Li},
  \& {Barden}}]{Yang2007}
{Yang}, X., {Mo}, H.~J., {van den Bosch}, F.~C., {et~al.} 2007, \apj, 671, 153

\bibitem[{{Zheng} {et~al.}(2007){Zheng}, {Coil}, \& {Zehavi}}]{Zheng2007}
{Zheng}, Z., {Coil}, A.~L., \& {Zehavi}, I. 2007, \apj, 667, 760

\bibitem[{{Zinger} {et~al.}(2020){Zinger}, {Pillepich}, {Nelson}, {Weinberger},
  {Pakmor}, {Springel}, {Hernquist}, {Marinacci}, \&
  {Vogelsberger}}]{Zinger2020}
{Zinger}, E., {Pillepich}, A., {Nelson}, D., {et~al.} 2020, \mnras, 499, 768

\bibitem[{{Zwicky}(1933)}]{Zwicky1933}
{Zwicky}, F. 1933, Helvetica Physica Acta, 6, 110

\end{thebibliography}

\appendix

\section{Guide to DfL}

\begin{table*}
\begin{center}
\caption{INPUT DATA FORMAT}
\label{tabA1}
\begin{tabular}{c c c l } 
 \hline
Column Name & Format & Units & Comments:    \\ 
  \hline
  \hline \\
GAL\_ID   & int   & [Unitless] & Unique Galaxy Identifier   \\ 
Z\_SPEC & float & [Unitless] & Spectroscopic Redshift \\
RA & float & [DEG] & Right Ascension of Galaxy \\
DEC & float & [DEG] & Declination of Galaxy \\
STELLAR\_MASS & float & [$\log_{10} M_{\odot}$] & Stellar Mass of Galaxy \\\\
\multicolumn{4}{c} {--- stellar mass uncertainty choices, optional ---} \\\\

STELLAR\_MASS\_ERROR & float & [$\log_{10} M_{\odot}$] & 1$\sigma$ Point Error on Stellar Mass\\
STELLAR\_MASS\_PDF\_BINS & array[float] & [$\log_{10} M_{\odot}$] & PDF Bin Centers\\
STELLAR\_MASS\_PDF\_VALS & array[float] & [Unitless] & PDF Probabilities \\\\

 \hline
\end{tabular}
\label{tab:input-data}
\end{center}
\end{table*}

\noindent In this appendix we provide a brief guide to the {\sc python} package, Dark from Light (DfL), which is publicly available for use\footnote{\url{https://github.com/abluck/Dark-from-Light}}.\\

\noindent This software provides a fast, efficient, and straightforward method to go from standard spectroscopic survey data on galaxies to a physically motivated group catalog, with halo mass estimates and virial properties of the groups. See Section \ref{sec3} for a full discussion of the pipeline mechanics. Here we focus on the practicalities for using and running DfL.\\

\noindent Required function arguments:\\

\noindent input\_data\_path : str\\
A string specifying a path to the input data file, either
in ASCII or FITS format. A template for input column names, formats,
and units is presented in Table~\ref{tab:input-data}.\\

\noindent output\_data\_path : str\\
A string specifying a path to the output file in FITS format.
A summary of output names, formats, and units of columns
appended to the input file is listed in Table~\ref{tab:output-file}.\\

\noindent model\_type: str\\
A string specifying regression model to use in Random Forest
halo mass regression. Choice among "HYBRID", "EAGLE", and "TNG",
where "HYBRID" yields an average prediction between "EAGLE" and "TNG".
This enables one to construct model dependent halo mass estimates which are optimized to either EAGLE or IllustrisTNG, or else opt simply for the average of the two.\\

\noindent seed: int\\
An integer specifying the seed for random number generation. Required
to ensure reproducibility.\\

\noindent include\_error: bool\\
A boolean switch indicating a choice between Case 1 and Cases 2 \& 3 
in Section \ref{sec3}.\\

\begin{table*}
\begin{center}
\caption{OUTPUT DATA FORMAT}
\label{tabA2}
\begin{tabular}{ c c c l } 
 \hline
Column Name  & Format & Units & Comments:    \\ 
  \hline
  \hline \\
groupID  & int   & [Unitless] & Unique Group Identifier   \\ 
group\_total\_Mstar   & float   & [$\log_{10} M_{\odot}$] & Total Group Stellar Mass   \\ 
group\_count   & float   & [Unitless] & Number of Galaxies in Group   \\ 
Mhalo   & float   & [$\log_{10} M_{\odot}$] & Predicted Group Halo Mass ($M_{200}$)   \\ 
Mhalo\_2sigma\_neg   & float   & [$\log_{10} M_{\odot}$] & 2$\sigma$ Negative Uncertainty   \\ 
Mhalo\_1sigma\_neg   & float   & [$\log_{10} M_{\odot}$] & 1$\sigma$ Negative Uncertainty   \\ 
Mhalo\_1sigma\_pos   & float   & [$\log_{10} M_{\odot}$] & 1$\sigma$ Positive Uncertainty   \\ 
Mhalo\_2sigma\_pos   & float   & [$\log_{10} M_{\odot}$] & 2$\sigma$ Positive Uncertainty   \\  
Mhalo\_pdf   & array[float]   & [Unitless] & Output Probabilities for Halo Mass PDF   \\ 
Mhalo\_pdf\_bins   & array[float]  & [$\log_{10} M_{\odot}$] & Output Bin Centers for Halo Mass PDF   \\ 
 Rvir   & float   & [pkpc] & Virial Radius of Group ($R_{200}$)   \\ 
 Vvir   & float   & [km/s] & Virial Velocity of Group ($V_{200}$)   \\ 
 Tvir   & float   & [$\log_{10} K$] & Virial Temperature of Group ($T_{200}$)   \\ 
galaxy\_class   & int   & [Unitless] & 1 = Cen.; 0 = Core Sat.; -1 = Ext. Sat.   \\
 num\_iteration   & int   & [Unitless] & No. of DfL Iterations   \\ 
 Mhalo\_central\_only   & float   & [$\log_{10} M_{\odot}$] & Halo Mass Prediction from Central Only   \\\\
 
 \hline
\end{tabular}
\label{tab:output-file}
\end{center}
\end{table*}

\noindent Optional arguments:\\

\noindent In addition to the five necessary function arguments,
a user can specify the following optional features:\\

\noindent hubble\_param0: float [default = 70.0]\\
Hubble parameter value at redshift $z=0$ in units of km/s/Mpc
for a spatially-flat $\Lambda$CDM cosmology\\

\noindent omega\_matter0: float [default = 0.3]\\
Mean redshift $z=0$ fractional energy density of all forms of matter
for a spatially-flat $\Lambda$CDM cosmology\\

\noindent verbose: bool [default is False]\\
A boolean verbosity switch.\\

\noindent convergence\_tol: float [default = 0.05]\\
Required tolerance for halo mass convergence in the iterative
part of the group finding algorithm in units of dex.\\

\noindent iteration\_limit: int [default = 10]\\
Maximum number of iterations allowed per halo finding loop.\\

\noindent num\_montecarlo = int [default = 100]\\
Number of random draws for Monte Carlo stellar mass error
propagation (Cases 2 and 3).\\

\noindent zrange: tuple[float, float] [default is None]\\
Redshift range to impose on the input data.\\

\section{Halo mass estimation with FOF-AM}

In this sub-section we present the FOF-AM method, which stands for abundance matching applied to total group stellar masses, extracted from friends-of-friends group finding. A comparison of performance results in TNG and EAGLE from FOF-AM to DfL is provided in Table \ref{tab2}. Here we discuss the details of this methodology and present additional training and testing plots. We first discuss the optimization of the FOF algorithm (Appendix~\ref{appB11}), next we discuss the AM method (Appendix~\ref{appB12}), and then we discuss combining the two techniques (Appendix~\ref{appB13}). Thereafter we present performance tests of FOF-AM in halo mass estimation, and compare to DfL (Appendix~\ref{appB2}).

\subsection{FOF-AM method}
\label{appB1}

The FOF-AM method is essentially a three stage process. First we optimize the linking lengths in FOF for application to observational-like simulated data from TNG and EAGLE. Second, we learn a one-to-one mapping between total group stellar masses and halo masses in the simulations, via matching on cumulative comoving number densities. Finally, we apply the FOF algorithm followed by the abundance matching procedure to the observational-like testing data, in order to extract both group information and halo mass estimates.

\subsubsection{FOF group finder}
\label{appB11}

\begin{figure*}
\begin{centering}
\includegraphics[width=0.99\textwidth]{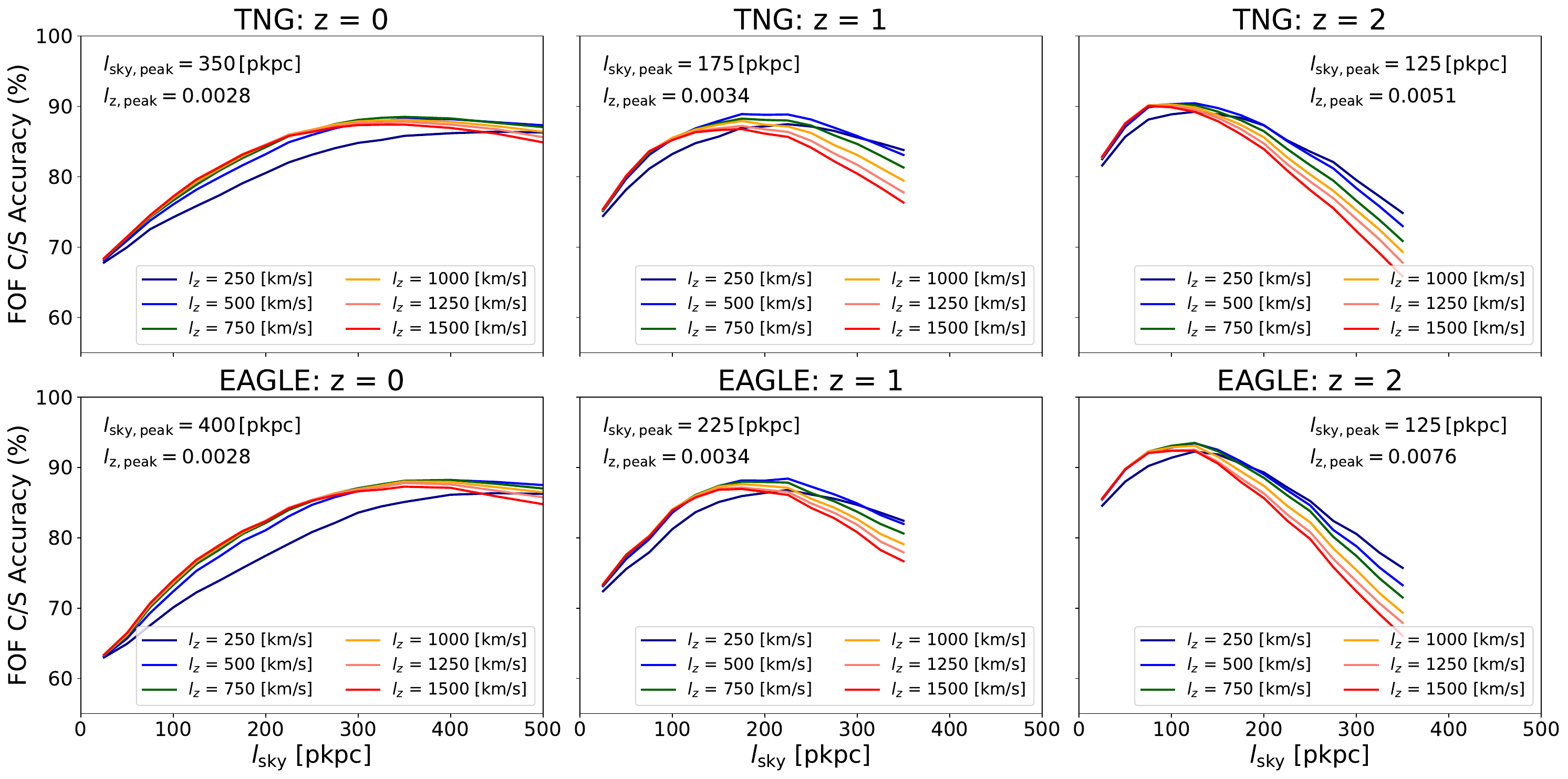}
\caption{Optimization of linking lengths for the friends-of-friends (FOF) group finding algorithm. The top row shows results for TNG, with the bottom row showing results for EAGLE. Each column displays a redshift snapshot (as displayed on the titles of each panel). Separate sky ($l_{\rm sky}$) and redshift ($l_{z}$) `lengths' are used as the maximum threshold for linking galaxies together in the FOF procedure. The $X$-axis displays varying $l_{\rm sky}$ values, which are measured in physical distances, with varying $l_{z}$ values shown by different colored lines (see legends), which are measured in velocity space. Optimization is performed with respect to the total accuracy of central - satellite classification (shown on the $Y$-axis). The optimal values are presented on each panel.}
\label{figB1}
\end{centering}
\end{figure*}

We follow an approach to FOF similar to \cite{Gerke2005, Knobel2009}; and \cite{Looser2021}. Galaxies are grouped together if they satisfy a linking-length criteria:

\begin{equation}
l_{\rm sky} < l_{\rm sky \,\, lim.} \,\,\,\,  \&  \,\,\,\, l_{z} < l_{\rm z \,\, lim.}.
\end{equation}

\noindent Different linking lengths are used on sky and in redshift space to account for the famous `fingers of god' effect (e.g., \citealt{Knobel2009}). Moreover, we use physical distances for sky and velocities for redshift linking lengths. These limits are usually determined as some fraction of the average separation between galaxies in a survey, which better enables application to other surveys with different depths. However, since we are running this on simulated data (which is stellar mass complete) we opt to optimize directly on physical distances and velocities. In essence, we test the potential of FOF to constrain groups in observer space (2D+$z$), compared to the full 6D phase space. 

In order to optimize the linking lengths we must choose a parameter to be optimized. Here we select the overall accuracy of central - satellite classification. We then systematically vary the on-sky physical and redshift linking lengths, running the FOF group finder for each configuration. We select the peak of the optimization curves as the optimal values to use in FOF-AM for comparison to DfL.

In Fig. \ref{figB1} we plot the FOF output central - satellite classification accuracy (as defined in eq. 28) as a function of $l_{\rm sky}$. We additionally split each curve into different values of $l_z$ (as labeled by the legends). We run this optimization procedure separately at different redshift snapshots (columns in Fig, B1) for TNG (top row) and EAGLE (bottom row). The peak of the optimization curve is then used to determine the optimal linking-lengths for each bespoke run of FOF-AM. The final sky and redshift limits used in our analysis are presented on each panel.

\subsubsection{AM technique}
\label{appB12}

\begin{figure*}
\begin{centering}
\includegraphics[width=0.9\textwidth]{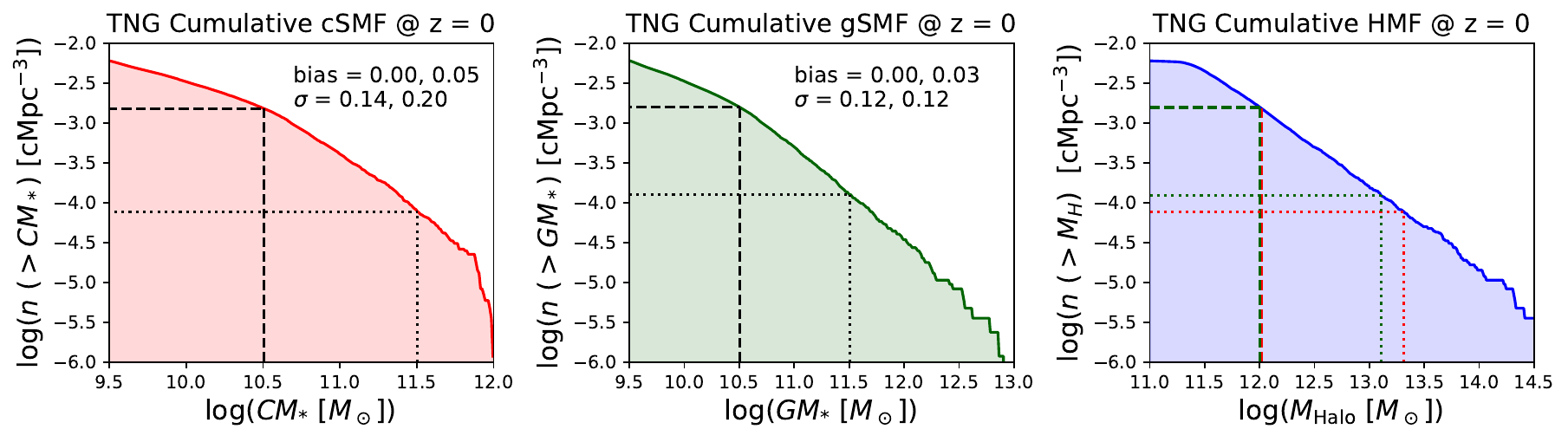}
\includegraphics[width=0.9\textwidth]{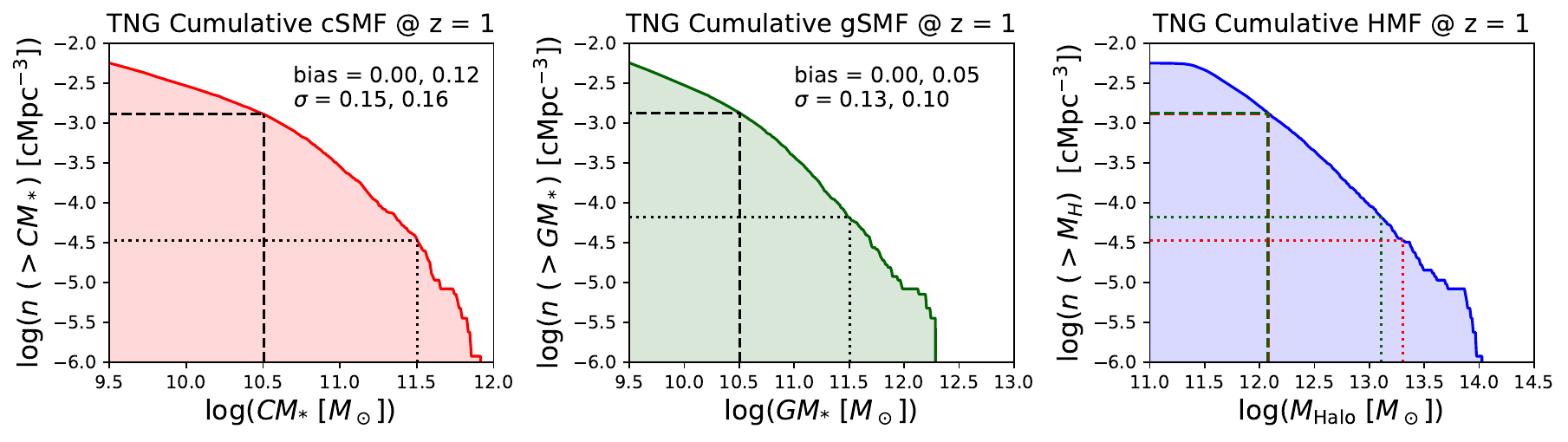}
\includegraphics[width=0.9\textwidth]{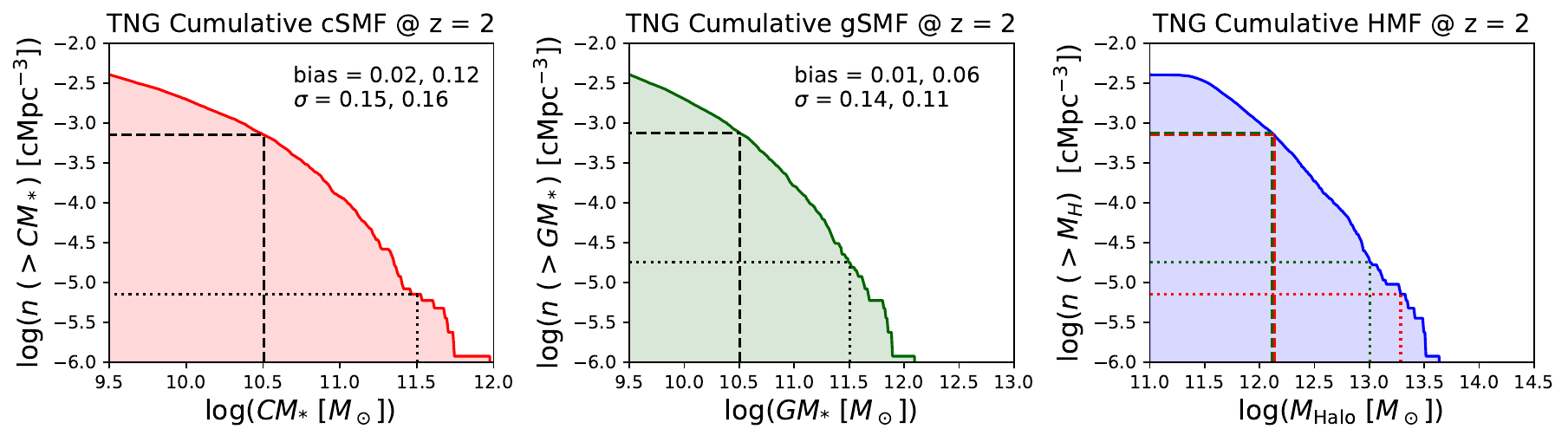}
\caption{Illustration of the abundance matching approach applied to centrals and groups for the TNG simulation. The rows show results from different redshift snapshots, and the columns indicate (from left to right): the cumulative central stellar mass function, the cumulative group total stellar mass function, and the cumulative halo mass function (explicitly computed for $M_{200}$). In abundance matching the cumulative number densities of either the central stellar mass or the group total stellar mass are matched to the cumulative number density of the halo mass function. Dashed and dotted lines indicate this matching procedure, and illustrate how the matching between central and group stellar mass is not in general identical (especially at high masses). In this method we train and test on different subsamples from within each simulation. Indicated on the stellar mass function panels are the bias and standard deviation in halo mass recovery from abundance matching within each approach. Additionally, the performance is given for all groups (first value) and for multi-galaxy groups (second value, after comma). Performing abundance matching on the total group stellar mass is superior to with the central stellar mass, and this is particularly evident for multi-galaxy groups (as one might expect).  }
\label{figB2}
\end{centering}
\end{figure*}

For the abundance matching technique, we follow an approach similar to \cite{Behroozi2010} and \cite{Moster2010}. In Fig. \ref{figB2} we show the cumulative comoving number densities for TNG galaxies as a function of central stellar mass (left panels), group total stellar mass (center panels), and halo mass (explicitly $M_{200}$, right panels). Results are shown for $z = 0$ (top row), $z = 1$ (middle row), and $z = 2$ (bottom row). We also perform an analogous procedure in EAGLE (not shown here for the sake of brevity). 

Leveraging the basic assumption of abundance matching, i.e. that the most massive galaxies (or groups) reside in the most massive haloes, one can link stellar and halo properties as follows. For central galaxies we define the AM halo mass as:

\begin{equation}
M_{\rm Halo-AM\,,\,i} = M_{\rm Halo \, List}\,[n(M_{\rm Halo\,List}) = n(CM_{*,i})]
\end{equation}

\noindent and for groups we define the AM halo mass as:

\begin{equation}
M_{\rm Halo-AM\,,\,i} = M_{\rm Halo\,List} \,[n(M_{\rm Halo\,List}) = n(GM_{*,i})],
\end{equation}

\noindent where $M_{\rm Halo-AM\,,\,i}$ indicates the AM estimate of halo mass for galaxy `$i$', $M_{\rm Halo\,List}$ indicates the list of halo masses from the cumulative distribution function (CDF) plot, and $n(M_{\rm Halo\,List})$ indicates the CDF number densities for each halo mass in the list. In eq. B2, $n(CM_{*,i})$ indicates the number density of the central galaxy stellar mass in question. In eq. B3, $n(GM_{*,i})$ indicates the number density of the group total stellar mass in question.

The AM procedure is much easier to understand visually than in abstract mathematics. For central galaxy abundance matching, to go from a stellar to a halo mass, simply read off the number density of the stellar mass from the CDF. Then read off the halo mass which has the same number density. Similarly, for total group stellar mass abundance matching, simply read off the number density of the total group stellar mass from the CDF. Then read off the halo mass which has the same number density. This procedure is illustrated with dashed and dotted lines in Fig. \ref{figB2}. Note that for low mass galaxies (and groups) the central and group AM procedure typically lead to similar (or identical) halo mass estimates. However, for high mass galaxies (and groups), the central and group AM halo mass predictions typically differ significantly.

On the panels of Fig. B2 we display the bias and standard deviation of halo mass recovery from AM for all groups (first number) and for multi-galaxy groups (second number, after comma). We show these results for central galaxy AM on the left panels, and for group AM on the center panels. It is clear that group abundance matching leads to superior results than central galaxy abundance matching, especially for multi-galaxy groups (see also, e.g., \citealt{Yang2007, Yang2009}). These results can be compared to those obtained in RF and ANN in Table \ref{tab1} in the main body of this paper.

\subsubsection{Combining FOF \& AM}
\label{appB13}

Having optimized the linking lengths for FOF (see Appendix \ref{appB11}) and having learned a mapping between central and group stellar mass to halo mass in simulations via AM (see Appendix \ref{appB12}), we are now in position to infer halo masses via FOF-AM. We first run FOF on the observational-like testing sample from each simulation at $z = 0, 1, 2$. We then extract the group total stellar mass from the FOF group catalogs. Thereafter, we use AM to convert the group total stellar masses into estimated halo masses. We also keep track of the central - satellite classification through FOF for further analyses. In the remainder of this appendix we explore the performance in halo mass estimation of FOF-AM, and compare to DfL.

\subsection{FOF-AM performance results}
\label{appB2}

\begin{figure*}
\begin{centering}
\includegraphics[width=0.95\textwidth]{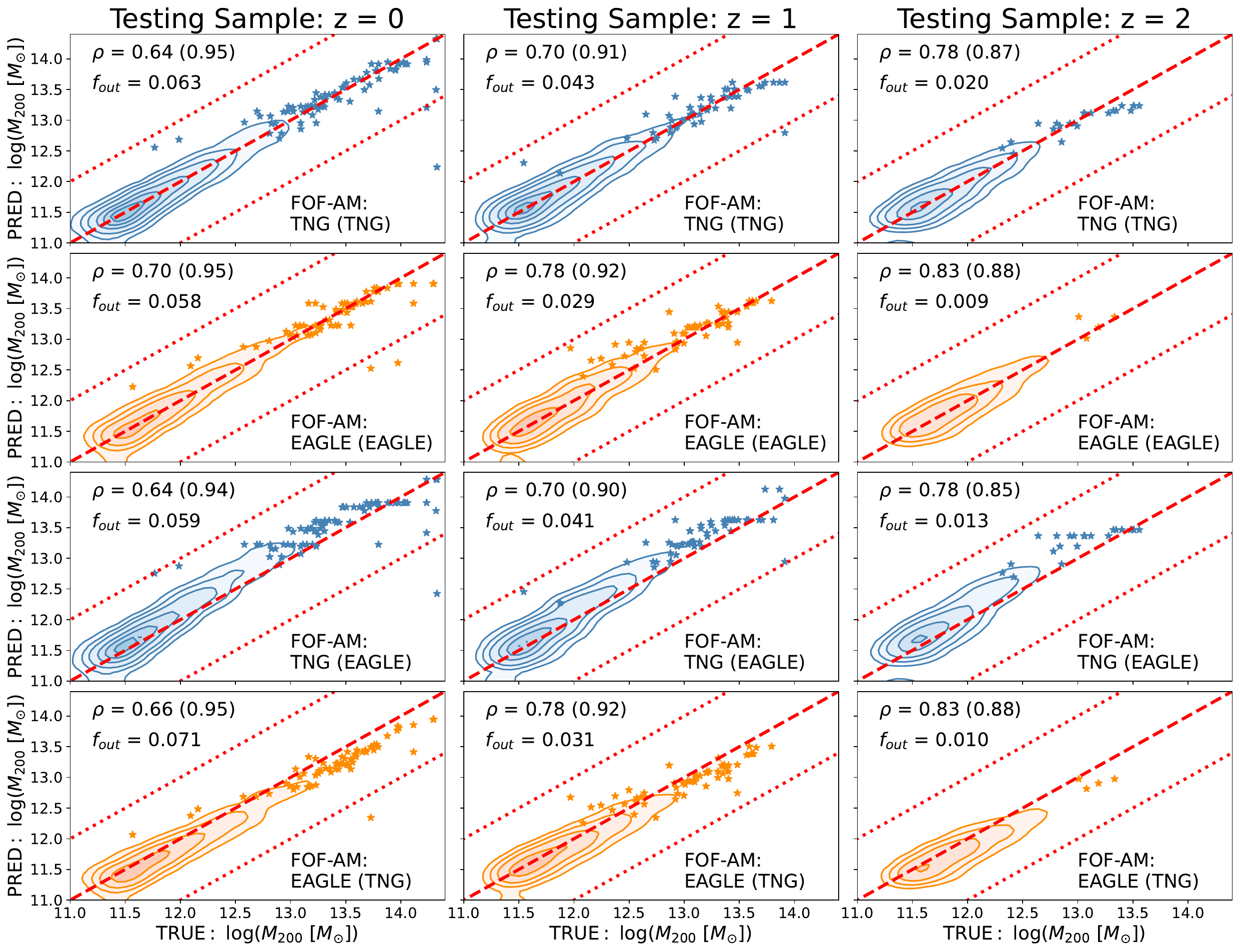}
\caption{Identical in structure to Fig. \ref{fig6}, but here showing results for halo mass estimation from the FOF-AM approach. Generally, a good recovery of halo masses is obtained by this approach, but the performance is significantly poorer than with DfL. Note also that similar biases are seen when changing the model - data pairing in FOF-AM to DfL. No uncertainties are shown in this figure since the abundance matching technique does not provide an obvious framework for error propagation when applied to groups (unlike in the DfL method).}
\label{figB3}
\end{centering}
\end{figure*}

\begin{figure*}
\begin{centering}
\includegraphics[width=0.95\textwidth]{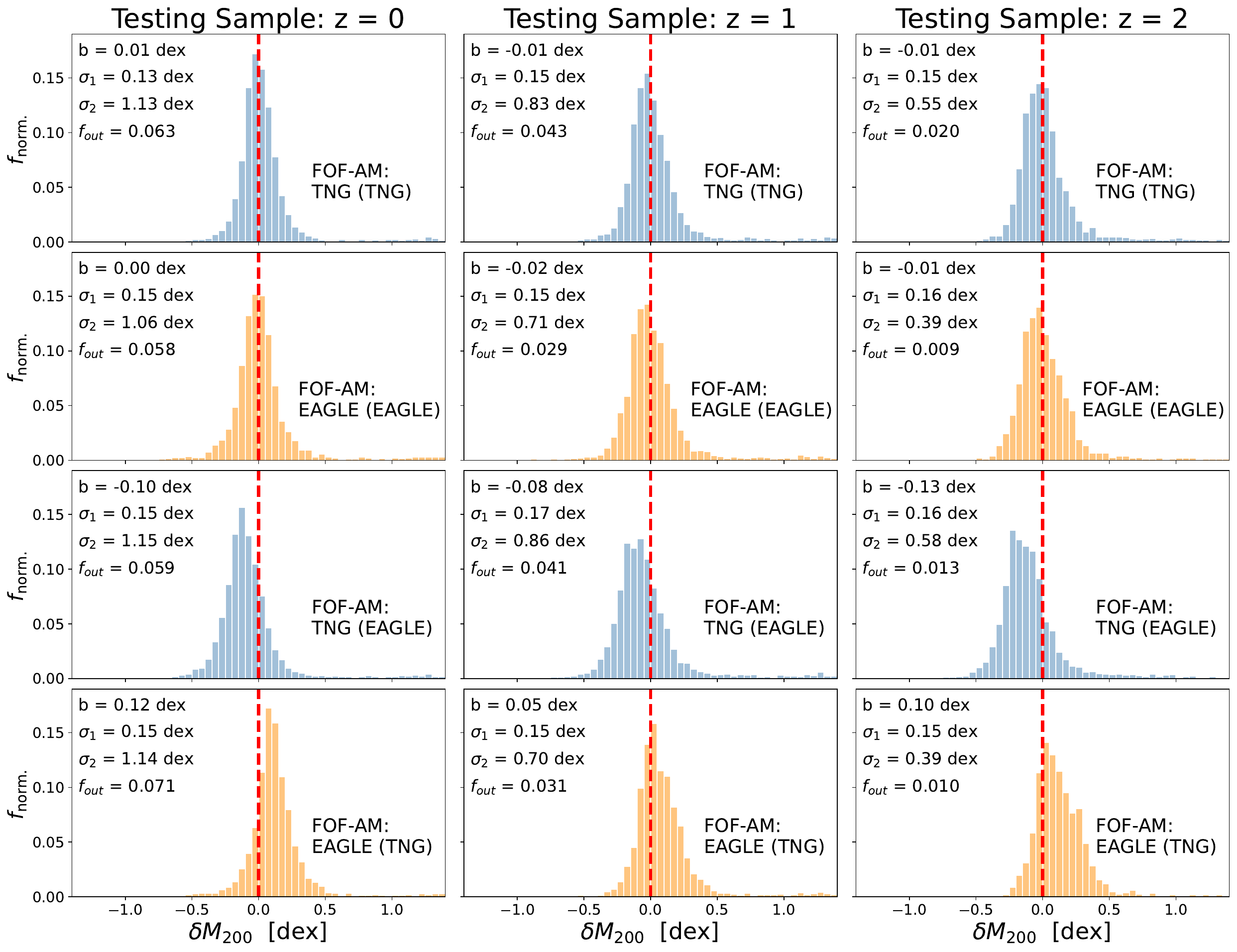}
\caption{Identical in structure to Fig. \ref{fig7}, but here showing results for halo mass estimation from the FOF-AM approach. Whilst many halo masses are accurately estimated via this technique, a significantly higher fraction of galaxies are found to be catastrophic outliers. This in turn leads to less accurate results across redshift ranges than seen in DfL (compare to Fig. 7, especially in terms of $\sigma_2$). }
\label{figB4}
\end{centering}
\end{figure*}

\begin{figure*}
\begin{centering}
\includegraphics[width=0.95\textwidth]{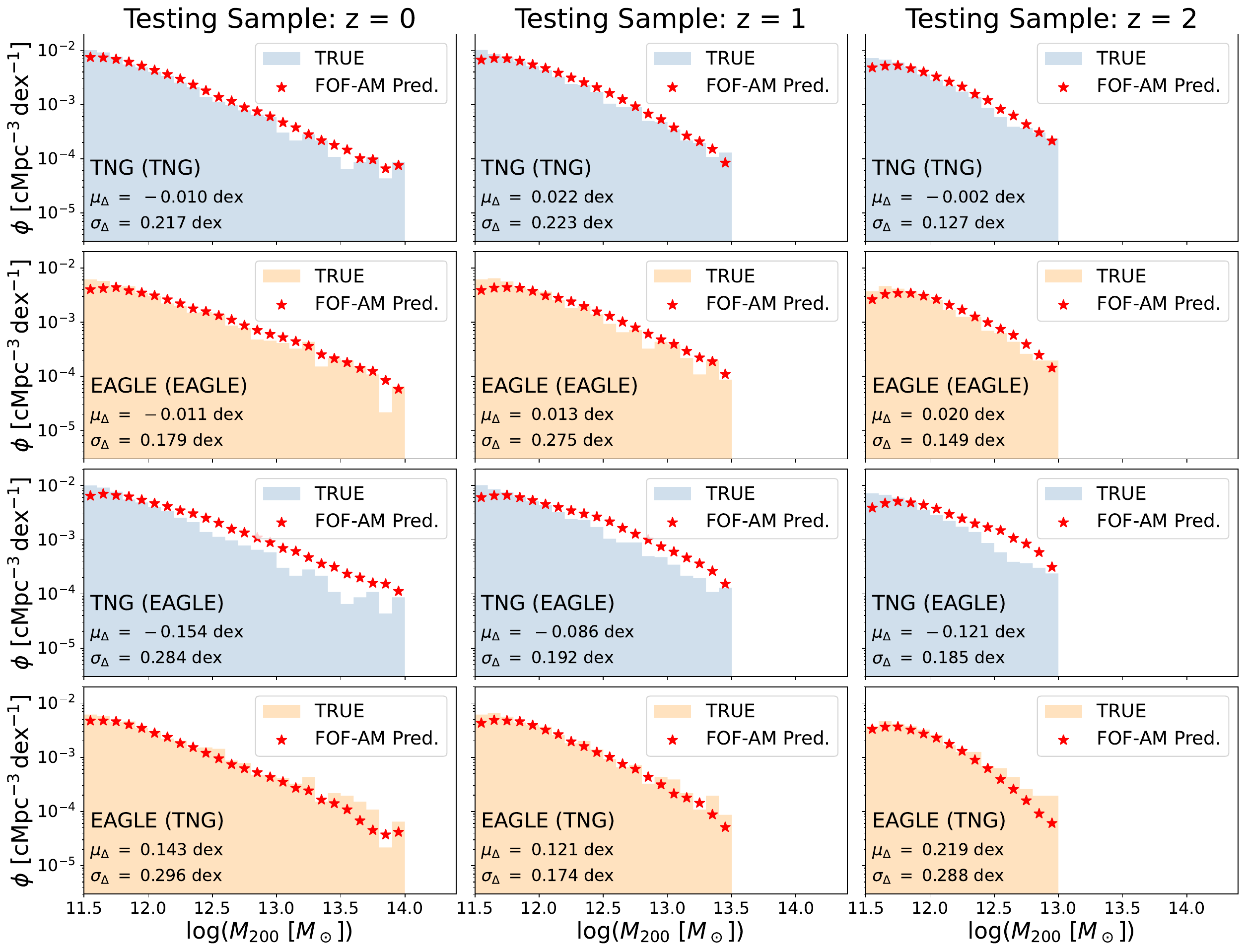}
\caption{Identical in structure to Fig. \ref{fig9}, but here showing results for halo mass estimation from the FOF-AM approach. Clearly, the FOF-AM approach recovers reasonable estimates of the halo mass functions across epochs. However, DfL achieves a significantly tighter reproduction (as quantified by the mean offsets and standard deviations, displayed on each panel, compared to Fig. \ref{fig9}).}
\label{figB5}
\end{centering}
\end{figure*}

\begin{figure*}
\begin{centering}
\includegraphics[width=0.95\textwidth]{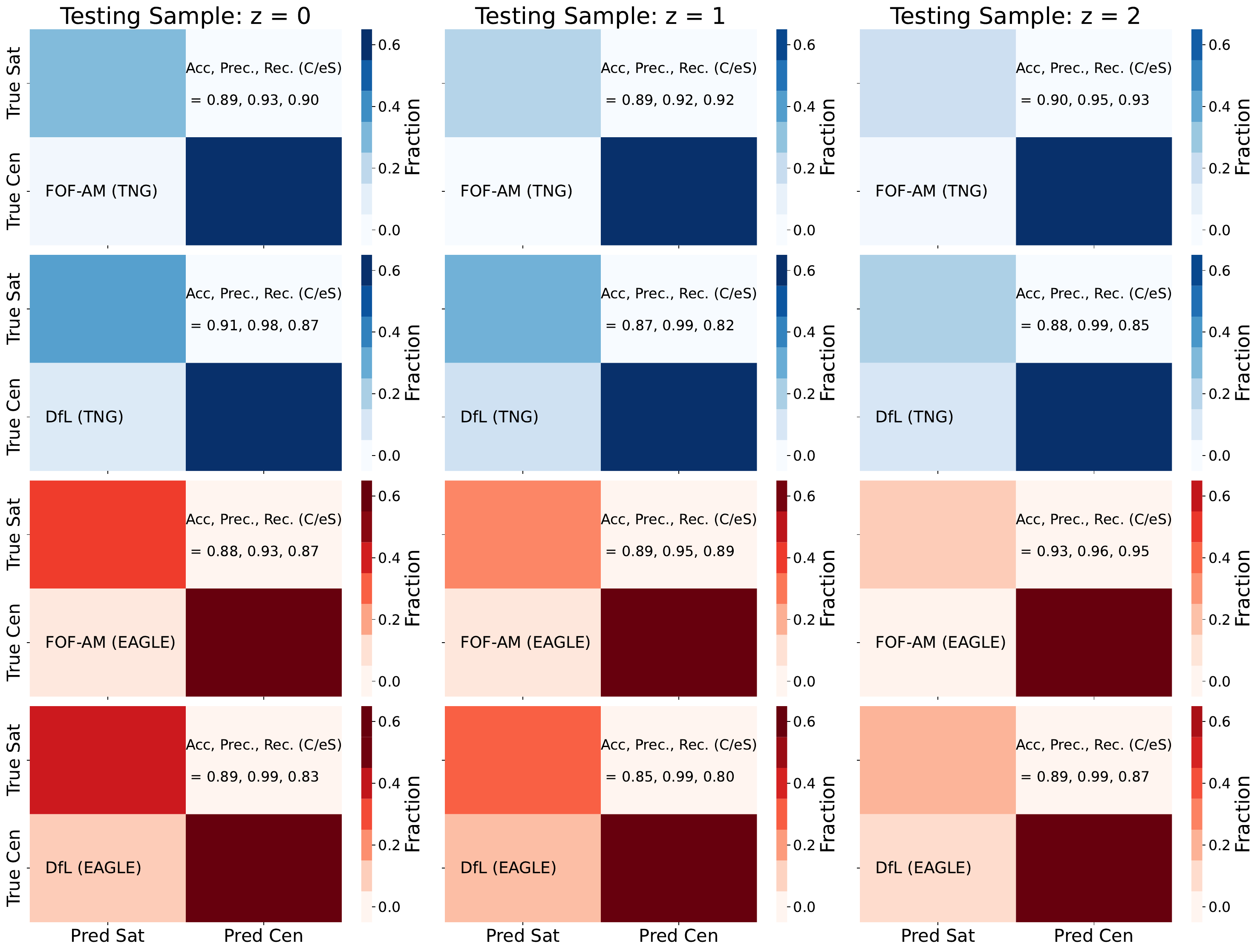}
\caption{Confusion matrices from FOF-AM compared to DfL for the same model and data pairings. Here the performance statistics are shown for the full sample of all galaxies, with galaxies in the extended satellite region treated as normal satellites. The overall accuracy of both methods are similar, but DfL achieves notably higher precision in the central galaxy sample, which is important for accurately identifying haloes. On the other hand, the recall in central galaxies is lower in DfL compared to FOF-AM, indicating a trade off between purity and completeness.}
\label{figB6}
\end{centering}
\end{figure*}

In Figs. \ref{figB3}, \ref{figB4} \& \ref{figB5} we reproduce Figs. \ref{fig6}, \ref{fig7} \& \ref{fig9} for the FOF-AM method. In Fig. \ref{figB3} we plot predicted vs. true halo masses derived from the FOF-AM methodology. As in DfL (see Fig. \ref{fig6}), we find relatively narrow contour distributions centered around the one-to-one relation. However, we note that the correlation strengths are significantly lower, and the catastrophic outlier fractions are significantly higher, than in DfL (see statistics presented on panels). However, the correlation strengths rise to similar values to that found in DfL in the case of excluding catastrophic outlier (statistics shown in parenthesis). This indicates that FOF-AM is similarly accurate to DfL for predicting halo masses for the majority of the sample, but is significantly worse for a relatively small sub-sample ($\sim$5\,\% of the total).

In Fig. \ref{figB4} we plot the distributions of $\delta M_{200}$, as defined in eq. 27 (but here for the FOF-AM method). The structure of Fig. \ref{figB4} is identical to Fig. \ref{fig7}. We find that the FOF-AM method yields essentially bias-free results for the same data and model pairings, as in DfL. Additionally, we find similar (although slightly broader) $\sigma_1$ values in FOF-AM, compared to DfL. Most interestingly, we find that the $\sigma_2$ values are much higher from FOF-AM than in DfL. This is a direct consequence of FOF-AM having higher catastrophic outlier fractions than DfL (see also Fig. \ref{figB3}). This is discussed in Section \ref{sec4}.

In Fig. \ref{figB5} we show the FOF-AM estimated halo mass functions compared to the simulation truth. As in DfL, visually there is a reasonable correspondence between the two distributions. However,  looking in more detail one can immediately see that DfL achieves a higher fidelity (as confirmed by the offset statistics displayed on each panel in comparison to Fig. \ref{fig9}).

Finally, in Fig. \ref{figB6} we compare the confusion matrices for FOF-AM against DfL directly, for the case of matched model and data pairings. In the case of central - satellite classification in DfL we found no significant differences with the mismatched model - data pairings, so we exclude that here for the sake of brevity. Here we present the full central - (all) satellite confusion matrices for DfL because this is a fairer comparison to FOF-AM, in which selecting sub-regions within proto-groups is not well motivated (see Section \ref{sec4} for further discussion on this).

In Fig. \ref{figB6} we find that the overall accuracy in central - satellite classification is similar between DfL and FOF-AM. However, the precision in central galaxy identification is notably higher in DfL compared to FOF-AM. Ultimately, this helps to explain the enhanced performance of DfL in the wings of the $\delta M_{200}$ distributions (commented on above). In the case of DfL, fewer groups are erroneously constructed than in FOF-AM. The price for this, however, is that DfL performs a little worse than FOF-AM in terms of recall (i.e., completeness in the central galaxy population). Nonetheless, we emphasize that this limitation has little impact on the recovery of halo mass functions, which are in excellent accord with the simulation truth in DfL (see back to Fig. \ref{fig9}).

In summary, DfL and FOF-AM achieve similar halo mass recovery in terms of bias and $\sigma_1$. However, DfL significantly outperforms FOF-AM in terms of $\sigma_2$ and the $f_{\rm out}$. Hence, in terms of halo mass estimation, DfL is overall superior to FOF-AM as a methodology. Considering central - satellite classification, DfL and FOF-AM perform similarly well in overall accuracy. DfL is superior to FOF-AM in precision, yet FOF-AM is superior to DfL in recall. However, given the ability to remove the extended satellite population in DfL (which is not a natural feature of FOF-AM), one can recover a much higher performance in accuracy, precision, and recall in DfL than in FOF-AM (see Fig.~\ref{fig10}). This does come at the price of reducing the completeness of the satellite population, yet for many analyses this is not especially problematic. 

A side-by-side comparison of performance in halo mass estimation between FOF-AM and DfL is presented in Table \ref{tab2}, in the main body of the paper.

\section{Machine learning reproducibility}

\begin{table*}
\begin{center}
\caption{RF hyper-parameters for DfL}
\label{tabC1}
\begin{tabular}{ l c c c c c c c c  } 
 \hline \\
Simulation  & Method & $N_{\rm trees}$ & Max Depth & Max Features & T : V & MSL & $\sigma$ & $\Delta \sigma$ \\ \\
  \hline
  \hline \\
TNG (Central) & RF Regression  & 250  & 1000  & None & 70 : 30  & 8  & 0.15  & 0.03    \\ 
TNG (Group) & RF Regression  & 250  & 1000  & `sqrt' &  70 : 30  & 5  & 0.12  & 0.02    \\ 
TNG (All) & RF Regression  & 250  & 1000  & `sqrt' & 70 : 30  & 5  & 0.11  & 0.03    \\ \\

EAGLE (Central) & RF Regression  & 250  & 1000  & None & 70 : 30  & 10  & 0.15  & 0.02    \\ 
EAGLE (Group) & RF Regression  & 250  & 1000  & `sqrt' & 70 : 30  & 5  & 0.13  & 0.03   \\ 
EAGLE (All) & RF Regression  & 250  & 1000  & `sqrt' & 70 : 30  & 5  & 0.12  & 0.03   \\  \\

 \hline
\end{tabular}
\end{center}
\end{table*}

\begin{table*}
\begin{center}
\caption{ANN hyper-parameters for comparison to DfL method}
\label{tabC2}
\begin{tabular}{ l c c c c c c c c c c  } 
 \hline \\
Simulation  & Method & Solver & Activation & Hidden Layers & Learning Rate & Early Stop &  T : V & $\sigma$ & $\Delta \sigma$ \\ \\
  \hline
  \hline \\
TNG (Central) & ANN Reg  & `lbfgs'  & $tanh$  & (10 : 50 : 10)  & 0.001  & True & 70 : 30 & 0.14  & 0.00    \\ 
TNG (Group) & ANN Reg  & `lbfgs'  & $tanh$  & (10 : 50 : 10)  & 0.001  & True & 70 : 30 & 0.12  & 0.00    \\ 
TNG (All) & ANN Reg  & `lbfgs'  & $tanh$  & (10 : 50 : 10)  & 0.001  & True & 70 : 30 & 0.12  & 0.01    \\ \\

EAGLE (Central) & ANN Reg.  & `lbfgs'  & $tanh$  & (10 : 50 : 10)  & 0.001  & True & 70 : 30 & 0.14  & 0.01    \\ 
EAGLE (Group) & ANN Reg.  & `lbfgs'  & $tanh$  & (10 : 50 : 10)  & 0.001  & True & 70 : 30 & 0.13  & 0.01    \\ 
EAGLE (All) & ANN Reg.  & `lbfgs'  & $tanh$  & (10 : 50 : 10)  & 0.001  & True & 70 : 30 & 0.12  & 0.00    \\ \\

 \hline
\end{tabular}
\end{center}
\end{table*}

\noindent To facilitate reproduction of our machine learning results by others, we provide in this appendix details on the hyper-parameters used in the RF (see Table \ref{tabC1}) and ANN (see Table \ref{tabC2}) analyses. All regression models are constructed using {\small SCIKIT-LEARN}\footnote{\url{https://scikit-learn.org}} (\citealt{Pedregosa2011})

In the group and all parameter runs we set: {\it max\_features}=`sqrt'. This ensures that trees within the forest are different to one another by both random sampling of the available features and through bootstrapped random sampling of the training data (with return). In the case of the central galaxy runs (which have only two parameters: stellar mass and redshift), we set {\it max\_features}=`None'. This ensures that both parameters are always available to decision trees in training. Nonetheless, the trees remain distinct due to the bootstrapped random sampling. 

In \cite{Bluck2022}, we have demonstrated that the all parameter (`None') mode represents the most effective RF architecture for disentangling inter-correlated `nuisance' parameters from the underlying causal relations in simple models, simulations, and observations. However, note that this is not the default method of RF classification in {\small SCIKIT-LEARN}. In this work, we opt to use the more conventional (`sqrt') setting for two reasons. First, we seek the most stable predictions, and this mode is less sensitive to individual data. Second, this mode yields a more thorough exploration of the predicted parameter space, which is advantageous for estimation and propagation of uncertainties.

The full list of hyper-parameters used in RF analyses are listed in Table \ref{tabC1}, with the full list of hyper-parameters used in ANN analyses listed in Table \ref{tabC2}. In the remainder of this appendix we explain what the table headings mean. The table columns are as follows:\\

\noindent Method: Type of machine learning used. Here either RF Regression (Table \ref{tabC1}) or ANN Regression (Table \ref{tabC2}).\\

\noindent $N_{\rm trees}$: The number of decision trees used in the RF model. \\

\noindent Max Depth: The number of decision forks permitted within each decision tree in the RF models. \\

\noindent Max Features: The maximum number of features sent to each individual tree in the models. Here we use `None' for centrals and `sqrt' for all other data sets. The latter indicates that the integer closest to the square root of the total features is used. \\

\noindent Min-Samples-Leaf (MSL):  The MSL hyper-parameter sets the minimum number of data in a given node of a given decision tree needed to attempt further splits in the data. We optimize MSL to maximize accuracy in the validation sample, whilst attempting to avoid significant over-fitting through comparison in performance to the validation sample. In all models and data sets we require that the difference in performance between training validation is less that 0.03 dex, which we achieve via systematically varying MSL. This parameter is used only in the RF models. \\

\noindent Solver: The chosen method to update weights in the ANN back-propagation process. Here we use `lbdgs' for all ANN runs, which is an optimizer in the quasi-Newtonian method family. We found the best ANN performance in preliminary testing with this mode. However, we note that other methods perform similarly (e.g., ADAM and `sgd'). \\

\noindent Activation: The activation function for neuron firing in the ANN. Here we use the $tanh$ function throughout the ANN models. This function is often faster to train with than other popular choices.\\

\noindent Hidden Layers: The hidden layers of the fully connected ANN. In preliminary testing we explored a wide variety in complexity of the ANN structure. We found that our chosen structure (10 : 50 : 10) performed well on all data sets and led to no significant over-fitting. Increasing either the depth or complexity of the network does not lead to higher performance on the validation samples. Equally, decreasing the depth or complexity does not significantly lower the difference in performance in training vs. validation. Hence, we are confident that this hidden layer structure is appropriate for the task at hand. \\

\noindent Learning Rate: The initial step size in gradient descent. Note that we utilize adaptive learning in all cases, so this is not fixed throughout the process of minimization.\\

\noindent Early Stop: In all ANN runs we enable early-stopping, which permits the ANN solver to consider whether there is any improvement in a randomly generated 10\% of the training sample (not used for training in that step), before deciding whether to continue in gradient descent further. This method is proven to reduce the potential for over-fitting, and clearly works well in these data (see $\Delta \sigma$ values in table \ref{tabC2}).\\

\noindent T : V (Train : Validate): The ratio of data sizes used in training and validating. Note that this is completely separate to the 50\% of the data which is converted into an observational-like form, and used to test DfL in Section 4. We emphasize again that the testing data is completely unseen in either training or validation.  \\

\noindent $\sigma$: The standard deviation of offsets in true-to-predicted halo masses in the validation sample (given in [dex]).\\

\noindent $\Delta \sigma$: The difference in standard deviation between training and validation samples (given in [dex]).\\

\end{document}